\documentstyle[amsart12,righttag,amscd,amssymb,epsf]{amsart}

\renewcommand{\le}{\leqslant}
\renewcommand{\ge}{\geqslant}

\newtheorem{theorem}{Theorem}[subsection]
\newtheorem{corollary}[theorem]{Corollary}
\newtheorem{lemma}[theorem]{Lemma}
\newtheorem{proposition}[theorem]{Proposition}

\theoremstyle{definition}
\newtheorem{definition}{Definition}[subsection]
\newtheorem{acknowledgement}{Acknowledgements}

\newtheorem{exercise}{Exercise}
\newtheorem{easyexercise}[exercise]{Easy exercise}

\theoremstyle{remark}
\newtheorem{remark}{Remark}[subsection]

\newtheorem{notation}{Notation}
\newtheorem{warning}{Warning}

\numberwithin{equation}{subsection}

\newcommand{\thmref}[1]{Theorem~\ref{#1}}
\newcommand{\secref}[1]{Section~\ref{#1}}
\newcommand{\lemref}[1]{Lemma~\ref{#1}}
\newcommand{\propref}[1]{Proposition~\ref{#1}}

\newcommand{\defref}[1]{Definition~\ref{#1}}
\newcommand{\corref}[1]{Corollary~\ref{#1}}

\newcommand{\figref}[1]{Figure~\ref{#1}}

\newcommand{\C}{\Bbb C}
\newcommand{\CC}{{\cal C}}
\newcommand{\F}{\Bbb F}
\newcommand{\G}{\Bbb G}
\newcommand{\K}{\Bbb K}
\newcommand{\Q}{\Bbb Q}
\newcommand{\R}{\Bbb R}
\newcommand{\Si}{\Sigma}
\newcommand{\SS}{\frak S}
\newcommand{\T}{\Bbb T}
\newcommand{\U}{\cal U}
\newcommand{\Z}{\Bbb Z}
\newcommand{\twoG}{2\Bbb G}

\newcommand{\al}{\alpha}
\renewcommand{\b}{\frak b}
\newcommand{\e}{\varepsilon}
\newcommand{\f}{{\bold f}}
\newcommand{\g}{\frak g}
\newcommand{\h}{\frak h}
\newcommand{\hh}{{\bold h}}
\newcommand{\esel}{\frak s\frak l}
\newcommand{\ka}{\kappa}
\newcommand{\si}{\sigma}
\newcommand{\so}{\frak o}
\newcommand{\espe}{\frak s\frak p}
\renewcommand{\u}{\bold u}

\newcommand{\one}{_{(1)}}
\newcommand{\two}{_{(2)}}
\newcommand{\tre}{_{(3)}}
\newcommand{\four}{_{(4)}}

\newcommand{\<}{\langle}
\renewcommand{\>}{\rangle}
\newcommand{\be}{\begin{equation}}
\newcommand{\jacobi}[2]{\left(\frac{#1}{#2}\right)}
\newcommand{\mt}[2]{{t_{#1}}^{#2}}

\renewcommand{\nmid}{\mbox{\,$\mid$\,\put(-7,1){$\scriptstyle\not$}}}
\newcommand{\nquad}{{\!\!\!\!\!\!}}

\newcommand{\nqqquad}{\nquad\nquad\nquad}
\newcommand{\op}{{\text{op}}}
\newcommand{\pa}{\partial}
\newcommand{\pti}{{}\spcheck}
\newcommand{\qqquad}{\quad\qquad}
\newcommand{\Ro}{\makebox[0pt]{\put(4,10){$\scriptscriptstyle\circ$}}R}
\newcommand{\tens}{\otimes}
\newcommand{\tr}{\triangleright}
\newcommand{\Surf}{Sur\!f}
\newcommand{\ueg}{u_\varepsilon(\frak g)}
\newcommand{\und}{\underline}
\newcommand{\uqg}{u_q(\frak g)}

\newcommand{\ad}{\operatorname{ad}}
\newcommand{\Ann}{\operatorname{Ann}}
\newcommand{\Aut}{\operatorname{Aut}}
\newcommand{\coev}{\operatorname{coev}}
\newcommand{\comod}{\operatorname{-comod}}
\newcommand{\Comod}{\operatorname{-Comod}}
\newcommand{\Cow}{\operatorname{Cow}}
\newcommand{\End}{\operatorname{End}}
\newcommand{\ev}{\operatorname{ev}}
\newcommand{\Hom}{\operatorname{Hom}}
\newcommand{\id}{\operatorname{id}}
\newcommand{\im}{\operatorname{Im}}
\newcommand{\Int}{\operatorname{Int}}
\newcommand{\Ker}{\operatorname{Ker}}
\newcommand{\modul}{\operatorname{-mod}}
\newcommand{\Ob}{\operatorname{Ob}}
\newcommand{\vect}{\operatorname{-vect}}
\newcommand{\Vect}{\operatorname{-Vect}}

\begin{document}

\title[3-invariants and representations of mapping class groups]
{Invariants of 3-manifolds and projective representations
of mapping class groups\\ via quantum groups at roots of unity}

\author[V. V. Lyubashenko]{Volodimir V. Lyubashenko}

\address
{Department of Mathematics\\ University of York\\
Heslington, York, YO1 5DD\\ England, U.K.}
\email{vvl1@@unix.york.ac.uk}
\thanks
{This work was supported in part by the SERC research grant GR/G 42976.}

\date {6 November 1994}

{\hfill{hep-th/9405167}}

\bigskip

\maketitle

\begin{abstract}
An example of a finite dimensional factorizable ribbon Hopf $\C$-algebra
is given by a quotient $H=u_q(\g)$ of the quantized universal enveloping
algebra $U_q(\g)$ at a root of unity $q$ of odd degree.
The mapping class group $M_{g,1}$ of a surface of genus $g$ with one
hole projectively acts by automorphisms in the $H$-module $H^{*\otimes g}$,
if $H^*$ is endowed with the coadjoint $H$-module structure.
There exists a projective representation of the mapping class group
$M_{g,n}$ of a surface of genus $g$ with $n$ holes labelled by finite
dimensional $H$-modules $X_1,\dots,X_n$ in the vector space
$\Hom_H(X_1\otimes\dots\otimes X_n,H^{*\otimes g})$.
An invariant of closed oriented 3-manifolds is constructed.
Modifications of these constructions for a class of ribbon Hopf algebras
satisfying weaker conditions than factorizability (including most of
$u_q(\g)$ at roots of unity $q$ of even degree) are described.
\end{abstract}

After works of Moore and Seiberg~\cite{MooSei}, Witten~\cite{Wit:Jones},
Reshetikhin and Turaev~\cite{ResTur:3}, Walker~\cite{Wal},
Kohno~\cite{Ko:inv,Ko:3man} and Turaev~\cite{Tur:q3} it became clear that any
semisimple abelian ribbon category with finite number of simple objects
satisfying some non-degeneracy condition gives rise to projective
representations of mapping class groups of surfaces as well as to
invariants of closed 3-manifolds. It was proposed in \cite{Lyu:r=m} to
get rid of semisimplicity and so extend the class of categories which
serves as the set of labels for a modular functor.

In this article we describe (eventually non-semisimple) ribbon Hopf
algebras $H$, whose modules form a category with the required properties,
thereby giving representations of mapping class groups. These algebras are
called 2-modular. All finite dimensional factorizable ribbon Hopf algebras
have those properties.

As a byproduct we obtain a projective representation of the mapping class
group $M_{g,1}$ of a surface of genus $g$ with one hole in the vector
space $H^{*\otimes g}$. If $H^*$ is endowed with the coadjoint $H$-module
structure, $M_{g,1}$ acts by automorphisms of the $H$-module. For genus 1
and factorizable Hopf algebras this representation was obtained by Majid
and the author~\cite{LyuMaj}. In the case of Drinfeld's doubles another
proof of modular relations for genus 1 was given by Kerler~\cite{Ker:mcg}.
The projective representations of $M_{1,1}$ thus obtained for finite
dimensional $H=u_q(\esel(2))$ are very close to those of Crivelli,
Felder and Wieczerkowski~\cite{CFW:mcg}, which come from conformal
field theories on the torus based on $SU(2)$.

Finite dimensional quotients $u_q(\g)$ of quantized universal enveloping
algebras $U_q(\g)$ at roots $q$ of unity are studied as an example.
We show that if the degree $l$ of the root $q$ of unity is odd, the
algebra is factorizable, if it is even, then $u_q(\g)$ will be 2-modular
or not, depending on arythmetical properties of $q$.

We describe also an intermediate class of categories and Hopf algebras
between factorizable and 2-modular ones. These categories and Hopf algebras
are called 3-modular and they give rise to invariants of closed oriented
3-manifolds. In the case of semisimple factorizable categories this is
the Reshetikhin--Turaev invariant \cite{ResTur:3}. In the case of
Hopf algebras it turns out to be the Hennings' invariant \cite{Hen:3}
(in the form of Kauffman and Radford \cite{KauRad:3}).

The results of the paper apply to both main known classes of ribbon
categories: semisimple ones and the categories of all modules over a
ribbon Hopf algebra. For the former we obtain already known results,
the latter gives new representations. Although the
language of abelian tensor categories is the most suitable for our
purposes, the reader is advised to restrict the consideration
to the categories of modules.

Conformal field theories is a powerful source of ribbon categories.
Kazhdan and Lusztig constructed non-trivial braided tensor subcategories
in the category of modules over an affine Lie algebra \cite{KazLus:ten},
motivated by CFT. These categories can be semisimple as well as not.
Futhermore, Gaberdiel \cite{Gab:chi} associates with each CFT a braided
tensor category, namely, the category of modules over the chiral
symmetry algebra with a non-standard tensor product. By the very nature
of CFT one expects \cite{MooSei} appearance of representations of
mapping class groups (this is obvious for TQFT). It turns out
\cite{Lyu:r=m} that such representations can be constructed from ribbon
categories even if they are not describing the fusion in some CFT.

Turaev proved that in semisimple case the modular functor extends to a
TQFT \cite{Tur:q3}. Moreover, he constructs the modular functor
(that is, representations of mapping class groups) as a part of a bigger
functor (TQFT), assigning linear maps to 3-cobordisms. In non-semisimple
case the word-by-word repetition of his approach is impossible, which
forces one to seek for a direct construction of the modular functor as
was done in \cite{Lyu:r=m}. Besides, some remnants of TQFT-structure
survive in non-semisimple case; this is under consideration now.

We recall basic facts about ribbon abelian categories in
\secref{Introduction}. The quantum Fourier transform is discussed in
\secref{transformations}. Ordinary ribbon Hopf algebras produce braided
Hopf algebras in \secref{Hopf}. We construct finite dimensional ribbon
Hopf algebras $u_q(\g)$ in \secref{groups} and single out factorizable
and 3-modular ones in \secref{Examples}. Representations of mapping class
groups are described in \secref{mapping}. New invariants of closed
3-manifolds are proposed in \secref{Invariants}.

\begin{acknowledgement}
I am grateful to R.~R. Hall, T.~H. Jackson and A.~Sudbery for useful
discussions of number theory questions, which arose in this paper.
I thank C.~De~Concini for discussions of 3-manifold invariants and
the structure of quantum groups at roots of unity.
\end{acknowledgement}

\section{Introduction}\label{Introduction}
\subsection{Notations and conventions}
$k$ denotes a field. In this paper a {\em Hopf algebra} means a
$k$-bialgebra with an invertible antipode.
Associative comultiplication is denoted $\Delta x= x\one\tens x\two$,
counity is denoted by $\e$, antipode in Hopf algebras is denoted $\gamma$.
If $H$ is a Hopf algebra, $H^\op$ denotes the same coalgebra $H$ with
opposite multiplication, $H_\op$ denotes the same algebra $H$ with the
opposite comultiplication. The category of $H$-modules is denoted $H$-Mod,
and its subcategory of finite dimensional $H$-modules is denoted $H\modul$.
In particular case $H=k$ we use $k\Vect$ and $k\vect$ respectfully.
The category of $H$-comodules is denoted $H\Comod$, and its subcategory
of finite dimensional $H$-comodules is denoted $H\comod$.
The left adjoint action of $x\in H$ in a Hopf algebra $H$ means
\[ \ad x.y = x\one y\gamma(x\two), \]
where $y\in H$.

Let $X$ be an $H$-module, denote $X^*$ the space of linear functionals
on $X$. Denote by $X\pti$ and $\pti X$ the two different structures of
$H$-module on $X^*$, the former being $(h.\xi)(x) = \xi(\gamma^{-1}(h).x)$,
the latter being $(h.\xi)(x) = \xi(\gamma(h).x)$ for $h\in H$,
$\xi\in X^*$, $x\in X$.
Iterating this definition we get $X\pti\pti$, $\pti\pti X$. Notice that
$(\pti X)\pti$ and $\pti(X\pti)$ are naturally identified with $X$, so
we can use the general notation $X^{(m}\pti{}^)$, $m\in\Z$, such that
\[ \dots,\ X^{(-2}\pti{}^) = \pti\pti X,\ X^{(-1}\pti{}^) = \pti X,\
X^{(0}\pti{}^) = X,\ X^{(1}\pti{}^) = X\pti,\ X^{(2}\pti{}^) = X\pti\pti,
\ \dots \]

$\g$ will denote a complex semi-simple Lie algebra of rank $n$ with
Borel and Cartan subalgebras $\b_+$, $\b_-$, $\h$. The root lattice,
generated by the simple roots $\alpha_1,\dots,\alpha_n$, will be
denoted $Q$. The weight lattice, generated by the fundamental weights
$\omega_1,\dots,\omega_n$ is denoted $P$. We write the group $Q$ also
in multiplicative notations $K_\alpha = \alpha \in Q$, using
$K_i=\alpha_i$ as generators. There is a perfect bilinear pairing
\[ \<,\> : Q\times P\to \Z ,\qquad \< \alpha_i,\omega_j \> = \delta_{ij}. \]
The Cartan matrix $a_{ij}$ determines an inclusion
\[ Q \hookrightarrow P ,\qquad  \alpha_j = \sum_{i=1}^r a_{ij} \omega_i ,\]
and the inner product
\[ (\,|\,) : Q\times Q \to\Z, \qquad (\al_i|\al_j) = d_i a_{ij} ,\]
where $d_i=1,2,3$.

$q$ will denote an indeterminate or a primitive $l^{\text{th}}$ root of
unity $q=\e\in\C$. This root of unity is assumed to satisfy $\e^{2m}\ne1$
for all $1\le m \le \max_i d_i$.
Let $l_i$ be the smallest positive integers such that $\e^{2d_i l_i}=1$.
We use the following notations for $q$-numbers:
\begin{alignat*}{2}
(n)_q &= {q^n-1 \over q-1}, &\qqquad [n]_q &= {q^n-q^{-n}\over q-q^{-1}},\\
(n)_q! &= \prod^n_{m=1}(m)_q, &\qqquad [n]_q! &= \prod^n_{m=1}[m]_q.
\end{alignat*}

$U_h(\g)$ (resp. $U_q(\g)$) is a topological $\C[[h]]$-algebra
(resp. $\Q(q)$-algebra), generated by $H_i$ (resp. $K_i^{\pm1}$), $E_i$,
$F_i$, the quantum group of Drinfeld~\cite{Dri:qua} and
Jimbo~\cite{Jim:qdif}. Equipped with the comultiplication
\begin{align*}
\Delta H_i &= H_i\tens1 + 1\tens H_i \\
\Delta E_i &= E_i\tens 1 + K_i^{-1}\tens E_i \\
\Delta F_i &= F_i\tens K_i + 1\tens F_i
\end{align*}
(in $U_h(\g)$ $K_i$ denotes $e^{hd_iH_i}$) these algebras become
Hopf algebras. The Lusztig's divided powers algebra $\Gamma(\g)$
\cite{Lus:roots1} is a $\Z[q,q^{-1}]$-subalgebra of $U_q(\g)$ generated
by $E_i^{(m)} = E_i^m/[m]_{q_i}!$, $F_i^{(m)} = F_i^m/[m]_{q_i}!$
and some Laurent polynomials of $K_i$, where $q_i=q^{d_i}$.

Choosing a reduced expression $s_{i_1}s_{i_2} \dots s_{i_N}$ of the longest
element $w_0$ of the Weyl group of $\g$, we get a total ordering of
the positive part $\Delta^+$ of the root system $\Delta$:
\[ \beta_1 = \alpha_{i_1},\ \beta_2 = s_{i_1}\alpha_2,\ \dots,\
\beta_N = s_{i_1} \dots s_{i_{N-1}} \alpha_{i_N} . \]
Following \cite{KirRes:mult,LevSoi:JGP,Lus:roots1} introduce the
corresponding root vectors in $\Gamma(\g)$
\[ E_{\beta_k} = T_{i_1} \dots T_{i_{k-1}} E_{i_k} ,\qquad
F_{\beta_k} = T_{i_1} \dots T_{i_{k-1}} F_{i_k} , \]
where $T_i: \Gamma(\g) \to \Gamma(\g)$ are Lusztig's automorphisms
\cite{Lus:roots1,Lus:book}. In the products like $\prod_\al E_\al^{m_\al}$
we always assume that $\al$ runs over $\Delta^+$ according to the above
total order. We use also $q_{\beta_k} = q_{i_k}$, $l_{\beta_k} = l_{i_k}$
and $E_\beta^{(m)} = E_\beta^m/[m]_{q_\beta}!$,
$F_\beta^{(m)} = F_\beta^m/[m]_{q_\beta}! \in \Gamma(\g)$.

An $R$-matrix will be often denoted $R=\sum_i R_i'\tens R_i''$.

\subsection{Ribbon abelian categories}\label{intribbon}
{\sl Ribbon} (also {\sl tortile} \cite{Shu}) category is the following
thing: a braided monoidal category $\CC$ \cite{JoyStr:tor} with the tensor
product $\tens$, the associativity $a:X\tens(Y\tens Z)\to (X\tens Y)\tens Z$,
the braiding (commutativity) $c:X\tens Y\to Y\tens X$ and a unity object $I$,
such that $\CC$ is rigid (for any object $X\in\CC$ there are dual objects
$\pti X$ and $X\pti$ with evaluations $\ev:\pti X\tens X\to I$,
$\ev:X\tens X\pti\to I$ and coevaluations $\coev:I\to X\tens\pti X$,
$\coev:I\to X\pti\tens X$) and possess a ribbon twist $\nu$. A ribbon twist
\cite{JoyStr:tor,ResTur:rg,Shu} $\nu=\nu_X:X\to X$ is a self-adjoint
($\nu_{X\pti}=\nu_X^t$) functorial automorphism such that
$c^2=\nu_X^{-1}\tens\nu_Y^{-1}\circ\nu_{X\tens Y}$.

Morphisms constructed from braidings and (co)evaluations are often
described by tangles. In conventions of \cite{Lyu:tan} we denote
{\allowdisplaybreaks
\begin{alignat*}2
\text{a morphism }& f:X\to Y  &&\qquad \text{by} \qquad
\unitlength=0.7mm
\raisebox{-8mm}{
\begin{picture}(9,27)
\put(4,10){\framebox(4,8)[cc]{}}
\put(1,14){\makebox(0,0)[cc]{$f$}}
\put(6,18){\line(0,1){8}}
\put(6,10){\line(0,-1){8}}
\put(9,1){\makebox(0,0)[cc]{$Y$}}
\put(9,27){\makebox(0,0)[cc]{$X$}}
\end{picture}
} ,\\
\text{the braiding }&c:X\tens Y \to Y\tens X &&\qquad \text{by} \qquad
\unitlength=0.8mm
\makebox[20mm][l]{
\raisebox{-7mm}[14mm][8mm]{
\put(5,0){\line(2,3){10}}
\put(5,15){\line(2,-3){4}}
\put(15,0){\line(-2,3){4}}
\put(5,17){\makebox(0,0)[cb]{$X$}}
\put(15,17){\makebox(0,0)[cb]{$Y$}}
}} ,\\
\text{the inverse braiding }&c^{-1}:X\tens Y \to Y\tens X
&&\qquad \text{by} \qquad
\unitlength=0.8mm
\makebox[20mm][l]{
\raisebox{-7mm}[14mm][8mm]{
\put(5,17){\makebox(0,0)[cb]{$X$}}
\put(15,17){\makebox(0,0)[cb]{$Y$}}
\put(5,15){\line(2,-3){10}}
\put(15,15){\line(-2,-3){4}}
\put(5,0){\line(2,3){4}}
}} ,\\
\text{the evaluation }&\ev:X\tens X\pti \to k &&\qquad \text{by} \qquad
\unitlength=1mm
\makebox[20mm][l]{
\raisebox{-5mm}[7mm][7mm]{
\put(10,7){\oval(10,14)[b]}
\put(5,9){\makebox(0,0)[cb]{$X$}}
\put(15,9){\makebox(0,0)[cb]{$X\pti$}}
}} ,\\
\text{the coevaluation }&\coev:k \to X\pti\tens X &&\qquad \text{by} \qquad
\unitlength=1mm
\makebox[20mm][l]{
\raisebox{-5mm}[8mm][7mm]{
\put(10,6){\oval(10,14)[t]}
\put(5,1){\makebox(0,0)[cb]{$X\pti$}}
\put(15,1){\makebox(0,0)[cb]{$X$}}
}} .
\end{alignat*}
}
Consistency of these notations is due to the functor $\Phi$ from the
category of $\CC$-colored tangles to the category $\CC$ itself
\cite{FreYet:coh}.

In a ribbon category there are functorial isomorphisms \cite{Lyu:tan}
\[ 
\unitlength=0.75mm
\linethickness{0.4pt}
\begin{picture}(146.33,35)
\put(23,18){\oval(10,10)[r]}
\put(23,20){\oval(6,6)[lt]}
\put(11,35){\makebox(0,0)[cc]{$X$}}
\put(11,1){\makebox(0,0)[cc]{$X\pti\pti$}}
\put(1,18){\makebox(0,0)[cc]{$u_1^2\ =$}}
\put(61,18){\oval(10,10)[r]}
\put(61,20){\oval(6,6)[lt]}
\put(49,35){\makebox(0,0)[cc]{$X$}}
\put(49,1){\makebox(0,0)[cc]{$X\pti\pti$}}
\put(39,18){\makebox(0,0)[cc]{, $u_{-1}^2\ =$}}
\put(92,20){\oval(6,6)[rt]}
\put(106,35){\makebox(0,0)[cc]{$X$}}
\put(106,1){\makebox(0,0)[cc]{$\pti\pti X$}}
\put(76,18){\makebox(0,0)[cc]{, $u_1^{-2}=$}}
\put(132,20){\oval(6,6)[rt]}
\put(146,35){\makebox(0,0)[cc]{$X$}}
\put(146,1){\makebox(0,0)[cc]{$\pti\pti X$}}
\put(116,18){\makebox(0,0)[cc]{, $u_{-1}^{-2}=$}}
\put(20,20){\line(-2,-3){9.33}}
\put(11,30){\line(2,-3){6.67}}
\put(58,16){\line(-2,3){9.33}}
\put(49,6){\line(3,5){6}}
\put(106,6){\line(-4,5){8}}
\put(135,20){\line(4,-5){11.33}}
\put(146,30){\line(-4,-5){8}}
\put(23,16.50){\oval(6,7)[lb]}
\put(61.50,16){\oval(7,6)[lb]}
\put(92,18){\oval(12,10)[l]}
\put(132,18){\oval(12,10)[l]}
\put(91.50,16){\oval(7,6)[rb]}
\put(95,16){\line(4,5){11.20}}
\put(131.50,16){\oval(7,6)[rb]}
\end{picture}
\]
\[u_0^2=u_1^2\circ\nu^{-1}=u_{-1}^2\circ\nu:X\to X\pti\pti,\qquad
u_0^{-2}=u_1^{-2}\circ\nu^{-1}=u_{-1}^{-2}\circ\nu:X\to\pti\pti X.\]
Changing the category $\CC$ by an equivalent one, we can (and we will)
assume that ${}\pti X =X\pti$, $X\pti\pti = {}\pti\pti X = X$ and
$u_0^2 = u_0^{-2} = \id_X$ (see \cite{Lyu:tan}).

\begin{warning}
In the category $\CC=H$-mod, where $H$ is a ribbon Hopf algebra, the
equation $X\pti = \pti X$ is not satisfied, nevertheless $X\pti$ is
canonically isomorphic to $\pti X$. We identify these modules via
$u_0^2: \pti X \to X\pti$ (see \secref{RibHop}).
\end{warning}

If in addition $\CC$ is additive, it is $k$-linear with $k=\End I$. We
assume in the following that $k$ is a field, in which each element has a
square root. (In fact we need a square root only for one element of $k$
which depends on $\CC$.) In this paper $\CC$ will be a noetherian
abelian category with finite dimensional $k$-vector spaces
$\Hom_{\CC}(A,B)$. (One more technical condition: isomorphism classes in
$\CC$ form a set.) In such case there exists a coend $F=\int X\tens X\pti$
as an object of a cocompletion $\hat{\CC}$ \cite{Lyu:tan} of $\CC$.
Recall that this coend can be defined via an exact sequence
\be\label{coend}
\bigoplus_{f:A\to B\in\CC} A\tens B\pti @>f\tens B\pti-A\tens f^t>>
\bigoplus_{L\in\CC} L\tens L\pti @>\oplus i_L>> F \to 0 ,
\end{equation}
where $f^t:B\pti \to A\pti$ is the transposed to a morphism $f:A\to B$.
For a general definition of a coend see \cite{Mac:cat}.

$F$ is a Hopf algebra in the category $\hat{\CC}$
(see \cite{Lyu:mod,LyuMaj,Maj:bra}). The multiplication $m_F:F\tens F \to F$
is described in \cite{Lyu:mod} by any of the following $\CC$-$F$-tangles
\be\label{mult1}
\nquad
\unitlength=0.7mm
\raisebox{-16mm}{
\begin{picture}(46,41)
\put(1,6){\line(0,1){30}}
\put(16,6){\line(1,2){15}}
\put(31,6){\line(1,2){15}}
\put(16,36){\line(1,-1){8}}
\put(28,24){\line(1,-1){6}}
\put(38,14){\line(1,-1){8}}
\put(1,38){\makebox(0,0)[cb]{$L$}}
\put(16,38){\makebox(0,0)[cb]{$L\pti$}}
\put(31,38){\makebox(0,0)[cb]{$M$}}
\put(46,38){\makebox(0,0)[cb]{$M\pti$}}
\put(1,1){\makebox(0,0)[cb]{$L$}}
\put(16,1){\makebox(0,0)[cb]{$M$}}
\put(31,1){\makebox(0,0)[cb]{$M\pti$}}
\put(46,1){\makebox(0,0)[cb]{$L\pti$}}
\end{picture}
} \quad \text{or} \quad
\begin{CD}
L\tens L\pti\tens (M\tens M\pti) @>i_L\tens i_M>> F\tens F \\
@VL\tens cVV                               @V\exists Vm_FV \\
L\tens M\tens(L\tens M)\pti      @>i_{L\tens M}>> F
\end{CD}
\end{equation}
\be\label{mult2}
\nquad
\unitlength=0.7mm
\raisebox{-16mm}{
\begin{picture}(46,38)
\put(1,38){\makebox(0,0)[cb]{$L$}}
\put(16,38){\makebox(0,0)[cb]{$L\pti$}}
\put(31,38){\makebox(0,0)[cb]{$M$}}
\put(46,38){\makebox(0,0)[cb]{$M\pti$}}
\put(1,1){\makebox(0,0)[cb]{$M$}}
\put(16,1){\makebox(0,0)[cb]{$L$}}
\put(31,1){\makebox(0,0)[cb]{$L\pti$}}
\put(46,1){\makebox(0,0)[cb]{$M\pti$}}
\put(31,36){\line(-1,-1){30}}
\put(46,6){\line(0,1){30}}
\put(16,36){\line(1,-2){4}}
\put(22,24){\line(1,-2){9}}
\put(16,6){\line(-1,2){4}}
\put(10,18){\line(-1,2){9}}
\end{picture}
} \quad \text{or} \quad
\begin{CD}
(L\tens L\pti)\tens M\tens M\pti @>i_L\tens i_M>> F\tens F \\
@Vc\tens M\pti VV                          @V\exists Vm_FV \\
M\tens L\tens(M\tens L)\pti      @>i_{M\tens L}>> F
\end{CD}
\end{equation}
The antipode $\gamma_F:F \to F$ is given by
\be\label{antgamma}
\unitlength=0.70mm
\linethickness{0.4pt}
\raisebox{-13mm}{
\begin{picture}(38,40)
\put(11,2){\line(0,1){11}}
\put(11,13){\line(6,5){17}}
\put(28,27){\line(0,1){12}}
\put(32.50,14.50){\oval(11,9)[r]}
\put(32,19){\line(-1,-4){4.33}}
\put(28,13){\line(-6,5){7}}
\put(18,21){\line(-6,5){7}}
\put(11,26.67){\line(0,1){12.33}}
\put(20,40){\makebox(0,0)[cc]{$F$}}
\put(20,1){\makebox(0,0)[cc]{$F$}}
\put(1,20){\makebox(0,0)[cc]{$\gamma_F =$}}
\end{picture}
}
\end{equation}

There is a Hopf pairing $\omega:F\tens F\to I$ \cite{Lyu:mod},
\be\label{omega}
\unitlength=1mm
\linethickness{0.4pt}
\raisebox{-9mm}{
\begin{picture}(61,19)
\put(40,5){\line(0,1){4}}
\put(40,9){\line(4,5){7.33}}
\put(61,9){\line(0,1){9}}
\put(40,13){\line(-4,5){4}}
\put(28,19){\makebox(0,0)[cc]{$F$}}
\put(54,19){\makebox(0,0)[cc]{$F$}}
\put(5,10){\makebox(0,0)[cc]{$\omega\ =$}}
\put(32,9){\oval(24,18)[b]}
\put(45,10){\oval(32,20)[rb]}
\put(20,8){\line(0,1){10}}
\end{picture}
}
\end{equation}
such that
\[\Ann\omega = \Ann^{\text{left}}\omega =
\Ann^{\text{right}}\omega\in\hat{\CC}.\]
The quotient $\f=F/\Ann\omega\in\hat{\CC}$ is also a Hopf algebra.

The  morphisms called monodromies
$\Omega_l=\Omega^l_{X,F} : X\tens F \to X\tens F$,
$\Omega_r=\Omega^r_{F,X} : F\tens X \to F\tens X$,
$\Omega=\Omega_{F,F} : F\tens F \to F\tens F$ are defined via tangles
\[
\unitlength=0.8mm
\begin{picture}(140,37)
\put(6,18){\makebox(0,0)[rc]{$\Omega_l =$}}
\put(28,33){\line(-4,-5){12}}
\put(16,18){\line(4,-5){4}}
\put(24,8){\line(4,-5){4}}
\put(16,33){\line(4,-5){4}}
\put(24,23){\line(4,-5){4}}
\put(28,18){\line(-4,-5){12}}
\put(36,3){\line(0,1){30}}
\put(12,35){\makebox(0,0)[cc]{$X$}}
\put(32,35){\makebox(0,0)[cc]{$F$}}
\put(32,1){\makebox(0,0)[cc]{$F$}}
\put(12,1){\makebox(0,0)[cc]{$X$}}
\put(56,18){\makebox(0,0)[rc]{,\quad $\Omega_r =$}}
\put(80,33){\line(-4,-5){12}}
\put(68,18){\line(4,-5){4}}
\put(76,8){\line(4,-5){4}}
\put(68,33){\line(4,-5){4}}
\put(76,23){\line(4,-5){4}}
\put(80,18){\line(-4,-5){12}}
\put(60,3){\line(0,1){30}}
\put(64,35){\makebox(0,0)[cc]{$F$}}
\put(84,35){\makebox(0,0)[cc]{$X$}}
\put(84,1){\makebox(0,0)[cc]{$X$}}
\put(64,1){\makebox(0,0)[cc]{$F$}}
\put(104,18){\makebox(0,0)[rc]{,\quad $\Omega =$}}
\put(128,33){\line(-4,-5){12}}
\put(116,18){\line(4,-5){4}}
\put(124,8){\line(4,-5){4}}
\put(116,33){\line(4,-5){4}}
\put(124,23){\line(4,-5){4}}
\put(128,18){\line(-4,-5){12}}
\put(108,3){\line(0,1){30}}
\put(136,3){\line(0,1){30}}
\put(112,35){\makebox(0,0)[cc]{$F$}}
\put(132,35){\makebox(0,0)[cc]{$F$}}
\put(132,1){\makebox(0,0)[cc]{$F$}}
\put(112,1){\makebox(0,0)[cc]{$F$}}
\put(140,17){\makebox(0,0)[cc]{.}}
\end{picture}
\]
They project to $\f$ as morphisms
\[ \Omega_l=\Omega^l_{X,\f} : X\tens \f \to X\tens \f ,\,\,\,
\Omega_r=\Omega^r_{\f,X} : \f\tens X \to \f\tens X, \,\,\,
\Omega=\Omega_{\f,\f} : \f\tens \f \to \f\tens \f \]
also called monodromies.

\subsection{2-modular categories}
The first modular axiom is \cite{Lyu:mod}

(M1) $\f$ is an object of $\CC$ (and not only of a cocompletion $\hat{\CC}$)

\noindent (more scrupulously, it means that there exists an exact sequence
$0\to\Ann\omega\to F\to \f \to 0$ in $\hat{\CC}$, where $\f$ is an
object from $\CC\subset\hat{\CC}$).

Being the coend $\int X\tens X\pti$, the object $F\in \hat{\CC}$ has an
automorphism $\und{\nu\tens1} \overset{\text{def}}= \int\nu\tens1$
(notations are from \cite{Lyu:mod}, see also \secref{3-modcase}).
The second modular axiom says \cite{Lyu:mod}

(M2) $\und{\nu\tens1}(\Ann\omega)\subset\Ann\omega$

\noindent (more scrupulously, there exist morphisms
$T':\Ann\omega\to\Ann\omega\in\hat{\CC}$,
$T:\f\to\f\in\CC$ such that the diagram
\[
\begin{array}{ccccrcrcc}
0 & \to & \Ann\omega & \to & F & \to & \f & \to & 0\\
&& T'\bigg\downarrow && \und{\nu\tens1} \bigg\downarrow &&
T\bigg\downarrow &&\\
0 & \to & \Ann\omega & \to & F & \to & \f & \to & 0
\end{array}
\]
commutes).

An equivalent form of (M2) is \cite{Lyu:mod}

(M$2'$) There exists a morphism $\theta:I\to \f$ such that for any
$X\in\CC$ the ribbon twist $\nu:X\to X$ coincides with the composition
\[ X \simeq I\tens X @>\theta\tens X>> \f\tens X @>\Omega_r>> \f\tens X
@>\e\tens X>> I\tens X \simeq X .\]

\begin{definition}
A noetherian abelian ribbon category $\CC$ with finite dimensional $k$-vector
spaces of morphisms $\Hom_{\CC}(A,B)$ is called {\sl 2-modular}, if axioms
(M1), (M2) are satisfied.
\end{definition}

Here 2- refers to the dimension of a surfaces which will be the main
application.

It was shown in \cite{Lyu:mod} that in the case of a modular category there
exists a morphism $\mu:I\to\f$, which is the integral of a dual Hopf algebra
$\pti\f\simeq\f$, and
\[ 
\unitlength=0.7mm
\linethickness{0.4pt}
\begin{picture}(143,39)
\put(1,20){\makebox(0,0)[cc]{$\nu^{-1}$}}
\put(9,16){\framebox(4,8)[cc]{}}
\put(27,24){\oval(32,16)[lt]}
\put(30,5){\line(0,-1){4}}
\put(30,11){\line(0,1){28}}
\put(23,32){\line(0,1){5}}
\put(18,36){\makebox(0,0)[cc]{$\mu$}}
\put(35,37){\makebox(0,0)[cc]{$X$}}
\put(42,20){\makebox(0,0)[cc]{$=$}}
\put(49,20){\makebox(0,0)[cc]{$\lambda^{-1}$}}
\put(56,16){\framebox(4,8)[cc]{}}
\put(58,24){\line(0,1){15}}
\put(58,16){\line(0,-1){15}}
\put(63,37){\makebox(0,0)[cc]{$X$}}
\put(65,20){\makebox(0,0)[cc]{$\nu$ ,}}
\put(79,20){\makebox(0,0)[cc]{$\nu$}}
\put(87,16){\framebox(4,8)[cc]{}}
\put(105,24){\oval(32,16)[lt]}
\put(108,5){\line(0,-1){4}}
\put(108,11){\line(0,1){28}}
\put(101,32){\line(0,1){5}}
\put(96,36){\makebox(0,0)[cc]{$\mu$}}
\put(113,37){\makebox(0,0)[cc]{$X$}}
\put(120,20){\makebox(0,0)[cc]{$=$}}
\put(127,20){\makebox(0,0)[cc]{$\lambda$}}
\put(134,16){\framebox(4,8)[cc]{}}
\put(136,24){\line(0,1){15}}
\put(136,16){\line(0,-1){15}}
\put(141,37){\makebox(0,0)[cc]{$X$}}
\put(143,20){\makebox(0,0)[cc]{$\nu^{-1}$}}
\put(23,16){\oval(24,18)[b]}
\put(101,16){\oval(24,18)[b]}
\put(113,26){\line(0,-1){11}}
\put(35,26){\line(0,-1){11}}
\end{picture}
\]
for some invertible constant $\lambda\in k^{\times}$. The pair
$(\mu,\lambda)$ is unique up to a sign. Morphisms $S,S^{-1}:\f\to\f$
\[
\unitlength=0.7mm
\raisebox{-15mm}{
\begin{picture}(50,43)
\put(1,22){\makebox(0,0)[cc]{$S\ =$}}
\put(26,23){\oval(20,18)[b]}
\put(40,19.50){\oval(20,19)[t]}
\put(50,20){\line(0,-1){18}}
\put(30,2){\line(0,1){8}}
\put(36,33){\line(0,1){9}}
\put(16,23){\line(0,1){19}}
\put(26,43){\makebox(0,0)[cc]{$\f$}}
\put(40,1){\makebox(0,0)[cc]{$\f$}}
\put(40,29){\line(0,1){8}}
\put(45,35){\makebox(0,0)[cc]{$\mu$}}
\end{picture}
}
\qqquad,\qqquad
\unitlength=0.7mm
\raisebox{-15mm}{
\begin{picture}(51,44)
\put(41,25){\oval(20,18)[b]}
\put(25.50,20){\oval(21,20)[t]}
\put(31,34){\line(0,1){10}}
\put(51,25){\line(0,1){19}}
\put(36,12){\line(0,-1){10}}
\put(15,20){\line(0,-1){18}}
\put(25,30){\line(0,1){8}}
\put(41,44){\makebox(0,0)[cc]{$\f$}}
\put(26,1){\makebox(0,0)[cc]{$\f$}}
\put(20,36){\makebox(0,0)[cc]{$\mu$}}
\put(1,22){\makebox(0,0)[cc]{$S^{-1} \ =$}}
\end{picture}
}
\]
are inverse to each other. Morphisms $S$ and $T$ (defined via (M2)) yield a
projective representation of a mapping class group of a torus with one hole:
\[ (ST)^3=\lambda S^2, \qquad  S^2=\gamma_\f^{-1},\]
\[ T\gamma_\f=\gamma_\f T, \qquad  \gamma_\f^2=\nu. \]
Here $\gamma_\f:\f\to \f$ is the antipode of the Hopf algebra $\f$, given
by the same tangle as \eqref{antgamma}.

\section{Modular transformations in $F$}\label{transformations}
Here we reproduce results of \cite{Lyu:mod} in special assumptions,
which permit to prove more. Let $\CC$ be a 2-modular category. Fix a
morphism $\al:I\to F$ such that
\[ \gamma_F\alpha = \al :I\to F \qquad \text{and} \qquad
\mu= (I @>\al>> F @>\pi>> \f) \]
(if there is one).
In this section we adopt the convention $AB= A\circ B$ for composition.

\subsection{The quantum Fourier transform}\label{qFt}
\begin{theorem}[cf Theorem 6.3 \cite{Lyu:mod}]
For any $X\in\CC$ we have
\be\label{nulambda}
\unitlength=0.7mm
\raisebox{-13mm}{
\begin{picture}(114,39)
\put(50,20){\makebox(0,0)[cc]{$\nu$}}
\put(58,16){\framebox(4,8)[cc]{}}
\put(76,24){\oval(32,16)[lt]}
\put(79,5){\line(0,-1){4}}
\put(79,11){\line(0,1){28}}
\put(72,32){\line(0,1){5}}
\put(67,36){\makebox(0,0)[cc]{$\alpha$}}
\put(84,37){\makebox(0,0)[cc]{$X$}}
\put(91,20){\makebox(0,0)[cc]{$=$}}
\put(98,20){\makebox(0,0)[cc]{$\lambda$}}
\put(105,16){\framebox(4,8)[cc]{}}
\put(107,24){\line(0,1){15}}
\put(107,16){\line(0,-1){15}}
\put(112,37){\makebox(0,0)[cc]{$X$}}
\put(114,20){\makebox(0,0)[cc]{$\nu^{-1}$}}
\put(72,16){\oval(24,18)[b]}
\put(84,26){\line(0,-1){11}}
\put(1,20){\makebox(0,0)[cc]{$\nu$}}
\put(9,16){\framebox(4,8)[cc]{}}
\put(23,32){\line(0,1){5}}
\put(18,36){\makebox(0,0)[cc]{$\alpha$}}
\put(35,37){\makebox(0,0)[cc]{$X$}}
\put(42,20){\makebox(0,0)[cc]{$=$}}
\put(27,16){\oval(32,18)[lb]}
\put(23,24){\oval(24,16)[t]}
\put(35,25){\line(0,-1){13}}
\put(30,29){\line(0,-1){28}}
\put(30,34){\line(0,1){5}}
\end{picture}
}
\end{equation}
\be\label{nu-lambda-}
\unitlength=0.7mm
\raisebox{-13mm}{
\begin{picture}(114,39)
\put(50,20){\makebox(0,0)[cc]{$\nu^{-1}$}}
\put(58,16){\framebox(4,8)[cc]{}}
\put(76,24){\oval(32,16)[lt]}
\put(79,5){\line(0,-1){4}}
\put(79,11){\line(0,1){28}}
\put(72,32){\line(0,1){5}}
\put(67,36){\makebox(0,0)[cc]{$\alpha$}}
\put(84,37){\makebox(0,0)[cc]{$X$}}
\put(91,20){\makebox(0,0)[cc]{$=$}}
\put(98,20){\makebox(0,0)[cc]{$\lambda^{-1}$}}
\put(105,16){\framebox(4,8)[cc]{}}
\put(107,24){\line(0,1){15}}
\put(107,16){\line(0,-1){15}}
\put(112,37){\makebox(0,0)[cc]{$X$}}
\put(114,20){\makebox(0,0)[cc]{$\nu$ .}}
\put(72,16){\oval(24,18)[b]}
\put(84,26){\line(0,-1){11}}
\put(2,20){\makebox(0,0)[cc]{$\nu^{-1}$}}
\put(10,16){\framebox(4,8)[cc]{}}
\put(24,32){\line(0,1){5}}
\put(19,36){\makebox(0,0)[cc]{$\alpha$}}
\put(36,37){\makebox(0,0)[cc]{$X$}}
\put(43,20){\makebox(0,0)[cc]{$=$}}
\put(28,16){\oval(32,18)[lb]}
\put(24,24){\oval(24,16)[t]}
\put(36,25){\line(0,-1){13}}
\put(31,29){\line(0,-1){28}}
\put(31,34){\line(0,1){5}}
\end{picture}
}
\end{equation}
\end{theorem}

\begin{notation}
Let $\beta: I\to F$ be an arbitrary morphism. Introduce morphisms $F\to F$
\[ S_{\mp}(\beta) = \bigl( F\simeq F\tens I @>F\tens\beta>> F\tens F
@>\Omega^{\pm1}>> F\tens F @>\e\tens F>> I\tens F \simeq F \bigr) ,\]
graphically depicted as
\[
\unitlength=0.7mm
\raisebox{-15mm}{
\begin{picture}(50,44)
\put(1,22){\makebox(0,0)[cc]{$S_-(\beta)\ =$}}
\put(26,23){\oval(20,18)[b]}
\put(40,19.50){\oval(20,19)[t]}
\put(50,20){\line(0,-1){18}}
\put(30,2){\line(0,1){8}}
\put(36,33){\line(0,1){9}}
\put(16,23){\line(0,1){19}}
\put(26,43){\makebox(0,0)[cc]{$F$}}
\put(40,1){\makebox(0,0)[cc]{$F$}}
\put(40,29){\line(0,1){8}}
\put(45,35){\makebox(0,0)[cc]{$\beta$}}
\end{picture}
}\quad, \qquad\qquad
\raisebox{-15mm}{
\begin{picture}(51,44)
\put(41,25){\oval(20,18)[b]}
\put(25.50,20){\oval(21,20)[t]}
\put(31,34){\line(0,1){10}}
\put(51,25){\line(0,1){19}}
\put(36,12){\line(0,-1){10}}
\put(15,20){\line(0,-1){18}}
\put(25,30){\line(0,1){8}}
\put(41,44){\makebox(0,0)[cc]{$F$}}
\put(26,1){\makebox(0,0)[cc]{$F$}}
\put(20,36){\makebox(0,0)[cc]{$\beta$}}
\put(1,22){\makebox(0,0)[cc]{$S_+(\beta)\ =$}}
\end{picture}
}
\]
We shall use the shorthand $S_* = S_\pm(\beta):F\to F$
(a sign is chosen arbitrarily) and
\[ S_+ = \lambda S_+(\al), \qqquad S_-=\lambda^{-1} S_-(\al) ,\]
\end{notation}

\begin{proposition}[cf Proposition 6.5 \cite{Lyu:mod}]\label{"6.5"}
We have
\[ S_-(\al)\gamma_F = S_+(\al) = \gamma_F S_-(\al) .\]
In particular, $S_\pm$ commute with $\gamma_F$.
\end{proposition}

\begin{theorem}[cf Theorem 6.7 \cite{Lyu:mod}]
We have
\[S_*T^{-1}S_+=TS_*T,\qquad S_*TS_-=T^{-1}S_*T^{-1}.\]
\end{theorem}

\begin{corollary}\label{"6.8"}
$S_-TS_-=T^{-1}S_-T^{-1}$.
\end{corollary}

\begin{lemma}[cf Lemma 6.9 \cite{Lyu:mod}]
There are identities
\[S_*TS_-S_+=S_*T,\qquad S_*S_+TS_-=TS_*,\]
\[S_*T^{-1}S_+S_-=S_*T^{-1},\qquad S_*S_-T^{-1}S_+=T^{-1}S_*.\]
\end{lemma}

\begin{lemma}[cf Lemma 6.10 \cite{Lyu:mod}]\label{"6.10"}
The morphism $T$ commutes with $S_*S_+$ and
$S_*S_-$. We have
\[S_*S_+S_-=S_*=S_*S_-S_+.\]
\end{lemma}

\begin{corollary}\label{"6.11"}
The morphisms $P_1=S_-S_+$ and $P_2=S_+S_-$ are
projections such that $P_1P_2=P_1$ and $P_2P_1=P_2$.
\end{corollary}

\begin{proposition}\label{kerim}
The following kernels and images coincide
\be\label{KKKKKA}
\Ker S_+ = \Ker S_- = \Ker P_1 = \Ker P_2 = \Ker \pi = \Ann\omega ,
\end{equation}
\be\label{IIII}
\im S_+ = \im S_- = \im P_1 = \im P_2 .
\end{equation}
In particular, $P_1=P_2$.
\end{proposition}

\begin{pf}
We get by \corref{"6.11"} $\Ker S_+\subset \Ker P_1$,
$\Ker S_-\subset\Ker P_2$, $\Ker P_1=\Ker P_2$. \lemref{"6.10"} gives
$\Ker P_1\subset \Ker S_\pm$, $\Ker P_2\subset \Ker S_\pm$. Therefore,
$\Ker S_+ = \Ker S_- = \Ker P_1 = \Ker P_2$.

Since
\[
\unitlength=1mm
\begin{picture}(103,31)
\put(59,16){\line(0,1){4}}
\put(59,20){\line(4,5){7.33}}
\put(59,24){\line(-4,5){4}}
\put(47,30){\makebox(0,0)[cc]{$\Ann\omega$}}
\put(73,30){\makebox(0,0)[cc]{$X$}}
\put(51,20){\oval(24,18)[b]}
\put(64,21){\oval(32,20)[rb]}
\put(39,19){\line(0,1){10}}
\put(21,24){\line(-4,5){4}}
\put(9,30){\makebox(0,0)[cc]{$\Ann\omega$}}
\put(13,20){\oval(24,18)[b]}
\put(1,19){\line(0,1){10}}
\put(23,29){\line(0,-1){13}}
\put(23,11){\line(0,-1){10}}
\put(26,30){\makebox(0,0)[cc]{$X$}}
\put(32,16){\makebox(0,0)[cc]{=}}
\put(87,20){\oval(14,14)[t]}
\put(94,21){\line(0,-1){20}}
\put(103,16){\makebox(0,0)[cc]{$=\ 0$}}
\end{picture}
\]
we have $\Ann \omega \subset \Ker S_-$. This identity also implies
\begin{align*}
& \bigl( F\simeq F\tens I @>F\tens\al>> F\tens F @>\Omega>> F\tens F
@>\e\tens F>> I\tens F \simeq F @>\pi>> \f \bigr) = \\
&= \bigl( F @>\pi>> \f \simeq \f\tens I @>\f\tens\al>> \f\tens F @>\Omega>>
\f\tens F @>\e\tens F>> I\tens F \simeq F @>\pi>> \f \bigr)  \\
&= \bigl( F @>\pi>> \f \simeq \f\tens I @>\f\tens\mu>> \f\tens\f @>\Omega>>
\f\tens\f @>\e\tens\f>> I\tens\f \simeq \f \bigr)  \\
&= \bigl( F @>\pi>> \f @>S>> \f \bigr)  ,
\end{align*}
hence $\pi S_-= S\pi$. Thus, $\Ker(\pi S_-) = \Ker(S\pi) = \Ker\pi$ and
$\Ker S_- \subset \Ker\pi = \Ann\omega$, whence \eqref{KKKKKA} follows.

\propref{"6.5"} implies $\im S_-(\al) = \im S_+(\al)$, so
$\im S_-= \im S_+$. \corref{"6.11"} gives that $\im S_- \supset \im P_1$,
$\im S_+ \supset \im P_2$. Since $S_-S_+S_- = S_-$ and $S_+S_-S_+ = S_+$
by \lemref{"6.10"}, we get $\im S_- \subset \im P_1$,
$\im S_+ \subset \im P_2$. Therefore, \eqref{IIII} holds.
\end{pf}

\begin{notation}
Denote $P = P_1 = P_2 = S_+S_- = S_-S_+$.
\end{notation}

\begin{theorem}\label{sl2onImP}
The morphisms $S_+(\al), S_-(\al), T, \gamma_F :F\to F$ commute with the
projection $P:F\to F$. The restrictions of $S_\pm(\al)$ to
$\Ker P=\Ann\omega$ vanishe. The restrictions to $\im P$ (which depends
on $\alpha$)
\[ S=S_-(\al)\big|_{\im P}, \qquad S^{-1} =S_+(\al)\big|_{\im P} ,\]
inverse to each other, are identified with $S^{\pm1} :\f\to\f$ by an
isomorphism $\im P @>\pi>> \f$. Therefore,
\be\label{STtre=Stwo}
(ST)^3 = \lambda S^2, \qquad S^2 = \gamma_F^{-1}
\end{equation}
on $\im P$.
\end{theorem}

\begin{pf}
\lemref{"6.10"} implies that $T$ commutes with $P$ and $PS_-=S_-=S_-P$,
$PS_+=S_+=S_+P$. \propref{"6.5"} implies $P\gamma_F = \gamma_F P$.
Other statements follow by \propref{kerim}, \corref{"6.8"} and
\propref{"6.5"}.
\end{pf}

A certain similarity of the properties of $S$ with the properties of the
ordinary Fourier transform \cite{LyuMaj:L2(R)} suggests the name of
{\em quantum Fourier transform} for this morphism.

\subsection{3-modular categories}
Here 3- refers to applications to 3-manifolds. Recall that $\mu:I\to \f$
is a two-sided integral and $\gamma_F \mu = \mu$. If $F\in\CC$ (analogue
of being finite dimensional), then it has an invertible object of left
integrals $\Int_l$ \cite{Lyu:mod}. The canonical projection $\pi:F\to\f$
sends $\Int_l$ to $\im\mu$.

\begin{definition}
A {\em 3-modular category} is a 2-modular category $\CC$ with an additional
property

(M3) $F\in\CC$ and it has a two-sided integral $\si:I\to F$ such that
$\pi \si=\mu$.
\end{definition}

The assumption $F\in\CC$ is not needed in this paper and can be
consistently omitted. However, it holds in all known examples.

\begin{proposition}
Suppose (M3) holds. Then $\gamma_F \sigma = \si$.
\end{proposition}

\begin{pf}
Clearly, $\gamma_F \si$ is another two-sided integral in $F$. Therefore,
it must be proportional to $\si$ \cite{Lyu:mod}. Projecting the equation
$\gamma_F \si = C\si$, $C\in k^\times$, to $\f$ we get
$\mu = \gamma_\f \mu = C\mu$, hence, $C=1$.
\end{pf}

\begin{definition}
A {\em perfect modular category} is a 2-modular category with a condition

(PM) $\Ann\omega =0$ (equivalently, $\pi:F\to\f$ is an isomorphism).
\end{definition}

A perfect modular category is a particular case of a 3-modular category
(and the easiest to deal with). The reader is advised to assume $\CC$
perfect for the first reading whenever appropriate.

\section{Ribbon and braided Hopf algebras}\label{Hopf}
In this section we reformulate the results obtained in the abstract
setting of ribbon abelian categories in the case of category of finite
dimensional modules $\CC=H$-mod over a Hopf algebra $H$. This job was
already partially done by Majid and the author \cite{LyuMaj}, so we
shall omit most of the proofs.

\subsection{Quasitriangular Hopf algebras}\label{qtrHa}
Let $H$ be a Hopf $k$-algebra with an invertible antipode. There are
several (more or less equivalent) ways to make $\CC=H$-mod into a braided
category, if it permits this structure. We choose the most direct one.
Assume that $H$ has an {\em $R$-matrix}, which is an element $R\in H\tens H$
(algebraic tensor product!) satisfying the relations of Drinfeld
\cite{Dri:qua}
\begin{align*}
(\Delta\tens1) R &= R^{13}R^{23} \\
(1\tens\Delta) R &= R^{13}R^{12} \\
\Delta^\op a &= R\Delta(a) R^{-1}
\end{align*}
for any $a\in H$, so $(H,R)$ is {\em quasitriangular}.

For a finite dimensional $H$-module $V$ as for any vector space there is
a canonical linear map $v_0^2: V\to V\pti\pti$ such that
$\<v,y\> = \<y,v_0^2(v)\>$ for $v\in V$, $y\in V\pti$. Its square gives
$v_0^4=(V @>v_0^2>> V\pti\pti @>v_0^2>> V^{(4}\pti{}^) )$. On the other
hand, in $\CC$ as in any rigid braided category there are morphisms
$u_0^{-4} = (V @>u_1^{-2}>> \pti\pti V @>u_{-1}^{-2}>> V^{(-4}\pti{}^) )$.
Composing them we get linear bijections
\[ g_V: V @>u_0^{-4}>> V^{(-4}\pti{}^) @>v_0^4>> V .\]
They are decomposable into a product of the two bijections
\begin{align*}
u_1:& V @>v_0^2>> V\pti\pti @>u_{-1}^{-2}>> V ,\\
u_4:& V @>v_0^2>> V\pti\pti @>u_1^{-2}>> V .
\end{align*}

\begin{theorem}[Drinfeld \cite{Dri:cocom}]
The maps $u_1,u_4,g$ are given by the action of the following elements
\[ u_1=\gamma(R'')R', \qquad u_4=\gamma^2(R')R''=\gamma(u_1)^{-1},
\qquad g=u_1u_4 .\]
The element $g$ is grouplike ($\e(g)=1$, $\Delta g=g\tens g$) and for any
$a\in H$ we have $gag^{-1} = \gamma^4(a)$.
\end{theorem}

To find $g$ we can use the following. Let $\hh_+$ (resp. $\hh_-$) be the
minimal subspace of $H$ such that $R\in \hh_+\tens H$ (resp.
$R\in H\tens\hh_-$). Then $\hh_+$, $\hh_-$ are Hopf subalgebras. The
finite dimensional subspace spanned by their products $\hh=\hh_+\hh_-$
coincides with $\hh_-\hh_+$, therefore, $\hh$ is also a Hopf subalgebra.
Moreover, it is a minimal quasitriangular Hopf subalgebra of $(H,R)$
\cite{Rad:min}. All elements $u_1,u_4,g$ belong to $\hh$.

Pick a basis $(a_i)\subset \hh_+$ and a basis $(b_i)\subset \hh_-$ for
which $R=\sum_i a_i\tens b_i$. Introduce a non-degenerate pairing
$\pi :\hh_-\tens \hh_+ \to k$, $\pi(b_i,a_j)=\delta_j^i$. Axioms of
$R$-matrix imply that $\pi :\hh_{-\op}\tens \hh_+ \to k$ is a Hopf pairing,
i.e.
\begin{align*}
\pi(ab,c) &=\pi(a,c_{(1)})\pi(b,c_{(2)}), \\
\pi(a,cd) &=\pi(a_{(1)},d)\pi(a_{(2)},c) .
\end{align*}

Now construct the double $D(\hh_+)$ generated by its Hopf subalgebras
$\hh_+,\hh_-$. The natural projection $j:D(\hh_+) \to\hh$ is a homomorphism
of quasitriangular Hopf algebras. Clearly, $g_H = g_\hh = j(g_{D(\hh_+)})$
is the relationship between the elements $g$ for the three algebras.
To find $g_{D(\hh_+)}$ use the following

\begin{theorem}[Drinfeld \cite{Dri:cocom},
Kauffman and Radford \cite{KauRad:rib}]
Let $\delta_+\in\hh_+$, $\delta_-\in\hh_-$ be left integrals, that is,
$x\delta_\pm = \e(x)\delta_\pm$ for any $x\in\hh_\pm$.
Then $g_{D(\hh_+)} = a_+^{-1}a_-$ for grouplike elements
(called moduli) $a_+\in\hh_+$, $a_-\in\hh_-$ such that
\begin{alignat*}2
\delta_+ y &= \pi(a_-,y) \delta_+ &&\qquad \text{for any } y\in\hh_+ ,\\
\delta_- y &= \pi(y,a_+) \delta_- &&\qquad \text{for any } y\in\hh_- .
\end{alignat*}
\end{theorem}

\subsection{Ribbon Hopf algebras}\label{RibHop}
Assume now that $\CC=H$-mod has a ribbon structure. Then there is a morphism
$u_0^{-2} = u_{-1}^{-2}\nu = u_1^{-2}\nu^{-1} : V\to\pti\pti V$ for any
finite dimensional $H$-module $V$. One can prove that the map
\[ \ka_V: V @>u_0^{-2}>> \pti\pti V @>v_0^2>> V \]
commutes with all morphisms and satisfies
$\ka_{X\tens Y} = \ka_X \tens \ka_Y$ and $\ka_X^2=g_X$. If $H$ is finite
dimensional, we deduce that $\ka_V$ is the action of a grouplike element
$\ka$ of $H$.

\begin{definition}[comp. \cite{ResTur:rg}]\label{ribbon}
A {\em ribbon Hopf algebra} $(H,R,\ka)$ is a quasitriangular Hopf algebra
$(H,R)$ and a grouplike element $\ka\in H$ such that
\begin{align*}
\ka^2 &= g \\
\ka a \ka^{-1} &= \gamma^2(a)
\end{align*}
for any $a\in H$.
\end{definition}

In the category of finite dimensional modules over a ribbon Hopf algebra
we have canonical isomorphisms $u_0^2:V\to V\pti\pti$,
$u_0^2(v) = v_0^2(\ka v) = \ka v_0^2(v)$, which we use to identify these
modules.

The following is essentially proved by Kauffman and Radford.

\begin{theorem}[cf \cite{KauRad:rib}]
If $(H,R,\ka)$ is a ribbon Hopf algebra, then the category $H$-mod is a
ribbon braided category with the ribbon twist given by multiplication
by the central element
\[ \nu = \gamma^2(R')R''\ka^{-1} =
R''\gamma^2(R')\ka = R''\ka R' = R'\ka^{-1} R'' .\]
The following holds
\[ \nu^{-1} = R'\gamma(R'')\ka = \gamma(R'')R'\ka^{-1} ,\]
\begin{align}
\e(\nu) &= 1 , \label{e(nu)} \\
\gamma(\nu) &= \nu , \notag \\
\Delta \nu &= (R^{21}R^{12}) \cdot \nu\tens\nu . \notag
\end{align}
\end{theorem}

\begin{remark}
\defref{ribbon} is equivalent to the definition of a ribbon Hopf algebra
of Reshetikhin and Turaev \cite{ResTur:rg}. The element $\nu^{-1}$
was denoted $v$ in \cite{ResTur:rg}.
\end{remark}

\subsection{A braided Hopf algebra}
Here we describe explicitly the braided Hopf algebra $F$ and its dual
algebra $U$ for the case of $\CC=H$-mod. Let $H$ be a ribbon Hopf algebra
and let $H^\circ$  be its dual \cite{Swe:book}. Assume that $H$ has enough
finite dimensional modules, so that the pairing $H\tens H^\circ \to k$
is non-degenerate. Define the Hopf algebra $Fun = (H^\circ)_\op$ as
$H^\circ$ with the opposite coproduct (note that usually the algebra of
functions is a subalgebra of $H^\circ$, not of $(H^\circ)_\op$).
We have an equivalence of categories $\CC=H\modul \simeq Fun\comod$.
The pairing $\<,\>: H\tens Fun \to k$ satisfies
\begin{align*}
\<x,fg\> &= \<x\one,f\> \<x\two,g\> \\
\<xy,f\> &= \<x,f\two\> \<y,f\one\>
\end{align*}
for $x,y\in H$, $f,g\in Fun$, where $\Delta f = f\one\tens f\two$ is the
coproduct in $Fun$.

Consider linear maps $i_L: L\tens L\pti \to Fun$,
$l_a\tens l_b\mapsto \mt{La}b$, where $\mt{La}b$ is the matrix element
of the $H$-module $L$ with a basis $(l_a)$, that is, $\mt{La}b$ is a linear
function on $H$ given by $\<u,\mt{La}b\> = \<u.l_a, l^b\>$ for $u\in H$.
The maps $i_L$ become homomorphisms of $H$-modules if $Fun$ is given
the coadjoint $H$-module structure
\[ u\tr f = \<u,f\one \gamma(f\tre)\> f\two \]
for $u\in H$, $f\in Fun$. The vector space $Fun$ with this $H$-module
structure will be denoted $F$.

\begin{theorem}[\cite{Del:tan,Sch:recon,Yet:recon}]
The family $(i_L : L\tens L\pti \to F)_{L\in\CC}$ is a coend of the
bifunctor $\CC\times\CC^\op \to \CC$, $(A,B) \mapsto A\tens B\pti$, so
we can write $F=\int^L L\tens L\pti$. In other words,
the sequence~\eqref{coend} is exact.
\end{theorem}

Being the coend, $F$ is a Hopf algebra in the category
$\widehat\CC = Fun\Comod \subset H$-Mod (braided Hopf algebra), as we have
seen in \secref{intribbon}. The Hopf structure of $F$ described by tangles
in \cite{Lyu:mod} converts to the following. As the coalgebra $F$ coincides
with $Fun$. The multiplication in $F$ is expressed via the multiplication in
$Fun$ as
\begin{align*}
m_F(f\tens g)
&= \rho(\gamma(f\two)\tens g\one \gamma(g\tre) ) f\one g\two \\
&= \rho(f\two\tens g\tre\gamma^{-1}(g\one) ) f\one g\two \\
&= \rho(f\one\gamma(f\tre) \tens g\one ) g\two f\two
\end{align*}
by \eqref{mult1} and \eqref{mult2}, where
\[ \rho(a\tens b) = \<R,a\tens b\> = \<R',a\> \<R'', b\> .\]
The unity of $F$ is the same as the unity of $Fun$. The antipode $\gamma_F$
of $F$ is expressed via the antipode $\gamma$ of $Fun$:
\[ \gamma_F(f) = \rho(f\one\tens \gamma(f\four)) \<\nu\ka^{-1}, f\two\>
\gamma(f\tre) \]
for $f\in F$. The inverse antipode is
\[ \gamma_F^{-1}(f) = \rho(f\four, f\one) \<\ka^{-1}\nu^{-1}, f\tre\>
\gamma^{-1}(f\two) .\]
All the structure maps $\Delta, \e, m_F, \gamma_F$ are homomorphisms of
$H$-modules.

\subsection{The dual braided Hopf algebra}
In order to define a Hopf algebra dual to $F$ we shall not consider the
rational part $^\circ F$ of the $H$-module $^*F$ of all linear functionals
on $F$. Instead we denote by $U$ the $H$-submodule $H\subset {}^*F$.
This amounts to consider the adjoint action
\[ \ad a.x = a\one x \gamma(a\two) \]
for $a\in H$, $x\in U$. Since $^*F$ is an algebra in the category
$H$-Mod (being dual to the coalgebra $F\in H$-Mod), so is $U$ with the
usual multiplication $m$ of $H$. We want to introduce a comulptiplication
$\nabla:U \to U\tens U$ which would be dual to $m_F$ in the proper sense:
\[ \< \nabla u, f\tens g\> \equiv \< u^{(1)} \tens u^{(2)}, f\tens g\>
\equiv \<u^{(2)}, f\> \<u^{(1)}, g\> = \<u, m_F(f\tens g)\> \]
for $u\in U$, $f,g\in F$. By dualising the formulae for $m_F$ one arrives
to the following
\begin{align}
\nabla u &= \ad R''.u\two \tens R' u\one \label{nablaone} \\
&= u\one \gamma(R'') \tens \ad R'.u\two . \label{nablatwo}
\end{align}
The counity of $U$ coincides with the counity of $H$. The operations
$m,\nabla,\e$ make $U$ into a braided Hopf algebra in $H$-Mod, the antipode
$\gamma_U:U\to U$ being
\[ \gamma_U(u) = \nu\ka^{-1} \gamma(R'') \gamma(u) R' \]
and the inverse antipode being
\begin{align*}
\gamma_U^{-1}(u) &= R' \ka^{-1}\nu^{-1} \gamma(u) R'' \\
&= \gamma^2(R') \gamma^{-1}(u) R'' \ka^{-1}\nu^{-1}.
\end{align*}
This is the unique Hopf algebra structure dual to $F$ on the $H$-module $H$.

\subsection{The algebra $\f$}
The algebra $Fun^*$ acts in every finite dimensional $H$-module $X$.
Consider special elements $l_V^-(v\tens w) \in Fun^*$ determined for any
$w\in W\pti$, $v\in V\in H$-mod as the operators in $X$:
\begin{align}
l_V^-(v\tens w) (x) &= (\ev\tens1)(1\tens c^2) (v\tens w\tens x) \notag \\
&= \sum_{i,j} \<v, R_j''R_i'w\> R_j'R_i'' x . \label{l(vw)}
\end{align}
The subspace $\u$ spanned by $l_V^-(v\tens w)$ is contained in $H$ and even
in $\hh$, so it is finite dimensional. As shown in \cite{Lyu:mod} $\u$ is a
braided Hopf subalgebra of $^\circ F$, therefore, it is closed under the
operations of $U$ and constitutes a finite dimensional braided Hopf
subalgebra of $U$.

The map
\[ l^-: \bigoplus_{V\in\CC} V\tens V\pti @>\oplus l_V^->> U \]
factors through the coend \eqref{coend}, therefore, determines a map
$l^-:F\to U$, which is a homomorphism of Hopf algebras in $H$-Mod
\cite{Lyu:mod}. The image of $l^-$ is $\u$. The form $\omega:F\tens F\to k$
\eqref{omega} can be presented as
\[ \omega(f\tens \mt{La}b) = \<l^-(f).l_a ,l^b\> = \<l^-(f), \mt{La}b\> ,\]
hence, $\Ann^{\text{left}} \omega = \Ker l^-$. Therefore, the braided
Hopf algebras $\f=F/\Ann\omega$ and $\u$ are isomorphic
\cite[Corollary 3.10]{Lyu:mod}. Thus the first modular axiom (M1) is
always satisfied for the considered algebras $H$.

By definition the subspace $\u\subset\hh$ is the minimal subspace such that
$R^{12}R^{21} \in \u\tens\hh$. Since
$(\gamma\tens \gamma)(R^{12}R^{21}) = R^{21}R^{12}$, the minimal subspace
$B\subset \hh$ such that $R^{12}R^{21} \in \hh\tens B$ is $\gamma(\u)$.
It does not coincide necessarily with $\u$, for $\u$ is not an ordinary
Hopf subalgebra. Repeating the reasoning we get $\gamma^2(\u) = \u$ and
conclude that $R^{12}R^{21} \in \u\tens\gamma(\u)$. Similarly
$(R^{12}R^{21})^{-1} \in \u\tens\gamma(\u)$.

\subsection{2-modular Hopf algebras}
We already know by \eqref{l(vw)} the image
\be\label{l-(f)=}
l^-(f) = \sum_{i,j} \<\gamma^{-1}(R_j''R_i') ,f\> R_j'R_i''
\end{equation}
for any $f\in F$. Notice that
$\gamma^{-1}(R_j''R_i') \tens R_j'R_i'' \in\u\tens\u$. The pairing
$\u\tens F \hookrightarrow U\tens F \to k$ factorizes through a perfect
pairing $\u\tens\f \to k$ due to
$\Ann^{\text{left}} \omega = \Ann^{\text{right}} \omega$. Therefore,
an element $x\in H$ is representable in the form $l^-(f)$ for some
$f\in\f$ iff $x\in \u$.

\begin{theorem}
A ribbon Hopf algebra $(H,R,\ka)$ is {\em 2-modular} (that is, $H$-mod
is 2-modular) if and only if $\nu\in\u$, or equivalently, $\nu^{-1}\in\u$.
\end{theorem}

\begin{pf}
If the axiom (M$2'$) holds, then $\nu=l^-(\theta(1))$ for some
$\theta:k\to\f$. Hence, $\nu\in\u$ by the above discussion.

If $\nu\in\u$, then for some $f\in\f$ we have $\nu=l^-(f)$. Since $\nu$
is central, the subspace $k\nu\subset\u$ is a trivial $H$-submodule.
Hence, its preimage $kf\subset\f$ by an isomorphism $l^-:\f\to\u$ is
also a trivial $H$-submodule. Now $\theta:k\to\f$, $1\mapsto f$ is the
homomorphism required in the axiom (M$2'$).

The condition $\und{\nu\tens1} (\Ann\omega) \subset \Ann\omega$ is
equivalent to $\und{\nu^{-1}\tens1} (\Ann\omega) \subset \Ann\omega$
\cite[Corollary 5.12]{Lyu:mod}. Therefore, $\nu\in\u$ if and  only if
$\nu^{-1}\in\u$.
\end{pf}

\begin{corollary}
If a ribbon Hopf algebra $(H,R,\ka)$ is 2-modular, then $\ka\in\hh$.
Therefore, its minimal quasitriangular subalgebra $(\hh,R)$ equiped with
$\ka$ will be also a 2-modular ribbon Hopf algebra.
\end{corollary}

\begin{remark}
Let $H$ be a ribbon Hopf algebra. The category $H$-mod is perfect modular
if and only if $F=\f$, or equivalently, $H=\u$, so $H$ is called
factorizable \cite{ResSTS:fac}.
\end{remark}

\subsection{3-modular Hopf algebras}
\begin{proposition}\label{alpha}
Let $\al\in Fun$ be an element. Denote the corresponding linear functional
$H\to k$ and the linear map $k\to Fun$, $1\mapsto\al$, also by $\al$.
The following conditions are equivalent:
\begin{enumerate}
\renewcommand{\labelenumi}{(\roman{enumi})}
\item $\alpha:k\to F$ is a morphism of $H$-modules;
\item $\al\one \gamma(\al\tre) \tens \al\two = 1\tens\al$;
\item $\al(x\one u\gamma(x\two)) = \e(x)\al(u)$ for any $x,u\in H$;
\item $\al:U\to k$ is a morphism of $H$-modules;
\item $\al(xy) = \al(y\gamma^2(x))$ for any $x,y\in H$.
\end{enumerate}
\end{proposition}

Proof is a straightforward check and it is left to the reader.

\begin{theorem}\label{two-sided}
An element $\si\in F$ is a two-sided integral on the algebra $U$ if and
only if the equivalent conditions (i)--(v) of
\propref{alpha} are satisfied and $\si$ is a left integral on the algebra
$H$, that is, $(1\tens\si) \Delta u = \si(u)$ for any $u\in H$.
\end{theorem}

\begin{pf}
Conditions (i) and (iv) are clearly necessary. Assume now that (iii) holds.
$\si\in F$ is a two-sided integral if and only if
\begin{alignat}3
&\qquad& m_F(f\tens\si) &= \e(f)\si &&= m_F(\si\tens f) \notag \\
&\Leftrightarrow \qquad& \<u,m_F(f\tens\si)\> &= \e(f)\<u,\si\>
&&= \<u,m_F(\si\tens f)\> \notag \\
&\Leftrightarrow \qquad& \<\nabla u,f\tens\si\> &=\e(f)\si(u)
&&= \<\nabla u,\si\tens f\> \notag \\
&\Leftrightarrow \qquad& \<(\si\tens1)\nabla u,f\> &= \e(f)\si(u)
&&= \<(1\tens\si)\nabla u, f\> \notag \\
&\Leftrightarrow \qquad& (\si\tens1)\nabla u &= \si(u)
&&= (1\tens\si)\nabla u \label{leftright}
\end{alignat}
for any $f\in F$, $u\in U$.

Now we calculate
\begin{align*}
(\si\tens1)\nabla u &= \si(\ad R''.u\two) R'u\one \\
&= \si(u\two) \e(R'') R'u\one \\
&= u\one \si(u\two) ,\\
(1\tens\si)\nabla u &= u\one \gamma(R'') \si(\ad R'.u\two) \\
&= u\one \gamma(R'') \e(R') \si(u\two) \\
&= u\one \si(u\two)
\end{align*}
by \eqref{nablaone} and \eqref{nablatwo}. Thus, all equations
\eqref{leftright} hold if one equation
\[ (1\tens\si) \Delta u = \si(u) \]
holds.
\end{pf}

Let the algebra $F$ have a two-sided integral $\si:I\to F$. Then
$\pi\circ \si:k\to \f$ is a two-sided integral, therefore it is
proportional to $\mu$. If the proportionality constant does not vanish,
then $\si$ can be rescaled to satisfy $\pi\circ\si =\mu$. Non-vanishing
of $\pi\circ\si$ is equivalent to non-vanishing of $l^-(\si)\in\u$ or of
$\gamma^{-1}(l^-(\si))\in\u$. Formula \eqref{l-(f)=} gives
\[ \gamma^{-1}(l^-(\si)) = \si(R_i'R_j'') R_i''R_j' .\]
So we get

\begin{theorem}\label{3-modthm}
A finite dimensional ribbon Hopf algebra $(H,R,\ka)$ is {\em 3-modular}
(that is, $H$-mod is 3-modular) if and only if the following conditions hold

(M2) $\nu\in\u$

(M3) $\int (xy) = \int(y\gamma^2(x)) $, $\int(R_i'R_j'') R_i''R_j' \ne0$\\
where $\int:H\to k$ is a left integral on the algebra $H$.
\end{theorem}

The property $\int (xy) = \int(y\gamma^2(x)) $ above is equivalent to
{\em unimodularity}\footnote{I am grateful to the referee for this remark.}
of $H$, which means that each left integral $\Lambda\in H$ is a right
integral as well. This follows from the Radford's formula
$\int(y\gamma^2(x)) = \int(x\one \alpha(x\two)y) $ \cite{Rad:tr},
where $\alpha\in G(H^*)$ is the modulus relating left and right
integrals for $H$.

\begin{proposition}\label{unimod}
A factorizable ribbon Hopf algebra is unimodular.
\end{proposition}

\begin{pf}
This a corollary of the above theorem. Or, notice simply that in
factorizable case the map $l^-: F \to U = H$ is an isomorphism of algebras,
preserving the counity, and $F$ is unimodular.
\end{pf}

For Drinfeld doubles unimodularity was proven by Hennings~\cite{Hen:3}
and Radford~\cite{Rad:min}.

\subsection{Some operators in $F$ and $U$}
\subsubsection{3-modular case}\label{3-modcase}
Assume that $(H,R,\ka)$ is 3-modular. We find explicitly the linear
maps $S,T:F\to F$ and their transposed maps $^tS,{}^tT:U\to U$.

Let $y\in H$. There are maps $F\to F$
\[
\unitlength=0.8mm
\begin{picture}(73,29)
\put(4,11){\framebox(4,8)[cc]{}}
\put(6,19){\line(0,1){8}}
\put(6,11){\line(0,-1){8}}
\put(18,3){\line(0,1){24}}
\put(12,29){\makebox(0,0)[cc]{$F$}}
\put(12,1){\makebox(0,0)[cc]{$F$}}
\put(1,15){\makebox(0,0)[rc]{$\und{y\tens1}\ =\ \ y$}}
\put(22,15){\makebox(0,0)[cc]{,}}
\put(64,11){\framebox(4,8)[cc]{}}
\put(73,15){\makebox(0,0)[cc]{$y$ .}}
\put(66,19){\line(0,1){8}}
\put(66,11){\line(0,-1){8}}
\put(54,3){\line(0,1){24}}
\put(60,29){\makebox(0,0)[cc]{$F$}}
\put(60,1){\makebox(0,0)[cc]{$F$}}
\put(50,15){\makebox(0,0)[rc]{$\und{1\tens y}\ =$}}
\end{picture}
\]
The first is obtained as the projection of
$(y\tens1)(l_a\tens l^b) = \<y,\mt{La}c\> l_c\tens l^b$ in the form
$\und{y\tens1} (\mt{La}b) = \<y,\mt{La}c\> \mt{Lc}b$. Therefore,
\begin{align*}
\und{y\tens1}(f) &= \<y, f\one\> f\two \\
\intertext{and similarly}
\und{1\tens y}(f) &= \<\gamma^{-1}(y), f\two\> f\one .
\end{align*}
The transposed operators in $U$ are
\[ u.{}^t\und{(y\tens1)} = uy, \qquad
u.{}^t\und{(1\tens y)} = \gamma^{-1}(y)u \]
for any $u\in U$. In particular, we have
$T=\und{\nu\tens1} = \und{1\tens\nu}$ and its transpose $^tT$
\begin{alignat}2
T(f) &= \<\nu,f\one\>f\two =
f\one \<\nu,f\two\> &\qquad& \text{for } f\in F \label{T(f)} \\
^tT(u) &= u\nu = \nu u &\qquad& \text{for } u\in U \label{tT(u)}
\end{alignat}

Similarly other maps are constructed by projection:
\begin{align}
\Omega_l: X\tens F &\to X\tens F  \notag \\
\Omega_l(x\tens f) &= \sum_{i,j} R_j''R_i'x \tens \<R_j'R_i'', f\one\> f\two
\label{Omegaleft}
\end{align}
\begin{align}
\Omega_r: F\tens X &\to F\tens X  \notag \\
\Omega_r(f\tens x) &= \sum_{i,j} \<\gamma^{-1}( R_j''R_i'),f\two\> f\one
\tens R_j'R_i''x  \label{Omegar=}
\end{align}
\begin{align}
\Omega: F\tens F &\to F\tens F  \notag \\
\Omega(h\tens f) &= \sum_{i,j} \<\gamma^{-1}( R_j''R_i'),h\two\> h\one
\tens \<R_j'R_i'', f\one\> f\two
\end{align}
\begin{align}
\Omega^{-1}: F\tens F &\to F\tens F  \notag \\
\Omega^{-1}(h\tens f) &= \sum_{i,j} \<\gamma^{-2}( R_j'')R_i',h\two\> h\one
\tens \<R_i''R_j', f\one\> f\two
\end{align}

Assume according to (M3) that $\si\in F$ is a two-sided integral with
$\pi\circ \si=\mu$. Then
\begin{align}
S_-(\si) :F &\to F  \notag \\
S_-(\si)(f) &= \rho(\si\one\tens\gamma(f\one))
\rho(f\two\tens\gamma(\si\tre)) \si\two  \notag \\
&= \sum_{i,j} \<\gamma^{-1}( R_j''R_i'),f\> \<R_j'R_i'',\si\one\> \si\two \\
&= \sum_{i,j} \<\gamma^{-1}( R_j''R_i'),\si\two\> \<R_j'R_i'',f\> \si\one
\end{align}
\begin{align}
^tS_-(\si): U &\to U \notag \\
^tS_-(\si)(u) &= \sum_{i,j} \si(\gamma^{-1}(R_i'')u\gamma(R_j')) R_i'R_j''
\notag \\
&= \sum_{i,j} \si(u\gamma(R_i''R_j')) R_i'R_j'' \label{tS-sigma} \\
&= \sum_{i,j} \si(\gamma^{-1}(R_i''R_j')u) R_i'R_j''
\end{align}
\begin{align}
S_+(\si) :F &\to F  \notag \\
S_+(\si)(f) &= \rho(f\one\tens\si\one)
\rho(\si\tre\tens f\two) \si\two  \notag \\
&= \sum_{i,j} \<\gamma^{-2}(R_j'')R_i',f\> \<R_i''R_j',\si\one\> \si\two \\
&= \sum_{i,j} \<\gamma^{-2}(R_j'')R_i',\si\two\> \<R_i''R_j',f\> \si\one
\label{S+sigma}
\end{align}
\begin{align}
^tS_+(\si): U &\to U \notag \\
^tS_+(\si)(u) &= \sum_{i,j} \si(R_j'uR_i'') R_j''R_i' \label{tS+} \\
&= \sum_{i,j} \si(uR_i''R_j') \gamma^{-2}(R_j'') R_i' \label{tS+1} \\
&= \sum_{i,j} \si(\gamma^{-2}(R_i'')R_j'u) R_j''R_i'  \label{tS+2}
\end{align}

We know by \thmref{sl2onImP} that $P= S_+(\si) S_-(\si)$ is a projection
in $F$ with $\Ker P=\Ann\omega$. Therefore,
\[ {}^tP = {}^tS_-(\si) {}^tS_+(\si) = {}^tS_+(\si) {}^tS_-(\si) =
{}^tS_+(\si)^2 \gamma_U^{-1} \]
is a projection in $U$ with $\im {}^tP =\u$.

Let us determine now the normalization for the integral $\si$. Using the
pairing $\<,\>: U\tens F \to k$ one compute
\[ \< 1, \gamma^{-1}_F S_+(\si)^2(1)\> = \e(P(1)) =\e(1) =1 \]
since $1-P(1) \in \Ann\omega \subset \Ker\e$. On the other hand
\[ \< 1, \gamma^{-1}_F S_+(\si)^2(1)\>
= \<{}^tS_+(\si)^2 \gamma_U^{-1}(1) ,1\> = \<{}^tS_+(\si)^2 (1) ,1\> \]
so we have to satisfy $\e({}^tS_+(\si)^2 (1)) =1$. Using \eqref{tS+} we get
\begin{align*}
^tS_+(\si)(1) &= \sum_{i,j} \si(R_j'R_i'') R_j''R_i'  \\
^tS_+(\si)^2(1) &= \sum_{i,j,k,l} \si(R_j'R_i'')
\si(R_k'( R_j''R_i')R_l'') R_k''R_l' \\
\e({}^tS_+(\si)(1)) &= \sum_{i,j} \si(R_j'R_i'')\si(R_j''R_i' )
\end{align*}
Therefore, $\si$ is normalized so that
\be\label{normasig}
(\si\tens\si) (R^{12}R^{21}) =1.
\end{equation}

\subsubsection{2-modular case}
Let us assume less now, namely, that $(H,R,\ka)$ is 2-modular. Then the
formulae \eqref{Omegaleft}--\eqref{S+sigma}, \eqref{tS+1}, \eqref{tS+2}
with $F$ replaced by $\f$, $U$ replaced by $\u$ and $\si$ replaced by
$\mu\in\f$ or $\mu:\u\to k$ give the correct operators $\Omega^l_{X,\f}$,
$\Omega^r_{\f,X}$, $(\Omega_{\f,\f})^{\pm1}$, $S=S_-(\mu):\f\to\f$,
$S^{-1}=S_+(\mu):\f\to\f$, $^tS:\u\to\u$ and $^tS^{-1}:\u\to\u$.
Notice that
\[ \sum_{i,j} \gamma^{-2}(R_j'')R_i' \tens R_i''R_j' =
(\gamma^{-1}\tens1) (R^{21}R^{12})^{-1} \in \u\tens\u .\]

To determine the normalisation constant for $\mu$ we use \eqref{tS-sigma}.
As above, using the perfect pairing $\<,\>:\u\tens \f \to k$ we get
\[ 1= \<1, \gamma_\f S^2(1)\> = \e({}^tS^2(1)) =
\sum_{i,j} \mu( R_i'R_j'') \mu(\gamma(R_i''R_j')) ,\]
hence,
\[
(\mu\tens\mu\gamma) (R^{12}R^{21}) =1.
\]
In 3-modular case we can use either this formula or \eqref{normasig} in
the form
\[ \sum_{i,j} \mu(\gamma^{-2}(R_i'')R_j') \mu( R_j''R_i') =1 .\]

\begin{proposition}
Let $(H,R,\ka)$ be a 2-modular Hopf algebra. Then
\[ ^tS(\nu^{\pm1}) = \lambda^{\pm1} \nu^{\mp1}, \qquad
{}^tS^{-1}(\nu^{\pm1}) = \lambda^{\pm1} \nu^{\mp1}, \]
\[ \lambda=\mu(\nu), \qquad \lambda^{-1}=\mu(\nu^{-1}) .\]
Moreover, if $(H,R,\ka)$ is 3-modular, then
\be\label{tS+tS-}
^tS_+(\si)(\nu^{\pm1}) = \lambda^{\pm1} \nu^{\mp1}, \qquad
{}^tS_-(\si)(\nu^{\pm1}) = \lambda^{\pm1} \nu^{\mp1},
\end{equation}
\[ \lambda=\si(\nu), \qquad \lambda^{-1}=\si(\nu^{-1}) .\]
\end{proposition}

The perfect modular case was considered in \cite{LyuMaj}.

\begin{pf}
Both cases being similar, we prove the second statement. Equations
\eqref{nulambda}, \eqref{nu-lambda-} mean that for any $H$-module $X$
and any vector $x\in X$
\[ (\e\tens1) \Omega_r(T^{\pm1}(\si)\tens x) = \lambda^{\pm1}\nu^{\mp1} x.\]
Using \eqref{T(f)} and \eqref{Omegar=}
\begin{align*}
(\e\tens1) \Omega_r(T^{\pm1}(\si)\tens x) &=
(\e\tens1) \Omega_r(\<\nu^{\pm1},\si\one\>\si\two \tens x) \\
&= \<\nu^{\pm1},\si\one\> \sum_{i,j} \<\gamma^{-1}( R_j''R_i'),\si\tre\>
\e(\si\two) R_j'R_i''x  \\
&= \sum_{i,j} \<\nu^{\pm1},\si\one\> \<\gamma^{-1}( R_j''R_i'),\si\two\>
R_j'R_i''x  \\
&= \sum_{i,j} \<\gamma^{-1}( R_j''R_i')\nu^{\pm1} ,\si\> R_j'R_i''x  \\
&= {}^tS_-(\si)(\nu^{\pm1}) x .
\end{align*}
Therefore, $^tS_-(\si)(\nu^{\pm1}) = \lambda^{\pm1}\nu^{\mp1}$. Applying
$\gamma_U$ we get the same result for $^tS_+(\si)$. Appying $\e$ to
\eqref{tS+tS-} we get $\si(\nu^{\pm1}) = \lambda^{\pm1}$ by \eqref{e(nu)}.
\end{pf}

\section{Quantum groups}\label{groups}
\subsection{$R$-matrices for quantum groups}
\begin{theorem}[\cite{KirRes:mult,KhoTol:R-mat,LevSoi:weyl}]\label{R=RR0}
The expression $R=\tilde R \Ro \in  U_h(\g)\widehat\tens U_h(\g)$ is an
$R$-matrix in the topological Hopf algebra $U_h(\g)$, where
\begin{align}
\tilde R &= \prod_{\beta\in(\beta_1,\dots,\beta_N)}
\exp_{q_\beta^{-2}} ((q_\beta-q_\beta^{-1}) E_\beta\tens F_\beta) ,
\label{tildeR=} \\
\Ro &= \exp( \frac h2 \sum c_{ij} H_i\tens H_j) ,  \notag
\end{align}
where $q_\beta = e^{h(\beta|\beta)/2}$ and $(c_{ij})$ is the inverse
matrix to $(d_ia_{ij})$.
\end{theorem}

\begin{proposition}[cf \cite{Tan:form}]\label{R0E1R-1}
The Cartan part of the $R$-matrix, $\Ro\in U_h(\h)\widehat\tens U_h(\h)$
satisfies
\begin{alignat*}2
\Ro \cdot E_i\tens1 \cdot \Ro^{-1} &= E_i \tens K_i, &\quad
\Ro \cdot F_i\tens1 \cdot \Ro^{-1} &= F_i \tens K_i^{-1}, \\
\Ro \cdot 1\tens E_i \cdot \Ro^{-1} &= K_i \tens E_i, &\quad
\Ro \cdot 1\tens F_i \cdot \Ro^{-1} &= K_i^{-1} \tens F_i .
\end{alignat*}
\end{proposition}

Consider an embedding of Hopf algebras
$U_q(\g) \hookrightarrow U_h(\g) \tens_{\C[[h]]} \C[h^{-1},h]]$,
$K_i \mapsto e^{hd_iH_i}$, with respect to a field extension
$\Q(q) \hookrightarrow \C[h^{-1},h]]$, $q \mapsto e^h$. The image of
Lusztig's divided power algebra $\Gamma(\g)$ is contained in $U_h(\g)$. Let
\begin{align*}
U_h(\g)^\al &= \{ x\in U_h(\g) \mid [H_i,x] = \al(H_i) x \} \\
\Gamma(\g)^\al &= \{ x\in \Gamma(\g) \mid
K_i x K_i^{-1} = q^{(\al_i|\al)} x \}
\end{align*}
be natural gradings, $\al\in Q$. Note that
$\Gamma(\g)^\al \subset U_h(\g)^\al$. We have
$R,\tilde R \in \break
\prod_{\al\in Q_+} (U_h(\b_+)^\al \widehat\tens U_h(\b_-)^{-\al})$.
Denote $\tilde R= \sum_{\al\in Q_+} \tilde R_\al$, then
$\tilde R_\al \in U_h(\b_+)^\al \tens U_h(\b_-)^{-\al}$.
Combining \thmref{R=RR0} and \propref{R0E1R-1} we get

\begin{proposition}[cf \cite{Tan:form}]\label{propRtilda}
In $U_h(\g)^{\widehat \tens3}$ (or $U_h(\g)^{\widehat \tens2}$) the
following equations are satisfied
\begin{align}
(\Delta \tens1) \tilde R &= \tilde R^{13} \cdot
\sum_{\al\in Q_+} K_{-\al}\tens \tilde R_\al \label{Delta1tildeR}\\
(1\tens\Delta) \tilde R &= \tilde R^{13} \cdot
\sum_{\al\in Q_+}  \tilde R_\al \tens K_{\al} \\
\Delta^\op x \cdot \tilde R &= \tilde R \cdot \tilde\Delta x
\label{DeltaopxtildeR}
\end{align}
where $x\in U_h(\g)$ and the new coproduct
$\tilde \Delta x = \Ro \Delta x \Ro^{-1}$ satisfies
\begin{align*}
\tilde\Delta H_i &= H_i\tens1 + 1\tens H_i \\
\tilde\Delta E_i &= E_i\tens K_i + 1\tens E_i \\
\tilde\Delta F_i &= F_i\tens1 + K_i^{-1}\tens F_i
\end{align*}
\end{proposition}

\begin{pf}
Since $\tilde R\Ro$ is the $R$-matrix we have
\begin{align*}
(\Delta \tens1) (\tilde R\Ro)
&= \tilde R^{13}\Ro^{13} \cdot \tilde R^{23} \Ro^{23} \\
&= \tilde R^{13} \cdot (\sum_{\al\in Q_+}
K_{-\al}\tens \tilde R_\al) \cdot \Ro^{13}\Ro^{23} \\
&= \tilde R^{13} \cdot (\sum_{\al\in Q_+}
K_{-\al}\tens \tilde R_\al) \cdot (\Delta\tens1)(\Ro)
\end{align*}
whence \eqref{Delta1tildeR}. Other properties are similar.
\end{pf}

Moreover, $\tilde R_\al \in \Gamma(\b_+)^\al \tens_{\Z[q,q^{-1}]}
\Gamma(\b_-)^{-\al} \tens_{\Z[q,q^{-1}]} \C[[h]]$ due to the formula
\be\label{q-exp}
\exp_{q_\beta^{-2}} ((q_\beta-q_\beta^{-1}) E_\beta\tens F_\beta) =
\sum_{m\ge0} q_\beta^{m(m-1)} (q_\beta-q_\beta^{-1})^m
(m)_{q_\beta^{-2}}! E_\beta^{(m)}\tens F_\beta^{(m)} .
\end{equation}
Since $\Gamma(\g)$ is a free $\Z[q,q^{-1}]$-module \cite{Lus:roots1},
the equations
\eqref{Delta1tildeR}--\eqref{DeltaopxtildeR} with $x\in \Gamma(\g)$
can be interpreted as follows. All  terms are elements of the
$\Z[q,q^{-1}]$-module $\prod_{\al,\beta,\gamma\in Q}
\Gamma(\g)^\al \tens \Gamma(\g)^\beta \tens \Gamma(\g)^\gamma$ (or of
$\prod_{\al,\beta\in Q} \Gamma(\g)^\al \tens \Gamma(\g)^\beta\ni\tilde R$)
and the equations state in particular that these elements can be multiplied.

Let $\e\in\C$ be a root of unity. Change the base by the homomorphism
$\Z[q,q^{-1}]\to\C$, $q\mapsto\e$ and denote
\[ \Gamma_\e(\g) = \Gamma(\g) \tens_{\Z[q,q^{-1}]} \C .\]
The equations \eqref{Delta1tildeR}--\eqref{DeltaopxtildeR} still hold
for $\Gamma_\e(\g)$ in place of $\Gamma(\g)$. But the sum \eqref{q-exp}
is finite, so $\tilde R_\al$ vanish for all $\al$ except finite
number and $\tilde R \in \Gamma_\e(\b_+) \tens \Gamma_\e(\b_-)$.
Therefore, \eqref{Delta1tildeR}--\eqref{DeltaopxtildeR} hold in
$\Gamma_\e(\g)^{\tens3}$ or $\Gamma_\e(\g)^{\tens2}$ with
$x\in\Gamma_\e(\g)$.

Reversing the proof of \propref{propRtilda} we see that if we find a
symmetric tensor $\Ro\in \Gamma_\e(\h)^{\tens2}$ satisfying
$(\Delta\tens1)\Ro = \Ro^{13} \Ro^{23}$ and
\begin{align}
\Ro \cdot 1\tens E_i^{(p)} \cdot \Ro^{-1}
&= K_i^p \tens E_i^{(p)} \label{wan} \\
\Ro \cdot 1\tens F_i^{(p)} \cdot \Ro^{-1}
&= K_i^{-p} \tens F_i^{(p)} \label{twu}
\end{align}
for all $p\in \Z_{>0}$, we could construct an $R$-matrix
$R=\tilde R \Ro$ for $\Gamma_\e(\g)$. It turns out that such element
is easy to find not in $\Gamma_\e(\g)$ but in its quotient (which
sometimes coincide with $\Gamma_\e(\g)$). Consider a symmetric bilinear
(bimultiplicative) form
\[ \pi: Q\times Q \to\C^\times, \qquad \pi(K_i,K_j) = q_i^{a_{ij}},
\qquad  \pi(K_\al,K_\beta) = q^{(\al|\beta)} .\]
Let $\Ann\pi=\{g\in Q \mid \forall h\quad \pi(g,h)=1 \}$ be its annihilator.
Since for $h\in Q$
\begin{align*}
h E_j^{(p)} &= \pi(K_j,h)^p E_j^{(p)} h \\
h F_j^{(p)} &= \pi(K_j,h)^{-p} F_j^{(p)} h
\end{align*}
the elements of $\Ann \pi$ lie in the centre of $\Gamma_\e(\g)$.
Introduce the Hopf $\C$-algebra
\[ \Gamma'_\e(\g) = \Gamma_\e(\g)/(g-1)_{g\in\Ann\pi} .\]
The subgroup generated by $K_i$ in $\Gamma'_\e(\g)$ is denoted
$\G=Q/\Ann\pi$. The form $\pi$ factorizes through a non-degenerate
form on $\G$ denoted also $\pi$ by abuse of notations.

Clearly, $K_i^{2l_i}=1$ in $\G$, where $l_i$ are minimal positive integers
such that $q_i^{2l_i}=1$. More elements from $\Ann\pi$ can be found
via the following

\begin{lemma}\label{rho(d)}
For any $d=1,2,3$ let $l(d)$ be the minimal positive integer $l$ such that
$q^{2dl}=1$ and let
$\rho(d)=\sum\begin{Sb}||\al||^2=2d\\\al\in\Delta^+\end{Sb} \al$.
Then for any $i$
\[ q_i^{l(d) \<\al_i,\rho(d)\>} =1 .\]
\end{lemma}

\begin{pf}
The equation to prove is
\[ q^{l(d) (\al_i|\rho(d))} =1 .\]
Assume first that $||\al||^2\ne 2d$. Let $s_i$ be the simple reflection
corresponding to $\al_i$. The well known equation
$s_i(\Delta^+-\al_i) = \Delta^+-\al_i$ implies
$s_i(\Delta^+_d) = \Delta^+_d$, where
$\Delta^+_d = \{\al\in\Delta^+ \mid ||\al||^2=2d \}$. Hence,
$s_i(\rho(d))=\rho(d)$ and
\[ (\al_i|\rho(d)) = (s_i(\al_i))|s_i(\rho(d)))
= - (\al_i|\rho(d)) \]
vanishes.

Assume now that $||\al||^2=2d$. Then
$s_i(\Delta^+_d-\al_i) = \Delta^+_d-\al_i$ and
$s_i(\rho(d)-\al_i)=\rho(d)-\al_i$. Hence,
\[ (\al_i|\rho(d)) = (s_i(\al_i))|s_i(\rho(d)))
= - (\al_i|\rho(d)-2\al_i) \]
implies $(\al_i|\rho(d)) = (\al_i|\al_i) = 2d$ and the lemma
follows.
\end{pf}

\begin{corollary}
For any $i$
\[ q_i^{ \<\al_i, \sum_{\al\in\Delta^+} l_\al \al \>} =1 .\]
\end{corollary}

\begin{corollary}\label{prodKallalpha}
For any $d=1,2,3$ we have $K_{\rho(d)}^{l(d)} =1$ in $\G$. In particular,
\[ \prod_{\al\in\Delta^+} K_\al^{l_\al} =1 .\]
\end{corollary}

The $R$-matrix from the following theorem was already obtained by
Rosso \cite{Ros} in the case of $l$ relatively prime to $\det(d_ia_{ij})$
using the Drinfeld's double.

\begin{theorem}\label{R=RRo}
The Hopf algebra $H= \Gamma'_\e(\g)$  is quasitriangular with the
$R$-matrix $R=\tilde R\Ro$ where $\tilde R$ is given by \eqref{tildeR=} and
\[ \Ro = \frac1{|\G|} \sum_{g,h\in\G} \pi(g,h)^{-1} g\tens h .\]
\end{theorem}

\begin{pf}
We check the property \eqref{wan} of $\Ro\in\Gamma'_\e(\g)^{\tens2}$
{\allowdisplaybreaks
\begin{align*}
\Ro\cdot1\tens E_i^{(p)}
&= \frac1{|\G|} \sum_{g,h\in\G} \pi(g,h)^{-1} g\tens hE_i^{(p)} \\
&= \frac1{|\G|} \sum_{g,h\in\G} \pi(g,h)^{-1} \pi(K_i,h)^p
g\tens E_i^{(p)} h \\
&= \frac1{|\G|} \sum_{g,h\in\G} \pi(gK_i^{-p},h)^{-1} g\tens E_i^{(p)} h \\
&= \frac1{|\G|} \sum_{g,h\in\G} \pi(f,h)^{-1} K_i^p f \tens E_i^{(p)} h \\
&= K_i^p \tens E_i^{(p)} \cdot \Ro
\end{align*}
Similarly for \eqref{twu}.

$\Ro$ is an $R$-matrix for $\Gamma'_\e(\h)$. Indeed,
\begin{align*}
\Ro^{13}\Ro^{23} &= \frac1{|\G|^2}
\sum_{g,h,k,l\in\G} \pi(g,h)^{-1} \pi(k,l)^{-1} g\tens k \tens hl \\
&= \frac1{|\G|^2} \sum_{g,k,f\in\G} g\tens k \tens f
\sum_{h\in\G} \pi(g,h)^{-1} \pi(k,fh^{-1})^{-1} \\
&= \frac1{|\G|^2} \sum_{g,k,f\in\G} \pi(k,f)^{-1} g\tens k \tens f
\sum_{h\in\G} \pi(kg^{-1},h) \\
&= \frac1{|\G|^2} \sum_{g,k,f\in\G} \pi(k,f)^{-1} g\tens k \tens f
|\G|\delta_{k,g} \\
&= \frac1{|\G|} \sum_{g,f\in\G} \pi(g,f)^{-1} g\tens g \tens f \\
&= (\Delta\tens1) \Ro
\end{align*}
}
and by symmetry $(1\tens\Delta) \Ro = \Ro^{13}\Ro^{12}$. It follows from the
above discussion that $\tilde R\Ro$ is an $R$-matrix for $\Gamma'_\e(\g)$.
\end{pf}

\subsection{The algebra $\uqg$}
Assume that for $q=\e$ we have $\e^{2m}\ne1$ for all $0<m\le d_i$.
Rewrite \thmref{R=RRo} as
\be\label{R=sumprod}
R = \sum_{0\le m_\al<l_\al} \prod_\al
\frac{(q_\al-q_\al^{-1})^{m_\al}}{(m_\al)_{q_\al^{-2}}!}
\prod_\al E_\al^{m_\al} \tens
\prod_\al F_\al^{m_\al} \cdot \Ro
\end{equation}
($\al$ runs over $\beta_1,\dots,\beta_N$). Non-degeneracy of
$\pi: \G\times\G \to \C^\times$ implies that the minimal subspaces
$A,B\subset \C[\G]$ such that $\Ro\in A\tens B$ are $A=B=\C[\G]$. Thus,
\eqref{R=sumprod} implies that the smallest subspaces
$\hh_+,\hh_-\subset \Gamma'_\e(\g)$ such that $R\in \hh_+\tens\hh_-$
have bases
$h\prod_{\beta\in(\beta_1,\dots,\beta_N)} E_\beta^{k_\beta}$, resp.
$h\prod_{\beta\in(\beta_1,\dots,\beta_N)} F_\beta^{m_\beta}$, where
$h\in\G$, $0\le k_\beta<l_\beta$. As discussed in \secref{qtrHa}
(cf. \cite{Rad:min}) $\hh_+$ and $\hh_-$ are Hopf subalgebras. They
coincide with the subalgebras
$u_q(\b_+) = \C\<K_i^{\pm1}, E_i\>_{1\le i\le n}$ and
$u_q(\b_-) = \C\<K_i^{\pm1}, F_i\>_{1\le i\le n}$ since $E_i\in\hh_+$,
$E_\beta\in u_q(\b_+)$ by e.g. \cite[Proposition 40.1.3]{Lus:book}.
Therefore, by general theory $\uqg =
\C\<K_i^{\pm1}, E_i,F_i\>_{1\le i\le n} = \hh = \hh_+\hh_- = \hh_-\hh_+$
is a quasitriangular Hopf subalgebra of $\Gamma_\e'(\g)$. It has the basis
$\prod_{\al\in(\beta_1,\dots,\beta_N)} E_\al^{k_\al} \cdot h \cdot
\prod_{\beta\in(\beta_1,\dots,\beta_N)} F_\beta^{m_\beta}$, where
$h\in\G$, $0\le k_\beta, m_\beta<l_\beta$. The $R$-matrix
$R\in u_q(\b_+)\tens u_q(\b_-)$ is the dual tensor
to the non-degenerate Hopf pairing
\begin{alignat*}2
\pi:\hh_-\times \hh_+^\op&\to \C \\
\pi(h,E_i) &=0 &&\quad \hbox{ for } \ h\in \G, \\
\pi(F_i,h) &=0 &&\quad \hbox{ for } \ h\in \G, \\
\pi(F_i,E_j) &=\delta_{ij} (q_i-q_i^{-1})^{-1} &&\quad
          \hbox{ for } \ 1\le i,j\le n.
\end{alignat*}

\subsubsection{A presentation of $u_q(\g)$}
A lemma of Levendorskii and Soibelman \cite{LevSoi:CMP} states: for any
positive roots $\al<\beta$ there are unique constants
$c_{n_1,\dots,n_j} \in \C(q)$, such that in $U_q(\g)\tens_{\Q(q)} \C(q)$
\be\label{LeSole}
E_\al E_\beta - q^{-(\al|\beta)} E_\beta E_\al = \sum_n c_{n_1,\dots,n_j}
E_{\gamma_1}^{n_1} \dots E_{\gamma_j}^{n_j} ,
\end{equation}
where $j$ is the number of all positive roots lying between $\al$ and
$\beta$, which are denoted $\al<\gamma_1<\dots<\gamma_j<\beta$.
When this lemma is combined with Lusztig's basis theorem for $\Gamma(\g)$
\cite[Propositions 41.1.4, 41.1.7]{Lus:book} we see that in fact
\[ c_{n_1,\dots,n_j} \in \K =
\Z[q,q^{-1},(q-q^{-1})^{-1}, [N(\g)]_q!^{-1}] , \]
where the constant $N(\g)\ge \max_i d_i$ is minimal possible.
An upper bound for it is given in Table~\ref{table}.
\begin{table}
\centering
\caption{An estimate for the constants $N(\g)$}\label{table}
\begin{tabular}{|c|c|c|c|c|c|c|c|c|c|}\hline
$\g$      &$A_n$&$B_n$&$C_n$&$D_n$&$E_6$&$E_7$&$E_8$&$F_4$&$G_2$ \\ \hline
$N(\g)\le$& $1$ & $4$ & $4$ & $2$ & $3$ & $4$ & $6$ & $8$ & $9$  \\ \hline
\end{tabular}
\end{table}
It is obtained as the product of $\max_id_i$ with the maximal coefficient
$c_i$ in decomposition $\al_0=\sum c_i\al_i$ of the highest root
$\alpha_0\in\Delta$. (Apply
$T_{i_k}^{-1} T_{i_{k-1}}^{-1} \dots T_{i_1}^{-1}$ to \eqref{LeSole},
if $E_\al = T_{i_1} \dots T_{i_{k-1}} E_{i_k}$.)
This estimate is rather rough, and can be essentially improved.
The conjectured value of $N(\g)$ is $\max_i d_i$.

Introduce a $\K$-subalgebra $\U_q(\g) \subset U_q(\g)$ generated by
$K_i,E_i,F_i$. It is closed under the automorphisms $T_i$ and all relations
\eqref{LeSole} make sense in $\U_q(\g)$, thus $\U_q(\g)$ has a
Poincar\'e--Birkhoff--Witt basis $ \prod_\al  E_\al^{m_\al} \cdot K_\lambda
\cdot \prod_\beta F_\beta^{n_\beta} $ (compare \cite{LevSoi:JGP,Lus:book}).
Clearly, $\U_q(\g)$ is a subalgebra of $\Gamma(\g) \tens_{\Z[q,q^{-1}]} \K$.

Assume in this subsection that $\e^{2m}\ne1$ for all $1\le m \le N(\g)$.
Then we have a homomorphism of Hopf algebras
\be\label{UtoGamma}
\phi: \U_q(\g) \hookrightarrow \Gamma(\g) \tens_{\Z[q,q^{-1}]} \K
\to \Gamma_\e(\g) \to \Gamma'_\e(\g)
\end{equation}
via the homomorphism $\K\to \C$, $q\mapsto \e$. Since $\C\im \phi = \uqg$
the relations  \eqref{LeSole} are valid also in $\uqg$.

The algebra $U_\e(\g) = \U_q(\g)\tens_{\K}\C$
has the usual generators and relations of $U_q(\g)$ with $q$ set to $\e$.
The obvious homomorphism $\tilde\phi: U_\e(\g) \to \Gamma_\e'(\g)$
induced by \eqref{UtoGamma} has $u_q(\g)$ as its image. Its kernel contains
$E_\al^{l_\al}$, $F_\al^{l_\al}$ and $h-1$ for
$h\in\Ann(\pi: Q\times Q \to \C^\times)$. Indeed,
$E_\al^{l_\al} = [l_i]_{q_i}! T_{i_1} \dots T_{i_{k-1}} (E_i^{(l_i)})$
in $U_q(\g)$ and $\Gamma(\g)$ for some sequence $(i_1,\dots,i_{k-1},i)$,
and the factorial vanishes in $\Gamma_\e'(\g)$. Let $I\subset U_\e(\g)$
be a two-sided ideal generated by these elements. By the explicit form
of bases we get $\dim U_\e(\g)/I \le \dim u_q(\g)$.
Since $\tilde\phi$ induces an epimorphism $U_\e(\g)/I \to u_q(\g)$,
this is in fact an isomorphism, and a presentation of $u_q(\g)$ by
generators and relations follows. Namely, the new relations
\[ E_\al^{l_\al} = 0, \qquad F_\al^{l_\al} = 0, \qquad h=1 \
\text{ for } \al\in\Delta^+, \ h\in\Ann(\pi: Q\times Q \to \C^\times) \]
are added to the standard presentation of $U_\e(\g)$.

\subsubsection{The special grouplike element for $\uqg$}
Let us find the grouplike element $g=u\gamma(u)^{-1}$ for the algebras $H$
and $\hh$, where $u=\sum_n\gamma(b^n)a_n$, $R=\sum_n a_n\tens b^n$. Clearly,
\begin{align*}
\delta_+ &= \sum_{h\in\G}h \cdot
\prod_{\beta\in(\beta_1,\dots,\beta_N)} E_\beta^{l_\beta-1} \in\hh_+ ,\\
\delta_- &= \sum_{h\in\G}h \cdot
\prod_{\beta\in(\beta_1,\dots,\beta_N)} F_\beta^{l_\beta-1} \in\hh_-
\end{align*}
are non-zero left integrals in the algebras $\hh_+$ and $\hh_-$. For any
$y\in\hh_+$ and $a_-=\prod_\beta K_\beta^{-l_\beta+1} = K_{2\rho} \in \hh_-$
(cf \corref{prodKallalpha}) we have
\[ \delta_+ y = \pi(a_-,y) \delta_+ .\]
Indeed, it suffices to check this for $y\in\G$
\begin{align*}
\delta_+ y &= \sum_{h\in\G}h \cdot \prod_\beta E_\beta^{l_\beta-1} y \\
&= \sum_{h\in\G} hy \cdot
\prod_\beta \pi(K_\beta,y)^{-l_\beta+1}  E_\beta^{l_\beta-1} \\
&=  \pi(\prod_\beta K_\beta^{-l_\beta+1} ,y) \delta_+
\end{align*}

Similarly, for any $y\in\hh_-$ and
$a_+=\prod_\beta K_\beta^{l_\beta-1} = K_{-2\rho} \in \hh_+$
(cf \corref{prodKallalpha}) we have
\[ \delta_- y = \pi(y,a_+) \delta_- .\]
By Drinfeld's theorem \cite{Dri:cocom} (see \secref{qtrHa}) we find
\[ g = a^{-1}_+ a_- = K_{4\rho} \in \hh .\]

\subsection{Ribbon structure of $\Gamma'_\e(\g)$ and $\uqg$}
By definition a ribbon structure is a choice of a group-like element
$\ka$ such that $\ka^2=K_{4\rho}$ and $\ka a=\gamma^2(a)\ka$
for all $a\in \Gamma'_\e(\g)$. The only group-like elements of
$\Gamma'_\e(\g)$ are $K_\al\in\G$. Commuting $\ka$ with $E_i$ we
get $\pi(\ka,K_i)=q^2_i$, wich holds for the only element
$\ka=K_{2\rho}$. Therefore, the quasitriangular Hopf algebra
$\Gamma'_\e(\g)$ (or $\uqg$) admits the unique ribbon structure
$\ka=K_{2\rho}$.

Let us find the ribbon twist element using the formula
$\nu=\sum_n a_nK_{2\rho}^{-1}b^n$, $R=\sum_n a_n\tens b^n$. Equation
\eqref{R=sumprod} can be rewritten as
\begin{align*}
R &= \sum_{0\le m_\al<l_\al} \Bigl(\prod_\al
\frac{(q_\al-q_\al^{-1})^{m_\al}}{(m_\al)_{q_\al^{-2}}!}
E_\al^{m_\al} \tens1 \Bigr) \cdot \Ro \cdot
\Bigl( \prod_\al K_\al^{m_\al}\tens F_\al^{m_\al} \Bigr) \\
&= \sum_{0\le m_\al<l_\al} \Bigl(\prod_\al
\frac{(q_\al-q_\al^{-1})^{m_\al}}{(m_\al)_{q_\al^{-2}}!}
E_\al^{m_\al} \Bigr) \Ro_a' \Bigl(\prod_\al K_\al^{m_\al} \Bigr) \tens
\Ro_a'' \Bigl( \prod_\al  F_\al^{m_\al} \Bigr) .
\end{align*}
where $\Ro = \sum_a \Ro_a'\tens \Ro_a''$. Therefore,
\be\label{nu=sumprod}
\nu = \sum_{0\le m_\al<l_\al} \Bigl(\prod_\al
\frac{(q_\al-q_\al^{-1})^{m_\al}}{(m_\al)_{q_\al^{-2}}!}
E_\al^{m_\al} \Bigr) \cdot \Ro_a' \Ro_a'' K_{-2\rho}
\Bigl(\prod_\al K_\al^{m_\al} \Bigr) \cdot
 \Bigl( \prod_\al  F_\al^{m_\al} \Bigr) .
\end{equation}

\subsubsection{The subalgebra $\u$}
Recall that the subalgebra $\u\subset\hh$ is defined as the smallest
subspace such that $R^{12}R^{21} \in \u\tens\hh$. Let us find it.

\thmref{R=RRo} gives also an alternative expression for
\[ R = \Ro\cdot \prod_{\beta\in(\beta_1,\dots,\beta_N)} \exp_{q_\beta^{-2}}
((q_\beta-q_\beta^{-1}) E_\beta K_\beta \tens K_\beta^{-1} F_\beta) \]
which yields
\begin{multline}
R^{12}R^{21} = \sum_{n,m} \Bigl(\prod_\al
\frac{(q_\al-q_\al^{-1})^{m_\al}}{(m_\al)_{q_\al^{-2}}!}
E_\al^{m_\al} \Bigr)  \Ro_a' \Ro_b''  \Bigl(\prod_\beta
(K_\beta^{-1}F_\beta)^{n_\beta} \Bigr) \tens \label{R12R21} \\
\tens\Bigl( \prod_\al  F_\al^{m_\al} \Bigr) \Ro_a''\Ro_b' \Bigl(\prod_\beta
\frac{(q_\beta-q_\beta^{-1})^{n_\beta}}{(n_\beta)_{q_\beta^{-2}}!}
(E_\beta K_\beta)^{n_\beta} \Bigr) .
\end{multline}

\begin{proposition}\label{basisu}
The algebra $\u$ has the basis
\[ \bigl( \prod_\al  E_\al^{m_\al} \bigr) K_\lambda
\bigl(\prod_\beta F_\beta^{n_\beta} \bigr) ,\]
where $K_\lambda = K_{2\mu} \cdot \prod_\beta K_\beta^{n_\beta}$ in $\G$
for some $\mu \in Q$.
\end{proposition}

This follows from the formula \eqref{R12R21} above and

\begin{lemma}
The smallest subspace $A\subset \C[\G]$ such that $\Ro^2\in A\tens\C[\G]$
is $A=\C[\twoG]$, where $\twoG = \{ x^2 \mid x\in \G \} \subset \G$
is the subgroup of squares.
\end{lemma}

\begin{pf}
Note that $\Ro$ is a symmetric tensor and
\begin{align*}
\Ro^2 &= \frac1{|\G|} \sum_{g,h\in\G} \pi(g,h)^{-1} g\tens h
\cdot \frac1{|\G|} \sum_{a,b\in\G} \pi(a,b)^{-1} a\tens b \\
&= \frac1{|\G|^2} \sum_{c,d\in\G} c\tens d
\sum_{g,b\in\G} \pi(g,bd^{-1}) \pi(gc^{-1},b) \\
&= \frac1{|\G|^2} \sum_{c,d\in\G} c\tens d
\sum_{g,b\in\G} \pi(g,b^2d^{-1}) \pi(c^{-1},b) \\
&= \frac1{|\G|} \sum_{c,d\in\G} c\tens d
\sum_{b\in\G} \delta_{d,b^2} \pi(c,b)^{-1} .
\end{align*}
This implies that $\Ro^2 \in \C[\twoG] \tens \C[\twoG]$. Introduce a
symmetric bilinear (bimultiplicative) pairing
\[ \pi_2: \twoG \times \twoG \to \C^\times, \qquad \pi_2(c,d) = \pi(c,b),
\text{ where } b^2=d . \]
Clearly, $\pi(c,b)$ has the same value for all $b\in\G$ such that $b^2=d$.
All such $b$ differ by an element of $X=\{ b\in\G \mid b^2=1 \}$. Therefore,
\begin{align}
\Ro^2 &= \frac1{|\G|} \sum_{c,d\in\twoG} \pi_2(c,d)^{-1} c\tens d
\sum_{b\in\G} \delta_{d,b^2} \notag \\
&= \frac{|X|}{|\G|} \sum_{c,d\in\twoG} \pi_2(c,d)^{-1} c\tens d \notag \\
&= \frac1{|\twoG|} \sum_{c,d\in\twoG} \pi_2(c,d)^{-1} c\tens d
\label{Rosquare}
\end{align}
and the statement follows.
\end{pf}

\begin{corollary}
$\uqg$ is factorizable if and only if $\twoG=\G$. In particular, it is
factorizable if the degree $l$ of the root $\e$ is odd.
\end{corollary}

\subsection{2-modular structure of $\uqg$ and $\Gamma'_\e(\g)$}
Now we can answer the question when $\uqg$ is 2-modular.

\begin{theorem}\label{2-modthm}
The ribbon Hopf algebras $\uqg$ and $\Gamma'_\e(\g)$ are $2$-modular,
that is $\nu\in \u$, if and only if for any $x\in\G$ such that $x^2=1$
we have
\be\label{pi=pi}
\pi(x,x) = \pi(x, K_{2\rho}) .
\end{equation}
\end{theorem}

\begin{remark}
Both sides of \eqref{pi=pi} are characters $X\to \{1,-1\}$, where $X$
is the subgroup $\{ x\in\G \mid x^2=1 \}$.
\end{remark}

\begin{pf}
Equation \eqref{nu=sumprod} together with \propref{basisu} imply that
$\nu\in\u$ if and only if $\Ro_a' \Ro_a'' K_{-2\rho} \in \C[\twoG]$. We have
\be\label{GausFou}
\Ro_a' \Ro_a'' K_{-2\rho} = \frac1{|\G|} \sum_{g,h\in\G} \pi(g,h)^{-1}
ghK_{-2\rho} = \frac1{|\G|} \sum_{b\in\G} \phi(b)b ,
\end{equation}
where the coefficient $\phi(b)$ is the Gaussian sum
\be\label{Gauss}
\phi(b) = \sum_{g\in\G} \pi(g,g) \pi(g,b^{-1}K_{-2\rho}) .
\end{equation}
Its absolute value is found by the standard procedure
\begin{align}
|\phi(b)|^2 &= \sum_{h,x\in\G} \pi(hx,hx) \pi(hx,b^{-1}K_{-2\rho})
\pi(h,h)^{-1} \pi(h,b^{-1}K_{-2\rho})^{-1} \notag \\
&= \sum_{x\in\G} \pi(x,x) \pi(x,b^{-1}K_{-2\rho})
\sum_{h\in\G}\pi(h,x^2) \notag \\
&= |\G| \sum_{x\in X} \pi(x,x) \pi(x,b^{-1}K_{-2\rho}) . \label{|phi|}
\end{align}

This formula suggests to consider a function on $Y=\G/\twoG$
\[ \psi(y) = \sum_{x\in X} \pi(x,x) \pi(x,y) .\]
Here the pairing $\pi$ restricts to a non-degenerate pairing
$\pi:X\times Y \to \{1,-1\}$. Thus the algebra $\uqg$ is 2-modular iff
$\psi(y)=\text{const} \delta_{y,[K_{-2\rho}]}$. This condition means that
the Fourier coefficients $\pi(x,x)$ of the function $\psi$ are
proportional to the Fourier coefficients $\pi(x,K_{2\rho})$ of the delta
function $\delta_{y,[K_{-2\rho}]}$. Equivalently,
$\pi(x,x) = \pi(x, K_{2\rho})$ for all $x\in X$ since the proportionality
constant is 1.
\end{pf}

\subsection{The integral on $\uqg$}
\begin{proposition}
The functional $\int: \uqg \to \C$
\[ \int \bigl( \prod_\al  E_\al^{m_\al} \cdot K_\lambda \cdot
\prod_\beta  F_\beta^{n_\beta} \bigr) =
\prod_\al \delta_{m_\al,l_\al-1} \cdot \delta_{K_\lambda,K_{-2\rho}}
\cdot \prod_\beta \delta_{n_\beta, l_\beta-1} \]
is a left integral  on the Hopf algebra $\uqg$. It is independent
of the choice of the reduced expression for $w_0$.
\end{proposition}

\begin{pf}
Notice that for $\tau=\sum_{\al\in\Delta^+} (l_\al-1) \al$
\begin{align*}
(1\tens\int) \Delta \bigl( \prod_\al  E_\al^{l_\al-1} \cdot K_\lambda
\cdot \prod_\beta  F_\beta^{l_\beta-1} \bigr) &= K_{-\tau} K_\lambda
\int \bigl( \prod_\al  E_\al^{l_\al-1} \cdot K_\lambda
\cdot \prod_\beta  F_\beta^{l_\beta-1} \bigr) \\
&= K_{2\rho} K_{-2\rho}
\int \bigl( \prod_\al  E_\al^{l_\al-1} \cdot K_\lambda
\cdot \prod_\beta  F_\beta^{l_\beta-1} \bigr)
\end{align*}
since $K_{-\tau} = K_{2\rho}$ by \corref{prodKallalpha}. This implies
$(1\tens\int) \Delta x = \int x$.

All left integrals are proportional to each other, whence all functionals
$\int$ defined for different reduced expressions are proportional to each
other. The elements of the highest grade
$\prod_\al  E_\al^{l_\al-1} \tens \prod_\beta F_\beta^{l_\beta-1}$
are also proportional to each other. Since
\[ \pi(\prod_\beta F_\beta^{l_\beta-1} \tens \prod_\al  E_\al^{l_\al-1})
= \prod_\al \left( (l_\al-1)_{q_\al^{-2}}!
(q_\al-q_\al^{-1})^{1-l_\al} \right) \]
(see \cite{KirRes:mult,KhoTol:R-mat,LevSoi:weyl},
compare with \eqref{R=sumprod})
these elements are all equal. Therefore all integrals are equal.
\end{pf}

\subsubsection{Invariance of the integral}\label{invainte}
When $\uqg$ is factorizable, it is unimodular by \propref{unimod}.
We will prove it also in non-factorizable case.

\begin{proposition}
The algebra $\uqg$ is unimodular.
\end{proposition}

\begin{pf}
A left integral $\delta_+\in u_q(\b_+)$ and a right integral
$\omega_-\in u_q(\b_-)$ are given by
\[ \delta_+ = \sum_{h\in\G}h \cdot \prod_{\alpha)} E_\alpha^{l_\alpha-1} ,
\quad \omega_- = \prod_{\beta} F_\beta^{l_\beta-1} \cdot \sum_{h\in\G}h. \]
by a result of Hennings~\cite{Hen:3} and Radford~\cite{Rad:min} the double
$D(u_q(\b_+))$ is unimodular with the two-sided integral $\delta_+\omega_-$.
Its projection to $u_q(\g)$ via the epimorphism
$j:D(u_q(\b_+)) \to u_q(\g)$ is
\[ \delta = j(\delta_+ \omega_-) = |\G| \sum_{h\in\G}h \cdot
\prod_{\alpha)} E_\alpha^{l_\alpha-1} \cdot
 \prod_{\beta} F_\beta^{l_\beta-1} . \]
This is a two-sided non-zero integral in $u_q(\g)$.
\end{pf}

\subsection{$3$-modular structure of $\uqg$}
\begin{proposition}\label{2mod=>3mod}
If $\uqg$ is a $2$-modular Hopf algebra then it is $3$-modular as well.
\end{proposition}

\begin{pf}
We have to check the remaining condition (M3) from \thmref{3-modthm}, namely
\[ (\int\tens1) (R^{12} R^{21}) \ne0 .\]
Only maximal powers will contribute to this expression (see \eqref{R12R21})
\begin{align*}
&(\int\tens1) (R^{12} R^{21}) = \\
&= \text{const} \int \{ (\prod_\al E_\al^{l_\al-1} )  \Ro_i' \Ro_j''
\prod_\beta (F_\beta K_\beta^{-1})^{l_\beta-1} \}
( \prod_\al  F_\al^{l_\al-1} ) \Ro_i''\Ro_j' \prod_\beta
( K_\beta E_\beta)^{l_\beta-1} \\
&= \text{const} \int \{ (\prod_\al E_\al^{l_\al-1} )  \Ro^{2\prime}_i
K_{2\rho} \prod_\beta F_\beta^{l_\beta-1} \} ( \prod_\al  F_\al^{l_\al-1} )
\Ro^{2\prime\prime}_i (\prod_\beta E_\beta^{l_\beta-1}) K_{-2\rho} \\
&= \frac{\text{const}}{|\twoG|} \sum_{c,d\in\twoG} \pi_2(c,d)^{-1}
\int \{ (\prod_\al E_\al^{l_\al-1} )  c K_{2\rho}
\prod_\beta F_\beta^{l_\beta-1} \} ( \prod_\al  F_\al^{l_\al-1} )
d (\prod_\beta E_\beta^{l_\beta-1}) K_{-2\rho} \\
&= \frac{\text{const}}{|\twoG|} ( \prod_\al  F_\al^{l_\al-1} )
\sum_{d\in\twoG} \pi(K_{2\rho},d)  d
(\prod_\beta E_\beta^{l_\beta-1}) K_{-2\rho}
\end{align*}
where we used \eqref{Rosquare}. Since
\[ \text{const} = \Bigl(\prod_\al
\frac{(q_\al-q_\al^{-1})^{l_\al-1}}{(l_\al-1)_{q_\al^{-2}}!} \Bigr)^2 \]
this obviously does not vanish.
\end{pf}

Now we find the unique (up to a sign) normalization of the integral which
will be used for constucting switching operators.

\begin{proposition}
Let $\int': \uqg\to\C$ be the renormalized left integral
\[ \int' = \sqrt{|\twoG|} q^{-2(\rho|\rho)} \prod_\al
\frac{(l_\al-1)_{q_\al^{-2}}!}{(q_\al-q_\al^{-1})^{l_\al-1}} \int \]
considered also as an element $\mu\in fun_q(G)$. If $\uqg$ is 3-modular,
then $P=S^2 \gamma_F:F\to F$ is a projection for $S=S_-(\mu)$.
\end{proposition}

\begin{pf}
The claim is equivalent to equation \eqref{normasig}
\[ (\int'\tens\int') (R^{12} R^{21})  =1 .\]
Substituting the expression for $(\int\tens1) (R^{12} R^{21})$ from
\propref{2mod=>3mod} we get
\begin{align*}
(\int\tens\int) (R^{12} R^{21}) &= \Bigl(\prod_\al
\frac{(q_\al-q_\al^{-1})^{l_\al-1}}{(l_\al-1)_{q_\al^{-2}}!} \Bigr)^2
\frac1{|\twoG|} \int ( \prod_\al  F_\al^{l_\al-1} )
(\prod_\beta E_\beta^{l_\beta-1}) K_{-2\rho} \\
&= \Bigl(\prod_\al
\frac{(q_\al-q_\al^{-1})^{l_\al-1}}{(l_\al-1)_{q_\al^{-2}}!} \Bigr)^2
\frac1{|\twoG|} q^{(2\rho|2\rho)}
\end{align*}
due to $( \prod_\al  F_\al^{l_\al-1} ) \cdot K_{-2\rho} =
q^{(2\rho|2\rho)} K_{-2\rho} \prod_\al  F_\al^{l_\al-1}$. This fixes
the normalization.
\end{pf}

\begin{proposition}
The exponential central charge $\lambda=\int'\nu$ for $2$-modular
$\uqg$ is a root of unity
\begin{align}
\lambda &= q^{-2(\rho|\rho)} \frac{\sqrt{|\twoG|}}{|\G|}
\sum_{g\in\G} \pi(g,g) \pi(g,K_{-2\rho}) \label{lambdapi} \\
&= q^{-2(\rho|\rho)} \frac{\sqrt{|\twoG|}}{l^n}
\sum_{\al\in Q/lQ} q^{(\al|\al) - (\al|2\rho)} , \label{lambdaq}
\end{align}
where $n$ is the rank of $\g$ and $l$ is the degree of the root of unity
$q=\e$.
\end{proposition}

\begin{pf}
{}From \eqref{nu=sumprod} we find
\[ \nu = \prod_\al
\frac{(q_\al-q_\al^{-1})^{l_\al-1}}{(l_\al-1)_{q_\al^{-2}}!}
\int \bigl( \prod_\al E_\al^{l_\al-1}) \cdot \Ro_a' \Ro_a'' K_{-4\rho}
\cdot \prod_\al  F_\al^{l_\al-1} \bigr) .\]
Substituting $\Ro_a'\Ro_a''K_{-2\rho} = \frac1{|\G|} \sum_{b\in\G} \phi(b)b$
by \eqref{GausFou} we find
\[ \lambda=\int'\nu = \sqrt{|\twoG|} q^{-2(\rho|\rho)} \frac1{|\G|}\phi(1),\]
where by \eqref{Gauss}
\be\label{phi(1)}
\phi(1) = \sum_{g\in\G} \pi(g,g) \pi(g,K_{-2\rho}) .
\end{equation}
Its absolute value is determined by \eqref{|phi|}
\begin{align}
|\phi(1)|^2 &= |\G| \sum_{x\in X} \pi(x,x) \pi(x,K_{-2\rho}) .
\label{|phi(1)|} \\
&= |\G| \cdot |X| \notag
\end{align}
since all summands equal 1 by \thmref{2-modthm}. Therefore,
\[ |\lambda|^2 = \frac{|\twoG|}{|\G|} |X| =1 \]
due to an exact sequence $0\to X\to \G @>2>> \twoG \to 0$.

By \eqref{lambdapi} $\lambda$ satisfies an algebraic equation of degree
$2l$. Since $\lambda$ is an algebraic number of absolute value 1, it is
a root of unity.

If we sum up in \eqref{phi(1)} not over $\G$, but over its covering group
$Q/lQ$, the sum will multiply by $l^r/|\G|$. This proves \eqref{lambdaq}.
\end{pf}

\begin{remark}
The condition \eqref{pi=pi} implies that
\[ \pi(gx,gx) \pi(gx,K_{-2\rho}) = \pi(g,g) \pi(g,K_{-2\rho}) \]
for all $g\in\G$, $x\in X$.
Therefore, we can introduce a function on $\twoG=\G/x$,
\[ s:\twoG\to\C^\times, \qquad s(h) = \pi(g,g) \pi(g,K_{-2\rho})
\quad \text{if } g^2=h \]
and write
\be\label{lam2G}
\lambda = q^{-2(\rho|\rho)} \frac1{\sqrt{|\twoG|}} \sum_{h\in\twoG} s(h) .
\end{equation}
One easily recognizes in \eqref{lambdapi}, \eqref{lambdaq} and \eqref{lam2G}
generalized quadratic Gauss sums.

Notice also that if $\uqg$ is not 2-modular, then $\int\nu=0$ by
\eqref{|phi(1)|}.
\end{remark}

\section{Examples of 3-modular algebras}\label{Examples}
We assume that $(a_{ij})$ is  the Cartan matrix of a simple Lie algebra
$\g$ and $q=\e$ is a primitive root of unity of degree $l$ such that
$\e^{2p}\ne1$ for $1\le p\le N(\g) $. We explore case by case
when $\uqg$ is 3-modular or equivalently 2-modular.

In general, $\G=Q/\Ann \pi$, $Q$ being the root lattice and
$\Ann\pi = Q\cap l\Cow$, where $\Cow = \Z\{\frac1{d_i} \omega_i\}$ is the
coweight lattice and $\omega_i$ is the basis of the weight lattice $P$.
This implies $\Ann \pi = Q\cap \Z\{l_i' \omega_i\}$, where $l_i'=l$
if $d_i \nmid l$ and $l_i'=l/d_i$ if $d_i | l$. The inclusion
$Q\hookrightarrow P$, $\al_j=\sum_{i=1}^n a_{ij} \omega_i$, determines
the annihilator:
\[ \Ann \pi = \{ u=\sum_j u_j \al_j \mid \forall i\quad
\sum_j a_{ij} u_j \equiv 0 \pmod{l_i'} \} \]
Similarly for $X=\Ker \{2:\G\to\G\}$
\[ X = \{ u=\sum_j u_j \al_j \mid \forall i\quad
2\sum_j a_{ij} u_j \equiv 0 \pmod{l_i'} \}/\Ann \pi .\]
2- or 3-modularity of $\uqg$ is equivalent to the property
\[ (u|u) -(2\rho|u) \equiv 0 \pmod l \]
for all $u\in X$.

If $l$ is odd, $X=0$ and $\uqg$ is perfect modular (we shall see that this
is not the only case).

\subsection{The algebra $u_\e(\esel(2))$}
This example was already considered in \cite{LyuMaj} for odd $l$. Set
$l_1=l$ for $l$ odd and $l_1=l/2$ for $l$ even, then it is the degree of
$\e^2$. Since $\pi(K,K)=\e^2$ for $K=K_1$, the group $\G$ is isomorphic to
$\Z/l_1\Z$. Several cases emerge.

\begin{enumerate}
\item {\it $l_1=2m+1$ is odd.} Then $u_\e(\esel(2))$ is perfect and
\begin{align*}
\lambda &= \e^{-1} \frac1{\sqrt{l_1}} \sum_{k=1}^{l_1} \e^{2k^2-2k} \\
&= \jacobi{2a}{l_1} e^{-\frac{\pi im}2} \e^{-1-2m^2}
\end{align*}
where $a$ is determined by $\e^2=e^{2\pi i a/l_1}$. Here for any relatively
prime integers $c$ and odd positive $b$ $\jacobi cb =\prod_j \jacobi a{p_j}$
denotes the Jacobi quadratic symbol, where $b=p_1 p_2 \dots p_r$ with prime
$p_j$ and $\jacobi a{p_j} = \pm1$ is the Legendre symbol (see e.g.
\cite{Lan:ANT}).

\item {\it $l_1=2(2m+1)$.}
Then $u_\e(\esel(2))$ is 3-modular since $X=\{1,K^{2m+1}\}$ and
\[ \pi(K^{2m+1},K^{2m+1}) \pi(K^{2m+1}, K^{-1}) = \e^{2(2m+1)2m} =1 .\]
$\twoG$ is isomorphic to $\Z/(2m+1)\Z$ and according to \eqref{lam2G}
\begin{align*}
\lambda &= \e^{-1} \frac1{\sqrt{2m+1}} \sum_{k=0}^{2m} \e^{2k^2-2k} \\
&= \jacobi{a}{2m+1} e^{-\frac{\pi im}2} \e^{2m^2+2m-1} 
\end{align*}
where $a$ is determined by $\e^2=e^{2\pi i a/l_1}$. The value of the
generalized Gauss quadratic sum was found here by a method of Lang
\cite{Lan:ANT} combined with results of Chandrasekharan \cite{Cha:iANT}.

\item {\it $l_1=4m$.}
Then $u_\e(\esel(2))$ is not 2-modular since $X=\{1,K^{2m}\}$ and
\[ \pi(K^{2m},K^{2m-1}) = \e^{4m(2m-1)} = \e^{\frac l2 (2m-1)} = -1 .\]
\end{enumerate}

\subsection{The Cartan matrix $A_n$}
The determinant of the Cartan matrix is $n+1$. Let $p=(l,n+1)$ be the
greatest common divisor. Then
\[ \Ann \pi = \{{\textstyle \frac lp}v \mid \forall i\quad
\sum_j a_{ij} v_j \equiv 0 \pmod{p} \} .\]
Various possibilities appear here as we already have seen for $n=1$.

\subsection{The Cartan matrix $B_n$}
Here $\g=\so(2n+1)$, $d_i=2$ for $1\le i<n$ and $d_n=1$.
Index of connection $\det A=2$. Let $l'=l$ for $l$ odd and $l'=l/2$
for $l$ even. Then
\[ \Ann \pi = \{ \sum_j u_j \al_j \mid \forall i<n \,\,\,
\sum_j a_{ij} u_j \equiv 0\! \pmod{l'},
\,\,\, \sum_j a_{nj} u_j \equiv 0 \pmod{l}  \} .\]
In particular, $2u= (\det A)u \equiv 0\! \pmod{l'}$ if $u\in\Ann\pi$.
Solving the equations we find that $\Ann\pi=l'Q$ and $\G=Q/l'Q$.

\begin{enumerate}
\item {\it $l'$ is odd.}
Then $\uqg$ is perfect modular. $l$ might be even in this case.

\item {\it $l'=2m$ is even.}
Then $l=4m$, $X=\F_2\{m\al_i\}$ and
\begin{align*}
||m\al_i||^2 - (2\rho|m\al_i) &= 4m(m-1) \equiv 0 \pmod l
\qquad\text{for } i<n \\
||m\al_n||^2 - (2\rho|m\al_n) &= 2m(m-1) .
\end{align*}
The algebra $\ueg$ is 2-modular iff $m$ is odd.
\end{enumerate}

\subsection{The Cartan matrix $C_n$}
Here $\g=\espe(2n)$, $d_i=1$ if $1\le i<n$ and $d_n=2$.
The index of connection $\det A=2$. Let $l'=l$ for $l$ odd and $l'=l/2$
for $l$ even. Then
\[ \Ann \pi = \{ \sum_j u_j \al_j \mid \forall i<n \,\,\,
\sum_j a_{ij} u_j \equiv 0\! \pmod{l},
\,\,\, \sum_j a_{nj} u_j \equiv 0\! \pmod{l'}  \} .\]

\begin{enumerate}
\item {\it $l$ is odd.}
Then $\uqg$ is perfect modular with $\G=Q/lQ$.

\item {\it $l=2(2m+1)$.} Then $\uqg$ is 3-modular and
\[\nquad \Ann\pi =
\begin{cases}
lQ+\Z(2m+1)\al_n  &\text{for $n$  odd} \\
lQ+\Z(2m+1)\al_n+\Z(2m+1)(\al_1+\al_3+\dots+\al_{n-1})
\,\, &\text{for $n$ even}
\end{cases}
\]
In both cases $X=\F_2\{ (2m+1)\al_i \}$.

\item {\it $l=4m$.} Then
\[ \Ann\pi = lQ + \Z2m\al_n +
\Z m(2\al_1+2\al_3+\dots+2\al_{2[\frac n2]-1}+n\al_n) .\]
A) {\em $n$ is odd.} Then $\uqg$ is 3-modular and
\[ X= \F_2\{2m\al_i, m\al_n\} .\]
B) {\em $n=2n'$ is even.} Then
\[ X= \F_2\{2m\al_i, m\al_n, m(\al_1+\al_3+\dots+\al_{n-1}+n'\al_n) \} \]
and
\begin{multline*}
 ||m(\al_1+\al_3+\dots+\al_{n-1}+n'\al_n)||^2 -
(2\rho|m(\al_1+\al_3+\dots+\al_{n-1}+n'\al_n)) \equiv \\
\equiv 2mn'(m-3) \pmod l .
\end{multline*}
Therefore, if $n'$ even or $m$ odd, then $\uqg$ is 3-modular. If $n'$ odd
and $m$ even, it is not.
\end{enumerate}

\subsection{The Cartan matrix $D_n$}
Here $\g=\so(2n)$ and $d_i=1$. The index of connection $\det A=4$.
Since $Au\equiv0 \pmod l$ implies $4u\equiv0 \pmod l$, we have for
$p=(4,l)$, $u=\frac lp b$, $b=(b_1,\dots,b_n)$, $Ab\equiv0 \pmod p$.
This enables one to find all $u\in \Ann\pi=Q\cap lP$.

\begin{enumerate}
\item {\it $l$ is odd.} $\uqg$ is perfect modular with $\G=Q/lQ$.

\item {\it $l=2(2m+1)$.} Then $\uqg$ is 3-modular with
\begin{align*}
\G &= Q/(lQ+\Z(2m+1)(\al_{n-1}+\al_n)) \quad \text{for $n$ odd or} \\
\G &=
Q/(lQ+\Z(2m+1)(\al_{n-1}+\al_n)+\Z(2m+1)(\al_1+\al_3+\dots+\al_{n-1}))
\end{align*}
for $n$ even. In both cases $X=\F_2\{ (2m+1)\al_i \}$.

\item {\it $l=4m$ and $n$ is odd.} $\uqg$ is 3-modular with
\[ \Ann\pi = lQ + \Z2m(\al_{n-1}+\al_n) +
\Z m(2\al_1+2\al_3+\dots+2\al_{n-2}-\al_{n-1}+\al_n) ,\]
\[ X=\F_2\{ 2m\al_i, m(\al_{n-1}+\al_n) \} \]
since
\[ ||m(\al_{n-1}+\al_n)||^2 - (2\rho|m(\al_{n-1}+\al_n))
= 4m^2 -4m \equiv 0 \pmod l .\]

\item {\it $l=4m$ and $n=2n'$.} Here
\[ \Ann\pi = lQ + \Z2m(\al_{n-1}+\al_n) +
\Z2m(\al_1+\al_3+\dots+\al_{n-3}+\al_{n-1}) ,\]
\[ X=\F_2\{ 2m\al_i, m(\al_{n-1}+\al_n),
m(\al_1+\al_3+\dots+\al_{n-3}+\al_{n-1}) \} .\]
Since
\[ ||m(\al_1+\al_3+\dots+\al_{n-1})||^2 -
(2\rho|m(\al_1+\al_3+\dots+\al_{n-1})) = 2 mn'(m-1) \]
the algebra $\uqg$ is 3-modular for $n'$ even or $m$ odd, and it is not
if $n'$ is odd and $m$ is even.
\end{enumerate}

\subsection{The Cartan matrix $E_6$}
Here $d_i=1$ and $\det A=3$. The equation $Au\equiv0 \pmod l$ implies
$3u\equiv0 \pmod l$.

\begin{enumerate}
\item {\it $l$ is odd.} Then $\uqg$ is perfect modular with
\[ \G =
\begin{cases}
Q/lQ , \qquad & \text{if } 3\nmid l \\
Q/(lQ + \Z p(\al_1-\al_3+\al_5-\al_6)) ,\qquad & \text{if } l=3p
\end{cases}
\]

\item {\it $l=2m$ and $3\nmid m$.}
Then $\uqg$ is 3-modular with $\G=Q/lQ$ and $X=\F_2\{m\al_i\}$.

\item {\it $l=6m$.} Again $\uqg$ is 3-modular with
\[ \G = Q/(lQ + \Z2m(\al_1-\al_3+\al_5-\al_6)) ,\]
 \[ X=\F_2\{3m\al_i, m(\al_1-\al_3+\al_5-\al_6) \} ,\]
since
\[ ||m(\al_1-\al_3+\al_5-\al_6)||^2 - (2\rho|m(\al_1-\al_3+\al_5-\al_6))
= 12 m^2 \equiv 0 \pmod l .\]
\end{enumerate}

\subsection{The Cartan matrix $E_7$}
Here $d_i=1$ and $\det A=2$. The equation $Au\equiv0 \pmod l$ implies
$2u\equiv0 \pmod l$.

\begin{enumerate}
\item {\it $l$ is odd.} Then $\uqg$ is perfect modular with $\G=Q/lQ$.

\item {\it $l=2(2m+1)$.} Then $\uqg$ is 3-modular with
\[ \G = Q/(lQ + \Z(2m+1)(\al_2+\al_5+\al_7)) ,\]
 \[ X=\F_2\{(2m+1)\al_i \} .\]

\item {\it $l=4m$.} Then
\[ \G = Q/(lQ + \Z2m(\al_2+\al_5+\al_7)) ,\]
 \[ X=\F_2\{2m\al_i, m(\al_2+\al_5+\al_7) \} .\]
Since
\[ ||m(\al_2+\al_5+\al_7)||^2 - (2\rho|m(\al_2+\al_5+\al_7)) = 6m(m-1) \]
the algebra $\uqg$ is 3-modular iff $m$ is odd.
\end{enumerate}

\subsection{The Cartan matrix $E_8$}
Here $d_i=1$ and $\det A=1$, therefore, $\G=Q/lQ$.

\begin{enumerate}
\item {\it $l$ is odd.} Then $\uqg$ is perfect modular.

\item {\it $l=2m$.} Then $\uqg$ is 3-modular with $X=\F_2\{m\al_i\}$.
\end{enumerate}

\subsection{The Cartan matrix $F_4$}
Here $d_1=d_2=2$, $d_3=d_4=1$ and $\det A=1$, which implies $Q=P$.
Let $l'=l$ for $l$ odd and $l'=l/2$ for $l$ even. Then
\[ \Ann\pi = \Z\{ l'\omega_1, l'\omega_2, l\omega_3, l\omega_4 \} ,\]
where
\begin{align*}
\omega_1 &= 2\al_1 + 3\al_2 + 4\al_3 + 2\al_4 \\
\omega_2 &= 3\al_1 + 6\al_2 + 8\al_3 + 4\al_4 \\
\omega_3 &= 2\al_1 + 4\al_2 + 6\al_3 + 3\al_4 \\
\omega_4 &=\,\al_1 + 2\al_2 + 3\al_3 + 2\al_4
\end{align*}
are the fundamental weights.

\begin{enumerate}
\item {\it $l$ is odd.}
Then $\uqg$ is perfect modular with $\G=Q/lQ$.

\item {\it $l=2l'$.}
Then $\uqg$ is 3-modular with $\G\simeq (\Z/l'\Z)^2 \times (\Z/l\Z)^2$
\[ \G =
Q/( l'\al_1, l'(3\al_2+4\al_3+2\al_4), l(\al_1+2\al_2+3\al_3), l\al_4) ,\]
\[ X=
\begin{cases}
\F_2\{ l'(\al_1+2\al_2+3\al_3), l'\al_4 \} \quad & \text{for $l'$ odd} \\
\F_2\{ m\al_1, m(3\al_2+4\al_3+2\al_4),
l'(\al_1+2\al_2+3\al_3), l'\al_4 \} \quad & \text{for } l'=2m
\end{cases}
\]
\end{enumerate}

\subsection{The Cartan matrix $G_2$}
Here $d_1=1$, $d_2=3$ and $\det A=1$, which implies $Q=P$.
Let $l'=l$ if $3\nmid l$ and $l'=l/3$ if $3| l$. Then
$\Ann\pi = \Z\{ l\omega_1, l'\omega_2 \}$, where
\begin{align*}
\omega_1 &= 2\al_1 +  \al_2 \\
\omega_2 &= 3\al_1 + 2\al_2
\end{align*}
are the fundamental weights. Therefore,
\[ \G = Q/ \Z\{ l\omega_1, l'\omega_2 \} \simeq \Z/l\Z \times \Z/l'\Z .\]

\begin{enumerate}
\item {\it $l$ is odd.} Then $\uqg$ is perfect modular.

\item {\it $l=2m$.} Then $\uqg$ is 3-modular with
\[ X = \F_2 \{ m\omega_1,{\textstyle\frac{l'}2} \omega_2 \} \]
since
\[ ||{\textstyle\frac{l'}2} \omega_2||^2 -
(2\rho|{\textstyle\frac{l'}2} \omega_2)
= 3l'({\textstyle\frac{l'}2} - 3) \equiv 0 \pmod l .\]
\end{enumerate}

\section{Representations of mapping class groups}\label{mapping}
According to \cite{Lyu:r=m} for any 2-modular category $\CC$ there are
projective representations of mapping class groups of $\CC$-lableled
surfaces in vector spaces $\Hom_\CC(-,-)$. Here we pay attention mainly
to the categories $\CC=H$-mod, where $H$ is a 2-modular Hopf algebra,
and we describe the representations explicitly. In the particular
case of closed surfaces or surfaces with one hole these representations
are closely related to
the representations of mapping class groups in a category of tangles
obtained by Matveev and Polyak~\cite{MatPol}.

By a {\em labeled surface} we shall understand the following: a compact
oriented surface  with a labeling of boundary circles
$L:\pi_0(\pa\Si) \to \Ob\CC$, $i\mapsto X_i$ and with a chosen point $x_i$
on $i^{\text{th}}$ boundary circle, i.e. a section
$x:\pi_0(\pa\Si) \to \pa\Si$ of the projection $\pa\Si \to \pi_0(\pa\Si)$
is fixed. By a homeomorphism of labeled surfaces we mean an orientation
and labeling preserving homeomorphism, which sends the set of chosen
points to itself. A {\em mapping class group} $MCG(\Si)$ is defined as
the group of homeomorphisms $\Si\to\Si$ modulo isotopy equivalence
relation. An isotopy is supposed not to move the chosen points.

We use the convention $AB= A\cdot B=B\circ A$ in this section.

\subsection{A central extension of the category of surfaces}
Projective representations of the mapping class groups are constructed in
\cite{Lyu:r=m} as follows. The category $\Surf$ of labeled surfaces and
their homeomorphisms is embedded (non canonically) into a category $ON$
of oriented nets, which are a sort of labeled oriented graphs with
1-,2-, or 3-valent vertices. Morphisms of $ON$ are generated by natural
maps of graphs and some extra generators called fusing, braiding, twists
and switches. In particular, the homeomorphisms $S,T:\T^2\to\T^2$ of the
torus go to generators $S,T$ in $ON$. The generators of $ON$ are subject
to several relations, including
\be\label{ST3=S2}
(ST)^3 =S^2 .
\end{equation}

A central extension $EN$ of the category $ON$ is defined \cite{Lyu:r=m}
by adding one more generator $C$ commuting with other generators and
deforming the relation \eqref{ST3=S2} to
\[ (ST)^3 = C S^2 .\]
Lift the central extension to $\Surf$ as in the diagram
\[ \begin{CD}
1 @>>> \Z @>>> E\Surf @>>> \Surf @>>> 1 \\
@.     @|      @VVV        @VV\phi V  @.\\
1 @>>> \Z @>>> EN     @>j>> ON   @>>> 1
\end{CD} \]
The meaning of this diagram is the following. Objects of $E\Surf$ are
objects of $\Surf$. For any two homeomorphic surfaces
$\Si_1,\Si_2\in\Ob\Surf$ let $E\Surf(\Si_1,\Si_2)$ be determined as a
central extension of $\Surf(\Si_1,\Si_2)$ as in pull-back
\[ \begin{CD}
1 @>>> \Z @>>> E\Surf(\Si_1,\Si_2)         @>>> \Surf(\Si_1,\Si_2) @>>> 1 \\
@.     @|      @VVV                              @VV\phi V  @.\\
1 @>>> \Z @>>> EN(\phi(\Si_1),\phi(\Si_2)) @>j>> ON(\phi(\Si_1),\phi(\Si_2))
@>>> 1
\end{CD} \]
Simply set $E\Surf(\Si_1,\Si_2)= \overset{-1}j(\phi(\Surf(\Si_1,\Si_2)))$
and set $E\Surf(\Si_1,\Si_2)$ empty if $\Si_1$ and $\Si_2$ are not
homeomorphic in $\Surf$.

The central extension $E\Surf$ of the groupoid $\Surf$ (as any other
central extension by $\Z$) can be described by a cohomology class in
$H^2(\Surf,\Z)$ uniquely up to equivalence of categories. Namely, choose
arbitrarily a lifting $\tilde f\in E\Surf$ for any morphism $f\in\Surf$
and for any composable $f,g\in\Surf$ set $\theta(f,g)=m$ if
$\tilde f^{-1} \widetilde{(fg)} \tilde g^{-1} = C^m$. For any
$f,g,h\in\Surf$ such that $fg$ and $gh$ exist we have
\[ \theta(f,g) + \theta(fg,h) = \theta(f,gh) + \theta(g,h) .\]
Restricting to $f,g\in MCG(\Si)$ we get a 2-cocycle
$\theta\in Z^2(MCG(\Si),\Z)$, whose cohomology class
$[\theta]\in H^2(MCG(\Si),\Z)$ does not depend on the lifting.

A functor $EN\to k$-vect sending $C$ to $\lambda\in k^\times$ was
constructed in \cite{Lyu:r=m}. So a summary of this work can be given by the
following commutative diagram of functors
\[ \begin{CD}
Z': @. E\Surf @>>> EN @>>> k\vect \\
@.    @VpVV       @VVjV          \\
  @. \Surf @>\phi>> ON
\end{CD} \]
Any section of the projection $p$ would give a projective representation
$Z:\Surf \to k$-vect satisfying $Z(fg)=\theta(f,g) Z(f) Z(g)$ for any
composable $f,g\in\Surf$ with a 2-cocycle $\theta$.

Here we shall describe the projective representation of
$M'_{g,n}= MCG(\Si_{g,n})$ in
$Z(\Si_{g,n}) = \Hom(X_1\tens\dots\tens X_n,\f^{\tens g})$ for various
genera $g$ and number of holes $n$. The group $M'_{g,n}$ depends on
coincidence of labels $X_1,\dots,X_n$. To describe the representations
uniformly we allow for homeomorphisms which do not preserve the labels,
obtaining a larger group $M_{g,n}\supset M'_{g,n}$. It has a projection
$p:M_{g,n} \to \SS_n$ to the symmetric group. We represent elements
$h\in M_{g,n}$ by operators
\[ Z(h): \Hom(X_1\tens\dots\tens X_n,\f^{\tens g}) \to
\Hom(X_{p(h)^{-1}(1)}\tens\dots\tens X_{p(h)^{-1}(n)}, \f^{\tens g}) \]
satisfying $Z(hf) = \lambda^k Z(h)Z(f)$  for some $k\in\Z$.

The reader might want to consider $\CC=H$-mod as the category of labels
(set $I=k$ in this case), although the results are valid for an arbitrary
2-modular category $\CC$.

\subsection{Sphere}
Consider a sphere $\Si_{0,n}$ with $n$ disks removed, boundary circles are
labeled by $X_1,X_2,\dots,X_n$. The mapping class group $M_{0,n}$ is
generated by the braiding homeomorphisms $\omega_i$ interchanging the
$i^{\text{th}}$ and $i+1^{\text{st}}$ holes and the inverse Dehn twists
$R_i$ performed in a collar neighbourhood of $i^{\text{th}}$ boundary
circle. The relations
\[ R_i R_j=R_j R_i, \]
\[ \omega_i R_j=R_j\omega_i \text{ if } j\ne i, i+1,\]
\[ \omega_i R_i=R_{i+1}\omega_i, \quad \omega_i R_{i+1}=R_i\omega_i, \]
\[ \omega_i\omega_j=\omega_j\omega_i \text{ if } |i-j|>1 \]
\[ \omega_i\omega_{i+1}\omega_i=\omega_{i+1}\omega_i\omega_{i+1} \]
\[ \omega_1\omega_2\dots\omega_{n-1}^2\dots\omega_2\omega_1=R_1^{-2} \]
\[ (\omega_1\dots\omega_{n-1})^n=R_1^{-1}\dots R_n^{-1} .\]
are defining relations of $M_{0,n}$.

The group $M_{0,n}$ is represented in
\[ Z(\Si_{0,n}) = \Hom(X_1\tens X_2\tens\dots\tens X_n, I) \]
by the operators
\begin{align*}
Z(\omega_i) &= \Hom(1\tens c_{X_{i+1}, X_i} \tens1, I) ,\\
Z(R_i) &= \Hom(1\tens \nu_{X_i} \tens1, I) .
\end{align*}
This is a usual representation not only projective.

\subsection{Torus}
Consider a torus $\Si_{1,n}$ with $n$ disks removed, boundary circles are
labeled by $X_1,X_2,\dots,X_n$. The mapping class group $M_{1,n}$ is
generated by the homeomorphisms $S$, $T=Tw_\delta$, $R_i$,
$a_j=Tw_{\gamma_j}^{-1}\,Tw_\delta$, where cycles $\gamma_j$ separate
$j-1^{\text{st}}$ and $j^{\text{th}}$ holes as shown at Figure~\ref5,
\def\epsfsize#1#2{0.579#1}
\begin{figure}[htbp]
\[ \epsfbox{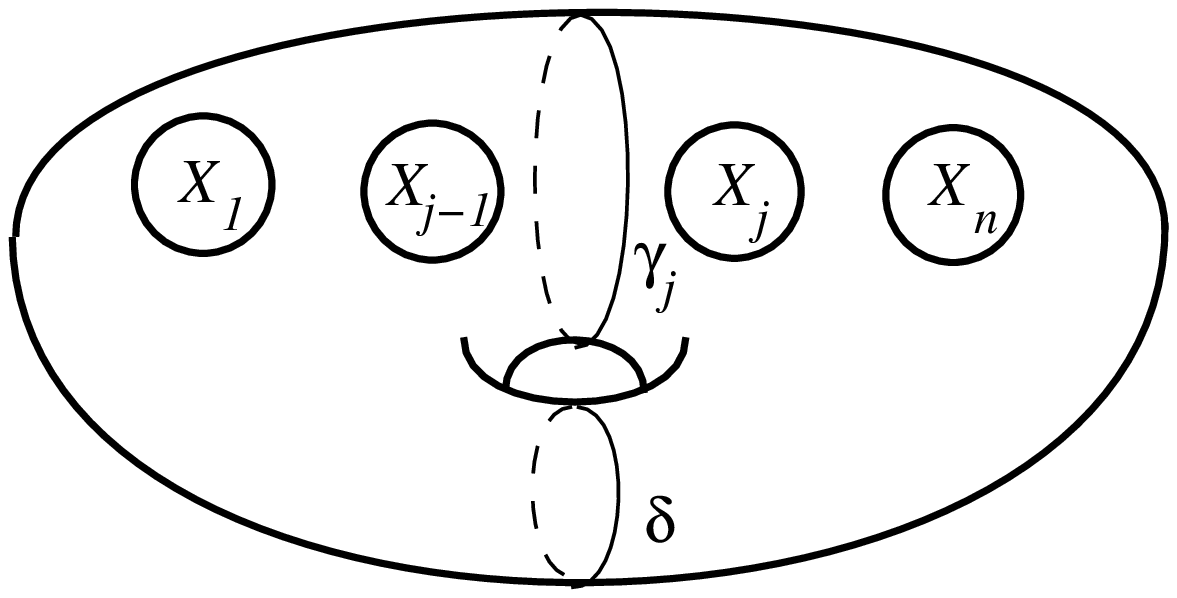} \]
\caption{\label5}
\end{figure}
and the braiding homeomorphisms $\omega_i$ interchanging the
$i^{\text{th}}$ and $i+1^{\text{st}}$ holes. Denote $b_j=S^{-1}a_jS$
and for $j<k$ denote
\begin{multline*}
B_{jk}^{-1} = (\omega_{k-1}\omega_{k-2}\dots\omega_j)
(\omega_k\omega_{k-1}\dots\omega_{j+1}) \dots
(\omega_{n-1}\omega_{n-2}\dots\omega_{n+j-k})  \\
(\omega_{j-k+n}\omega_{j-k+n-1}\dots\omega_j)
(\omega_{j-k+n+1}\omega_{j-k+n}\dots\omega_{j+1})  \dots
(\omega_{n-1}\omega_{n-2}\dots\omega_{k-1}) .
\end{multline*}
The relations
\[ (ST)^3=S^2, \]
\[ S^4=B_{12}B_{23}B_{34}\dots B_{n-1,n} R_1^{-1}\dots R_n^{-1} \]
\[  S^{-1}b_j S=b_j a_j^{-1} b_j^{-1} \]
\[ T^{-1}a_j T=a_j, \quad T^{-1}b_j T=b_j a_j \]
\[ \omega_i S=S\omega_i, \quad \omega_i T=T\omega_i \]
\[ \omega_i R_i=R_{i+1}\omega_i, \quad \omega_i R_{i+1}=R_i\omega_i \]
\[ \omega_i R_j=R_j\omega_i \text{  for  } j\ne i, i+1 \]
\[ \omega_i\omega_j=\omega_j\omega_i \text{ for } |i-j|>2 \]
\[ \omega_i\omega_{i+1}\omega_i=\omega_{i+1}\omega_i\omega_{i+1} \]
\[ a_1=b_1=1 \]
\[ a_j a_k=a_k a_j, \quad b_j b_k=b_k b_j \]
\[ a_ja_k^{-1}b_k a_j^{-1} a_k b_k^{-1}=B_{jk} \text{ for } j<k \]
\[ a_k^{-1}b_j^{-1}b_k a_k b_j b_k^{-1}=B_{jk} \text{ for }  j<k \]
\[ a_iB_{jk}=B_{jk}a_i,\qquad b_i B_{jk}=B_{jk}b_i \text{ for }  i\le j<k \]
are defining relations of $M_{1,n}$ (\cite{MooSei},
extending \cite{Bir:mcg}).

Set
\begin{align*}
Z(\Si_{1,n}) &= \Hom(X_1\tens \dots\tens X_n, \f) ,\\
Z(S) &= \Hom(X_1\tens \dots\tens X_n, S) ,\\
Z(T) &= \Hom(X_1\tens \dots\tens X_n, T) ,\\
Z(R_i) &= \Hom(X_1\tens \dots\tens \nu_{X_i} \tens\dots\tens X_n, \f) ,\\
Z(\omega_i) &=
\Hom(X_1\tens\dots\tens c_{X_{i+1}, X_i} \tens\dots\tens X_n, \f) ,
\end{align*}
\begin{multline*}
Z(a_j) = \bigl(
\Hom(X_1\tens\dots\tens X_{j-1}\tens X_j\tens\dots\tens X_n, \f) \\
@>\sim>>
\Hom(X_1\tens\dots\tens X_{j-1}, \f\tens(\pti X_n\tens\dots\tens\pti X_j))
@>\Hom(X_1\tens\dots\tens X_{j-1}, \Omega_r^{-1})>> \\
\Hom(X_1\tens\dots\tens X_{j-1}, \f\tens(\pti X_n\tens\dots\tens\pti X_j))
@>\sim>> \Hom(X_1\tens\dots\tens X_{j-1}\tens(X_j\tens\dots\tens X_n), \f) \\
@>\Hom(X_1\tens\dots\tens X_{j-1}\tens\nu^{-1}_{X_j\tens\dots\tens X_n},\f)>>
\Hom(X_1\tens\dots\tens X_{j-1}\tens X_j\tens\dots\tens X_n, \f) \bigr).
\end{multline*}
Here $S$ and $T$ in the right hand side mean $S:\f\to\f$, $T:\f\to\f$
described in \secref{qFt}. Equation~\eqref{STtre=Stwo} implies that
\[ (Z(S) Z(T))^3 = \lambda Z(S)^2 .\]
Moreover, the above is a projective representation of $M_{1,n}$
\cite{Lyu:r=m}.

Notice that in particular case $n=0$ we have
\[ Z(S)^4 = 1 : \Hom(I,\f) \to \Hom(I,\f) \]
since $S^4 = \nu^{-1} :\f\to\f$.

\begin{remark}
The map $Z(b_j) \overset{\text{def}}= Z(S)^{-1} Z(a_j) Z(S)$ has another
presentation, which is an important part of the proof \cite{Lyu:r=m}.
Consider for simplicity the case $n=2$. Introduce a map $B_2$ as the
composition
\begin{multline*}
\Hom(X_1\tens X_2,F) @>\Hom(c_{X_2,X_1},F)>> \Hom(X_2\tens X_1,F) @>\sim>>
\Hom(X_1,X_2\pti\tens \int^N N\tens N\pti) \\
@>-\tens\coev_{X_1}>>
\Hom(X_1,\int^N X_2\pti\tens N\tens N\pti\tens X_2\pti\pti\tens X_2\pti) \\
@>\sim>>
\Hom(X_1,\int^N (X_2\pti\tens N)\tens (X_2\pti\tens N)\pti\tens X_2\pti)
\to \Hom(X_1,\int^PP\tens P\pti\tens X_2\pti) \\
@>\sim>> \Hom(X_1\tens X_2\pti\pti,F) @>\Hom(X_1\tens u_0^2,F)>>
\Hom(X_1\tens X_2,F).
\end{multline*}
It covers $Z(b_2)$ in the sense that the diagram
\be\label{B2Z(b2)}
\begin{CD}
\Hom(X_1\tens X_2,F)  @>B_2>>    \Hom(X_1\tens X_2,F) \\
@VVV                             @VVV                 \\
\Hom(X_1\tens X_2,\f) @>Z(b_2)>> \Hom(X_1\tens X_2,\f)
\end{CD}
\end{equation}
is commutative. Notice that in perfect modular case $F=\f$ and $Z(b_2)$
simply equals $B_2$.

If $\CC$ is semisimple, the vertical arrows in \eqref{B2Z(b2)} are
surjective. The same holds if $\CC=H$-mod for finite dimensional $H$ and
one of the $H$-modules $X_1$ or $X_2$ is projective (then $X_1\tens X_2$
is also projective). In these assumptions we can represent $B_2$ in a
different way as a composition of bijections
\begin{multline*}
\Hom(X_1\tens X_2,F) @>\Hom(c_{X_2,X_1},F)>>
\Hom(X_2\tens X_1, \int^N N\tens N\pti) \\
\simeq \Hom(X_2\tens X_1, \int^N N\tens\pti N)
\simeq \int^N \Hom(X_2\tens(X_1\tens N), N) \\
\simeq \int^{N,P} \Hom(X_2\tens P,N) \tens \Hom(X_1\tens N,P)
\simeq \int^P \Hom(X_1\tens(X_2\tens P), P) \\
\simeq \int^P \Hom(X_1\tens X_2, P\tens\pti P) \simeq
\Hom(X_1\tens X_2, \int^P P\tens P\pti) = \Hom(X_1\tens X_2,F) .
\end{multline*}
Knowing $Z(b_2)$ by \eqref{B2Z(b2)} for projective $X_1$ or $X_2$, one
recovers this map for arbitrary $X_1,X_2$ using short projective
resolutions. In the calculations above we used
\end{remark}

\begin{lemma}[\cite{Lyu:r=m}]\label{lemcoend}
Let $F:\CC\to k\Vect$, $G:\CC^{\op}\to k\Vect$ be functors. Then
\[\int^X F(X)\tens \Hom(X,B)\to F(B),\qquad v\tens f\mapsto F(f).v\]
\[\int^X \Hom(B,X)\tens G(X)\to G(B),\qquad f\tens v\mapsto G(f).v\]
are isomorphisms of vector spaces.
\end{lemma}

\subsection{Closed surfaces and surfaces with one hole}
\subsubsection{Surfaces with one hole}
Let $\Si=\Si_{g,1}$ be a surface of genus $g$ with one disk removed and
the boundary labeled by an $H$-module $X$. Lickorish \cite{Lic:gen}
proved that its mapping class group $M_{g,1}$ is generated by inverse
Dehn twists in a neighbourhood of the following cycles: $a_k$, $b_k$,
$d_k$, $e_k$ ($a_1=d_1=e_1$) (see Figure~\ref6).
\def\epsfsize#1#2{0.579#1}
\begin{figure}[htbp]
\[ \epsfbox{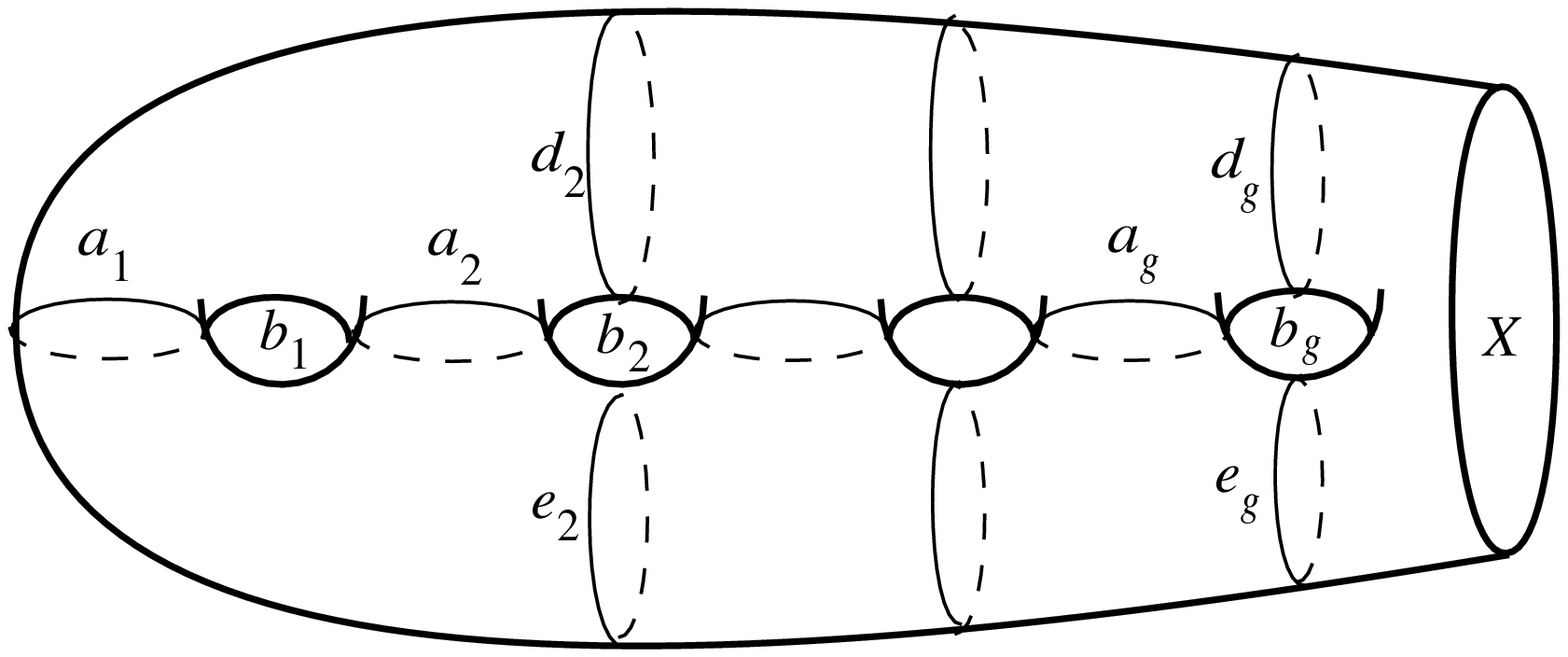} \]
\caption{\label6}
\end{figure}
These Dehn twists are denoted by the same letters as cycles. Wajnryb
\cite{Waj} found a system of defining relations for $M_{g,1}$. An
equivalent system of relations is the following:

(A) $a_ib_ia_i=b_ia_ib_i, \ a_{i+1}b_ia_{i+1}=b_ia_{i+1}b_i,\
 d_ib_id_i=b_id_ib_i, \ e_ib_ie_i=b_ie_ib_i $
and all other pairs of generators commute.

(B) $(a_1b_1a_2)^4=d_2e_2 $

(C) $\ e_kG_k =G_k d_k $, where
$G_k=b_k a_k\dots b_1 a_1 a_1 b_1\dots a_k b_k $.

(D) \ $J_k d_{k+1}= d_kJ_k$ , where
\[ J_k=b_ka_ka_{k+1}b_kb_{k+1}a_{k+1}a_k^{-1}b_k^{-1}d_kb_ka_kb_{k-1}d_{k-1}
       (a_k^{-1}b_{k-1}^{-1}b_k^{-1}a_k^{-1}b_{k-1}^{-1}b_k^{-1})
       d_k^{-1}b_k^{-1}a_{k+1}^{-1}b_{k+1}^{-1} \]

(E) $ hb_1^{-1}a_1^{-1}a_2^{-1}b_1^{-1}hb_1a_2a_1b_1d_2=d_3a_3a_2a_1  $,
where
\[ h=b_2^{-1}a_2^{-1}a_3^{-1}b_2^{-1}d_2b_2a_3a_2b_2. \]

To construct $Z(\Si)$ we follow the recipee of \cite{Lyu:r=m}. First of all
we choose an {\em oriented net} (roughly this is an oriented graph, for
precise definition see \cite{Lyu:r=m}) which encodes the structure of the
surface. Several graphs can be chosen, for instance, Figures~\ref1--\ref3.
\begin{figure}[htbp]
\[
\unitlength=0.8mm
\begin{picture}(140,41)
\put(17,8){\oval(14,14)[]}
\put(24,8){\circle*{1}}
\put(10,8){\vector(0,-1){0}}
\put(42,35){\vector(-1,0){0}}
\put(67,35){\vector(-1,0){0}}
\put(92,35){\vector(-1,0){0}}
\put(117,35){\vector(-1,0){0}}
\put(4,8){\makebox(0,0)[cc]{$D_1$}}
\put(42,41){\makebox(0,0)[ct]{$D_2$}}
\put(92,41){\makebox(0,0)[ct]{$D_{g-1}$}}
\put(117,41){\makebox(0,0)[ct]{$D_g$}}
\put(30,4){\makebox(0,0)[ct]{$C_2$}}
\put(55,4){\makebox(0,0)[ct]{$C_3$}}
\put(105,4){\makebox(0,0)[ct]{$C_g$}}
\put(42,28){\oval(14,14)[]}
\put(67,28){\oval(14,14)[]}
\put(92,28){\oval(14,14)[]}
\put(117,28){\oval(14,14)[]}
\put(24,8){\line(1,0){111}}
\put(42,21){\line(0,-1){13}}
\put(67,8){\line(0,1){13}}
\put(92,21){\line(0,-1){13}}
\put(117,8){\line(0,1){13}}
\put(33,8){\vector(-1,0){0}}
\put(54,8){\vector(-1,0){0}}
\put(79,8){\vector(-1,0){0}}
\put(104,8){\vector(-1,0){0}}
\put(129,8){\vector(-1,0){0}}
\put(117,15){\vector(0,1){0}}
\put(92,15){\vector(0,1){0}}
\put(67,15){\vector(0,1){0}}
\put(42,15){\vector(0,1){0}}
\put(45,16){\makebox(0,0)[lt]{$L_2$}}
\put(95,16){\makebox(0,0)[lt]{$L_{g-1}$}}
\put(120,16){\makebox(0,0)[lt]{$L_g$}}
\put(140,8){\makebox(0,0)[cc]{$X$}}
\put(42,8){\circle*{1}}
\put(67,8){\circle*{1}}
\put(92,8){\circle*{1}}
\put(117,8){\circle*{1}}
\put(117,21){\circle*{1}}
\put(92,21){\circle*{1}}
\put(67,21){\circle*{1}}
\put(42,21){\circle*{1}}
\end{picture}
\]
\caption{\label1}
\end{figure}
\begin{figure}[htbp]
\[
\unitlength=0.8mm
\begin{picture}(140,26)
\put(17,13){\oval(14,14)[]}
\put(42,13){\oval(14,14)[]}
\put(67,13){\oval(14,14)[]}
\put(92,13){\oval(14,14)[]}
\put(117,13){\oval(14,14)[]}
\put(24,13){\line(1,0){11}}
\put(49,13){\line(1,0){11}}
\put(74,13){\line(1,0){11}}
\put(99,13){\line(1,0){11}}
\put(124,13){\line(1,0){11}}
\put(24,13){\circle*{1}}
\put(35,13){\circle*{1}}
\put(49,13){\circle*{1}}
\put(60,13){\circle*{1}}
\put(74,13){\circle*{1}}
\put(85,13){\circle*{1}}
\put(99,13){\circle*{1}}
\put(110,13){\circle*{1}}
\put(124,13){\circle*{1}}
\put(10,13){\vector(0,-1){0}}
\put(29,13){\vector(-1,0){0}}
\put(54,13){\vector(-1,0){0}}
\put(79,13){\vector(-1,0){0}}
\put(104,13){\vector(-1,0){0}}
\put(129,13){\vector(-1,0){0}}
\put(42,20){\vector(-1,0){0}}
\put(42,6){\vector(-1,0){0}}
\put(67,6){\vector(-1,0){0}}
\put(67,20){\vector(-1,0){0}}
\put(92,20){\vector(-1,0){0}}
\put(92,6){\vector(-1,0){0}}
\put(117,6){\vector(-1,0){0}}
\put(117,20){\vector(-1,0){0}}
\put(4,13){\makebox(0,0)[cc]{$D_1$}}
\put(42,3){\makebox(0,0)[ct]{$E_2$}}
\put(92,3){\makebox(0,0)[ct]{$E_{g-1}$}}
\put(117,3){\makebox(0,0)[ct]{$E_g$}}
\put(42,26){\makebox(0,0)[ct]{$D_2$}}
\put(92,26){\makebox(0,0)[ct]{$D_{g-1}$}}
\put(117,26){\makebox(0,0)[ct]{$D_g$}}
\put(30,18){\makebox(0,0)[ct]{$C_2$}}
\put(55,18){\makebox(0,0)[ct]{$C_3$}}
\put(105,18){\makebox(0,0)[ct]{$C_g$}}
\put(140,13){\makebox(0,0)[cc]{$X$}}
\end{picture}
\]
\caption{\label2}
\end{figure}
\begin{figure}[htbp]
\[
\unitlength=0.8mm
\begin{picture}(140,26)
\put(124,13){\line(1,0){11}}
\put(124,13){\circle*{1}}
\put(10,13){\vector(0,-1){0}}
\put(129,13){\vector(-1,0){0}}
\put(42,20){\vector(-1,0){0}}
\put(42,6){\vector(-1,0){0}}
\put(67,6){\vector(-1,0){0}}
\put(67,20){\vector(-1,0){0}}
\put(92,20){\vector(-1,0){0}}
\put(92,6){\vector(-1,0){0}}
\put(117,6){\vector(-1,0){0}}
\put(117,20){\vector(-1,0){0}}
\put(4,13){\makebox(0,0)[cc]{$D_1$}}
\put(42,3){\makebox(0,0)[ct]{$E_2$}}
\put(92,3){\makebox(0,0)[ct]{$E_{g-1}$}}
\put(117,3){\makebox(0,0)[ct]{$E_g$}}
\put(42,26){\makebox(0,0)[ct]{$D_2$}}
\put(92,26){\makebox(0,0)[ct]{$D_{g-1}$}}
\put(117,26){\makebox(0,0)[ct]{$D_g$}}
\put(140,13){\makebox(0,0)[cc]{$X$}}
\put(30,13){\oval(40,14)[l]}
\put(104,13){\oval(40,14)[r]}
\put(29,6){\framebox(75,14)[cc]{}}
\put(54,6){\line(0,1){14}}
\put(79,6){\line(0,1){14}}
\put(29,13){\vector(0,1){0}}
\put(54,13){\vector(0,1){0}}
\put(79,13){\vector(0,1){0}}
\put(104,13){\vector(0,1){0}}
\put(104,6){\circle*{1}}
\put(79,6){\circle*{1}}
\put(54,6){\circle*{1}}
\put(29,6){\circle*{1}}
\put(29,20){\circle*{1}}
\put(54,20){\circle*{1}}
\put(79,20){\circle*{1}}
\put(104,20){\circle*{1}}
\put(32,13){\makebox(0,0)[lc]{$A_2$}}
\put(57,13){\makebox(0,0)[lc]{$A_3$}}
\put(107,13){\makebox(0,0)[lc]{$A_g$}}
\end{picture}
\]
\caption{\label3}
\end{figure}
They are all isomorphic in the category of oriented nets \cite{Lyu:r=m}.
Next step is to compute the functor $\CC^\op \to k$-vect corresponding
to these graphs, obtained by taking coends over internal edges. The
functor is obtained in different forms, which are isomorphic due to
associativity isomorphisms in $\CC$. By \lemref{lemcoend} we find that
to Figure~\ref1 corresponds
\begin{multline*}
\int^{C_i,D_i,L_i} \Hom(X,L_g\tens C_g) \tens \Hom(C_g,L_{g-1}\tens C_{g-1})
\tens\dots\tens \Hom(C_3,L_2\tens C_2) \tens \\
\tens \Hom(L_g\tens D_g,D_g) \tens\dots\tens \Hom(L_2\tens D_2,D_2)
\tens \Hom(C_2\tens D_1,D_1) \simeq
\end{multline*}
\begin{multline*}
\simeq \int^{D_i,L_i} \Hom(X,L_g\tens L_{g-1} \tens\dots\tens L_2\tens C_2)
\tens \Hom(L_g,D_g\tens\pti D_g) \tens\dots\tens \\
\tens \Hom(L_2,D_2\tens\pti D_2) \tens \Hom(C_2,D_1\tens\pti D_1) \simeq
\end{multline*}
\be\label{CDL}
\simeq \int^{D_i} \Hom(X,(D_g\tens D_g\pti) \tens\dots\tens
(D_2\tens D_2\pti) \tens (D_1\tens D_1\pti)) .
\end{equation}
To Figure \ref2 corresponds
\begin{multline*}
\int^{C_i,D_i,E_i} \Hom(C_2\tens D_1,D_1) \tens \Hom(D_2\tens E_2,C_2) \tens
\Hom(C_3,D_2\tens E_2) \tens\dots\tens \\
\tens \Hom(C_g,D_{g-1}\tens E_{g-1}) \tens \Hom(D_g\tens E_g,C_g) \tens
\Hom(X,D_g\tens E_g) \simeq
\end{multline*}
\begin{multline*}
\simeq \int^{C_i,D_i} \Hom(C_2,D_1\tens D_1\pti) \tens
\Hom(C_3,D_2\tens D_2\pti\tens C_2) \tens\dots\tens \\
\tens \Hom(C_g,D_{g-1}\tens D_{g-1}\pti\tens C_{g-1}) \tens
\Hom(X,D_g\tens D_g\pti\tens C_g) \simeq
\end{multline*}
\be\label{CDE}
\simeq \int^{D_i} \Hom(X,(D_g\tens D_g\pti) \tens\dots\tens
(D_2\tens D_2\pti) \tens (D_1\tens D_1\pti)) .
\end{equation}
To Figure \ref3 corresponds
\begin{multline*}
\int^{A_i,D_i,E_i} \Hom(E_2\tens D_1,A_2) \tens \Hom(D_2\tens A_2,D_1) \tens
\Hom(E_3,A_3\tens E_2) \tens\dots\tens \\
\tens \Hom(E_g,A_g\tens E_{g-1}) \tens \Hom(D_g\tens A_g,D_{g-1}) \tens
\Hom(X,D_g\tens E_g) \simeq
\end{multline*}
\begin{multline*}
\simeq \int^{D_i,E_i} \Hom(E_2\tens D_1,D_2\pti\tens D_1) \tens
\Hom(E_3,D_3\pti\tens D_2\tens E_2) \tens\dots\tens \\
\tens \Hom(E_g,D_g\pti\tens D_{g-1}\tens E_{g-1}) \tens
\Hom(X,D_g\tens E_g) \simeq
\end{multline*}
\be\label{ADE}
\simeq \int^{D_i} \Hom(X,(D_g\tens D_g\pti) \tens\dots\tens
(D_2\tens D_2\pti) \tens (D_1\tens D_1\pti)) .
\end{equation}

If $X$ is projective this space is
\[ \Hom(X, F\tens \dots \tens F\tens F) .\]
The same answer will be obtained if we calculate the coend in the category
of left exact functors $\CC^\op \to k$-vect. The final step is taking
a quotient and setting
\[ Z(\Si_{g,1})= \Hom(X, \f^{\tens g}) .\]

The generators $a_k,d_k,e_k$ of $M_{g,1}$ are represented by applying
ribbon twist to internal variables $A_k,D_k,E_k$. Thus from any
presentation \eqref{CDL}--\eqref{ADE} we get
\[ Z(d_k)= \Hom(X, \f^{\tens g-k} \tens T \tens \f^{\tens k-1}) .\]
{}From the third presentation \eqref{ADE} we get that $a_k$ acts by
applying ribbon twist to $D_k\pti \tens D_{k-1}$, which induces
$T\tens T\cdot\Omega : F\tens F \to F\tens F$. Hence,
\[ Z(a_k) = \Hom(X,\f^{\tens g-k} \tens T\tens T \tens \f^{\tens k-2})
\cdot \Hom(X,\f^{\tens g-k} \tens \Omega \tens \f^{\tens k-2}) .\]
{}From the second presentation \eqref{CDE} we get that $e_k$ acts by
applying ribbon twist to $D_k\pti \tens C_k$ or, equivalently, to
$D_k\pti\tens D_{k-1}\tens D_{k-1}\pti\tens\dots\tens D_1\tens D_1\pti$.
This induces
$\Omega^r\cdot T\tens \nu : F\tens F^{\tens k-1} \to F\tens F^{\tens k-1}$,
hence,
\[ Z(e_k) = \Hom(X,\f^{\tens g-k} \tens \Omega^r_{\f,\f^{\tens k-1}})
\cdot \Hom(X,\f^{\tens g-k} \tens T\tens \nu_{\f^{\tens k-1}}) .\]
Finally, $b_k$ is conjugate to $d_k$ by a homeomorphism of the type $S$,
and we set
\begin{align*}
Z(b_k) &= \Hom(X, \f^{\tens g-k} \tens STS^{-1} \tens \f^{\tens k-1}) \\
&= \Hom(X, \f^{\tens g-k} \tens S^{-1}TS \tens \f^{\tens k-1}) .
\end{align*}

As shown in \cite{Lyu:r=m} these operators define a projective
representation $M_{g,1} \to \break PGL(Z(\Si_{g,1}))$ with the
2-cocycle whose values are powers of $\lambda$. Therefore, this
representation is induced by a projective representation
$z: M_{g,1} \to  \Aut_\CC(\f^{\tens g})/\{\lambda^n\}_{n\in\Z}$
\begin{align*}
a_k &\longmapsto
\f^{\tens g-k} \tens (T\tens T \cdot \Omega)\tens \f^{\tens k-2} ,\\
b_k &\longmapsto \f^{\tens g-k} \tens STS^{-1} \tens \f^{\tens k-1} ,\\
d_k &\longmapsto \f^{\tens g-k} \tens T \tens \f^{\tens k-1} ,\\
e_k &\longmapsto \f^{\tens g-k} \tens (\Omega^r_{\f,\f^{\tens k-1}}
\cdot T\tens \nu_{\f^{\tens k-1}}) .
\end{align*}
Indeed, set $X=\f^{\tens g}$ and apply both parts of relations (A)-(E)
to the vector $\id_X\in Z(\Si_{g,1})$.

\subsubsection{Closed surfaces}
Now let $\Si_{g,0}$ be a closed surface of genus $g$. The mapping class
group $M_{g,0}$ is the quotient of $M_{g,1}$ by two extra relations
\cite{Waj}:

(F)\ $H_g^2=1$ , where
\[ H_k=b_ga_g \dots b_2a_2b_1a_1a_1b_1a_2b_2 \dots a_g b_g d_g e_g .\]

(G)\ $d_g=e_g$ .

\noindent Setting
\begin{align*}
Z(\Si_{g,0}) &= \Hom(\C, \f^{\tens g}) \\
Z(a_k) &=
\Hom(\C, \f^{\tens g-k} \tens (T\tens T\cdot\Omega)\tens \f^{\tens k-2}) ,\\
Z(b_k) &= \Hom(\C, \f^{\tens g-k} \tens STS^{-1} \tens \f^{\tens k-1}) ,\\
Z(d_k) &= \Hom(\C, \f^{\tens g-k} \tens T \tens \f^{\tens k-1}) ,\\
Z(e_k) &= \Hom(\C, \f^{\tens g-k} \tens (\Omega^r_{\f,\f^{\tens k-1}}
\cdot T\tens \nu_{\f^{\tens k-1}})) ,
\end{align*}
we get a projective representation $M_{g,0} \to PGL(Z(\Si_{g,0}))$
\cite{Lyu:r=m}.

\subsubsection{Exercises}
The reader might want to check some of the above results straightforwardly.
Several exercises will help in doing this. They present minor
generalizations of some of the quoted results and specify the power of
$\lambda$ involved in relations.

\begin{easyexercise}
All maps $z(a_i), z(d_j), z(e_k) : \f^{\tens g} \to \f^{\tens g}$ commute.
$z(b_i)$ commutes with all maps except $z(a_i)$, $z(a_{i+1})$, $z(d_i)$,
$z(e_i)$. Also $z(d_i) z(b_i) z(d_i) = z(b_i) z(d_i) z(b_i)$ (see
relation (A)).
\end{easyexercise}

\begin{exercise}
For any $Y\in\CC$ we have
\begin{multline*}
\Omega^r_{\f,Y} \cdot S^{-1}\tens\nu_Y \cdot \Omega^r_{\f,Y} \\
= \lambda (T^{-1}S^{-1}T^{-1})\tens Y \cdot \Omega^r_{\f,Y} \cdot
(S^{-1}T^{-1})\tens Y : \f\tens Y \to \f\tens Y ,
\end{multline*}
\begin{multline*}
\Omega^l_{Y,\f} \cdot \nu_Y\tens S^{-1} \cdot \Omega^l_{Y,\f} \\
= \lambda Y\tens(T^{-1}S^{-1}T^{-1}) \cdot \Omega^l_{Y,\f} \cdot
Y\tens(S^{-1}T^{-1}) : Y\tens\f \to Y\tens\f .
\end{multline*}
\end{exercise}

\begin{easyexercise}
Deduce from the previous exercise that
$z(a_i) z(b_i) z(a_i) = z(b_i) z(a_i) \break
z(b_i)$, $z(a_{i+1}) z(b_i) z(a_{i+1}) = z(b_i) z(a_{i+1}) z(b_i)$,
$z(e_i) z(b_i) z(e_i) = z(b_i) z(e_i) z(b_i)$ (see relation (A)).
\end{easyexercise}

\begin{exercise}
For any $Y\in\CC$ we have
\[ (\nu_Y\tens S^{-1} \cdot \Omega^l_{Y,\f})^4 =
\nu_Y\tens\f \cdot \nu_{Y\tens\f} : Y\tens\f \to Y\tens\f .\]
What does it mean for $Y=I$?
\end{exercise}

\begin{easyexercise}
Deduce from the previous exercise that
\[ (z(d_1) z(b_1) z(a_2))^4 = \lambda^4 z(d_2) z(e_2) \]
(relation (B)).
\end{easyexercise}

\begin{exercise}
For any $Y\in\CC$ we have
\[ (T^{-1}S^{-1})\tens Y \cdot \Omega^r_{\f,Y} \cdot (S^{-1}TS)\tens Y
\cdot (\Omega^r_{\f,Y})^{-1} \cdot (ST)\tens Y =
\Omega^r_{\f,Y} \cdot T\tens\nu_Y : \f\tens Y \to \f\tens Y .\]
\end{exercise}

\begin{exercise}
Deduce from the previous exercise that $z(e_2)=z(G_2)z(d_2)z(G_2)^{-1}$,
where $z(G_2) \overset{\text{def}}= z(b_2) z(a_2) z(d_1)^2 z(a_2) z(b_2)$.
\end{exercise}

\begin{easyexercise}
Show that
\[ Z(d_g) = Z(e_g) : \Hom(\C,\f^{\tens g}) \to \Hom(\C,\f^{\tens g}) .\]
\end{easyexercise}

\subsection{General case}
Let $\Si_{g,n}$ be a surface of genus $g$ with $n$ disks removed, boundary
circles are labelled by $X_1,\dots,X_n$. Figure \ref4 suggests an embedding
$\Si_{g,1} \hookrightarrow \Si_{g,n}$, which induces a homomorphism
$M_{g,1}\to M_{g,n}$. The mapping class group $M_{g,n}$ is generated by
images $a_k,b_k,d_k,e_k$ of generators of $M_{g,1}$, the braidings
$\omega_i$, the twists $R_i$, new generators $S_l$, which are
homeomorphisms of the type $S$ inside the $\T^2-D^2$ region $F_l$ and
identity outside, and inverse Dehn twists $t_{j,k}$ in tubular
neighbourhood of the cycles $t_{j,k}$
\def\epsfsize#1#2{0.579#1}
\begin{figure}[htbp]
\[ \epsfbox{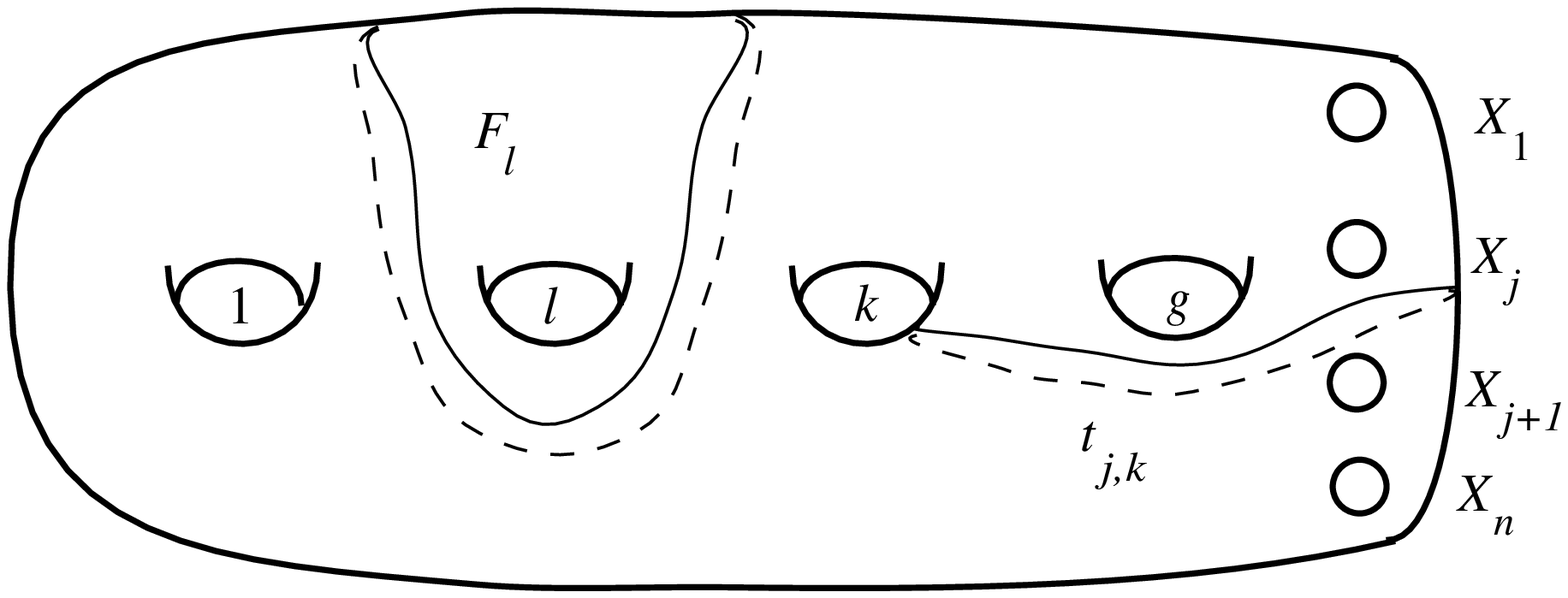} \]
\caption{\label4}
\end{figure}
This is not a minimal system of generators, since $S_k$ can be expressed
through $b_k$ and $d_k$. All of these generators are easily representable.

Set
{\allowdisplaybreaks
\begin{align*}
Z(\Si_{g,n}) &= \Hom(X_1\tens X_2\tens\dots\tens X_n, \f^{\tens g}) \\
Z(a_k) &= \Hom(X_1\tens\dots\tens X_n,
\f^{\tens g-k} \tens (T\tens T\cdot\Omega)\tens \f^{\tens k-2}) ,\\
Z(b_k) &= \Hom(X_1\tens\dots\tens X_n,
\f^{\tens g-k} \tens STS^{-1} \tens \f^{\tens k-1}) ,\\
Z(d_k) &= \Hom(X_1\tens\dots\tens X_n,
\f^{\tens g-k} \tens T \tens \f^{\tens k-1}) ,\\
Z(e_k) &= \Hom(X_1\tens\dots\tens X_n, \f^{\tens g-k} \tens
(\Omega^r_{\f,\f^{\tens k-1}}\cdot T\tens \nu_{\f^{\tens k-1}})) ,\\
Z(\omega_i) &= \Hom(X_1\tens\dots\tens c_{X_{i+1}, X_i} \tens\dots\tens X_n,
\f^{\tens g}) ,\\
Z(R_i) &= \Hom(X_1\tens\dots\tens \nu_{X_i} \tens\dots\tens X_n,
\f^{\tens g}) ,\\
Z(S_k) &= \Hom(X_1\tens\dots\tens X_n,
\f^{\tens g-k} \tens S \tens \f^{\tens k-1}) .
\end{align*}
}
Set $Z(t_{j,k})$ to be the composition
\[ \qquad
\begin{CD}
\Hom(X_1\tens\dots\tens X_j\tens X_{j+1}\tens\dots\tens X_n,\f^{\tens g}) \\
@V\wr VV \\
\Hom(X_1\tens\dots\tens X_{j}, \f^{\tens g-k}\tens\f\tens\f^{\tens k-1}
\tens\pti X_n\tens\dots\tens\pti X_{j+1}) \\
@V\Hom(X_1\tens\dots\tens X_{j}, \f^{\tens g-k}\tens
\Omega^r_{\f,\f^{\tens k-1}\tens\pti X_n\tens\dots\tens\pti X_{j+1}})VV \\
\Hom(X_1\tens\dots\tens X_{j}, \f^{\tens g-k}\tens\f\tens\f^{\tens k-1}
\tens\pti X_n\tens\dots\tens\pti X_{j+1}) \\
@V\Hom(X_1\tens\dots\tens X_{j}, \f^{\tens g-k}\tens T \tens
\nu_{\f^{\tens k-1} \tens\pti X_n\tens\dots\tens\pti X_{j+1}})VV \\
\Hom(X_1\tens\dots\tens X_{j}, \f^{\tens g-k}\tens\f\tens\f^{\tens k-1}
\tens\pti X_n\tens\dots\tens\pti X_{j+1}) \\
@V\wr VV \\
\Hom(X_1\tens\dots\tens X_j\tens X_{j+1}\tens\dots\tens X_n,\f^{\tens g})
\end{CD}
\]
Then the above is a projective representation of $M_{g,n}$ \cite{Lyu:r=m}.

I don't know any defining system of relations of $M_{g,n}$. Perhaps, it
was never written explicitly. Proof of the result \cite{Lyu:r=m} used the
exact sequence \cite{Bir:mcg,Scott}
\[ \begin{CD}
1 @>>> \bar B_{g,n} @>>> M_{g,n} @>>> M_{g,0} @>>> 1, \\
1 @>>> \Z^n @>>> \bar B_{g,n} @>>> B_{g,n} @>>> 1,
\end{CD} \]
where $B_{g,n}$ is the braid group of a surface of genus $g$, whose
presentation was given by Scott \cite{Scott}.

\section{Invariants of closed 3-manifolds}\label{Invariants}
I have nothing to add to the method of obtaining invariants of closed
3-manifolds via Kirby calculus invented by Reshetikhin and
Turaev~\cite{ResTur:3}. It was used afterwards by many authors including
Lickorish~\cite{Lic:TL,Lic:calTL}, Kirby and Melvin~\cite{KirMel}.
Turaev~\cite{Tur:q3} have shown that a semisimple modular category
serves well to define a 3-manifold invariant. In this paper I propose
to use (eventually non-semisimple) 3-modular categories $\CC$ as
starting data for the method. When $\CC=H$-mod this invariant
coincides with the one defined by Hennings~\cite{Hen:3} translated to
unoriented setting by Kauffman and Radford~\cite{KauRad:3}.

\subsection{3-manifolds from links in $S^3$}
It is well known that any closed connected oriented 3-manifold can be
obtained via surgery along a framed tame link in $S^3$ \cite{Lic:3}.
Namely, given a framed tame link
$L=L_1\sqcup L_2\sqcup \dots \sqcup L_m \subset S^3 = \pa B^4$ we construct
a compact oriented 4-manifold $W_L$ glueing $m$ 2-handles $D^2\times D^2$
($D^2$ is a disk) to the ball $B^4$ along closed tubular neighbourhoods
$U_i\subset S^3$ of $L_i$. The part of the boundary
$S^1\times D^2 = \pa D^2\times D^2$ of $i^{\text{th}}$ 2-handle is
identified with $U_i\simeq L_i\times D^2$ so that the linking number of
$S^1\times1 \subset \pa D^2\times\pa D^2$ with $L_i$ coincides with the
self-linking number of $L_i$. The boundary $M_L=\pa W_L$ is an oriented
compact closed connected 3-manifold.

Kirby proved \cite{Kir:cal} that two manifolds $M_L$ and $M_{L'}$ obtained
by surgery from two framed links $L,L'\subset S^3$ are homeomorphic iff
the link $L'$ can be obtained from $L$ by a finite sequence of
transformations, which we call Kirby moves:

\begin{enumerate}
\item The elimination or insertion of an unknotted component
labelled $\pm1$, unlinked with other components.
\item The band (or handle slide) move---making a connected sum of one
of the components
$L_i$ with a parallel copy $\tilde L_j$ of another component $L_j$.
\end{enumerate}

Fenn and Rourke \cite{FenRou:cal} proved a similar result with a
subset of transformations introduced by Kirby, which we call
Kirby--Fenn--Rourke moves (see Figures~\ref7--\ref8).
\def\epsfsize#1#2{0.3#1}
\begin{figure}[htbp]
\[ \epsfbox{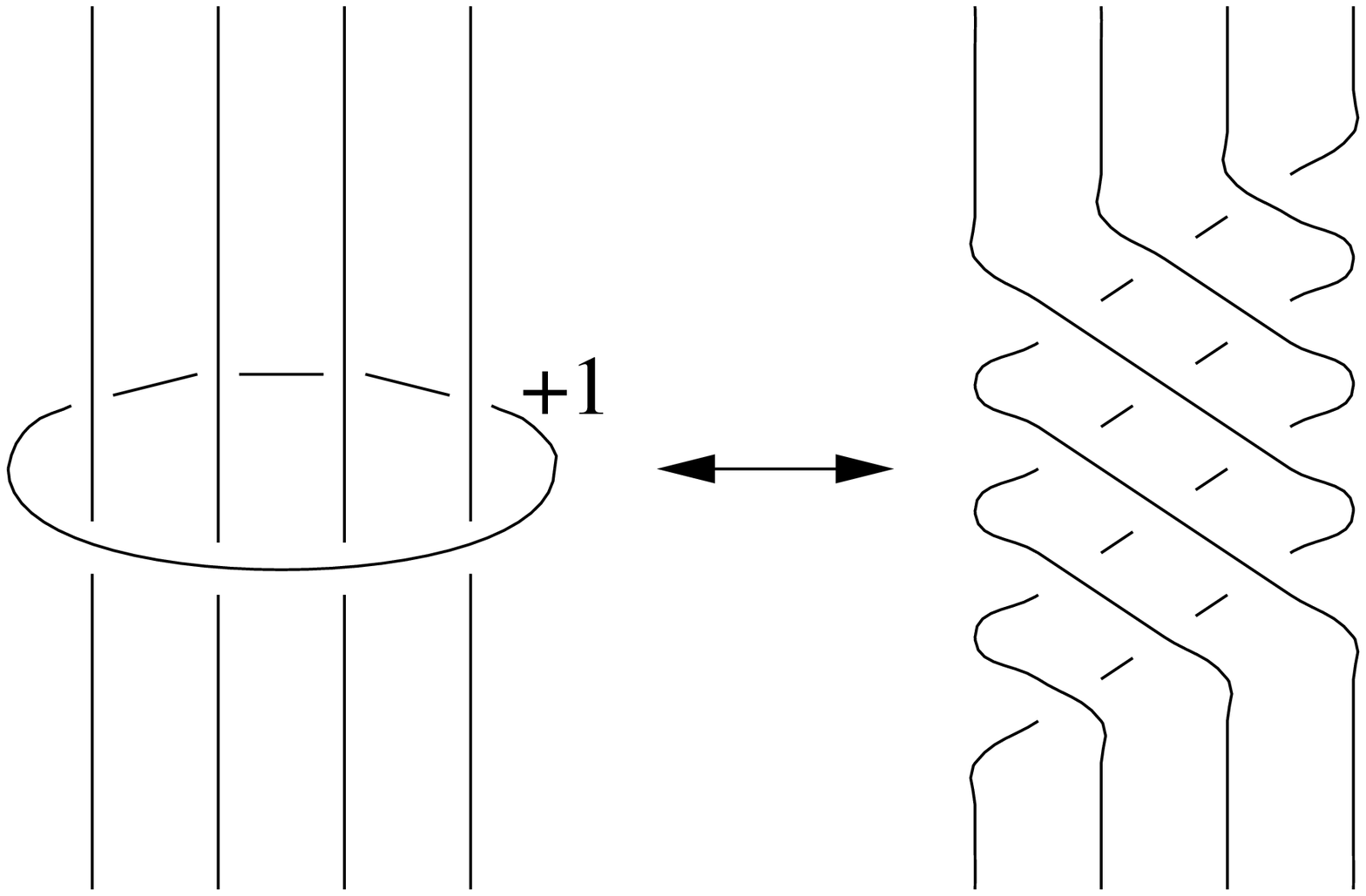} \]
\caption{\label7}
\end{figure}
\begin{figure}[htbp]
\[ \epsfbox{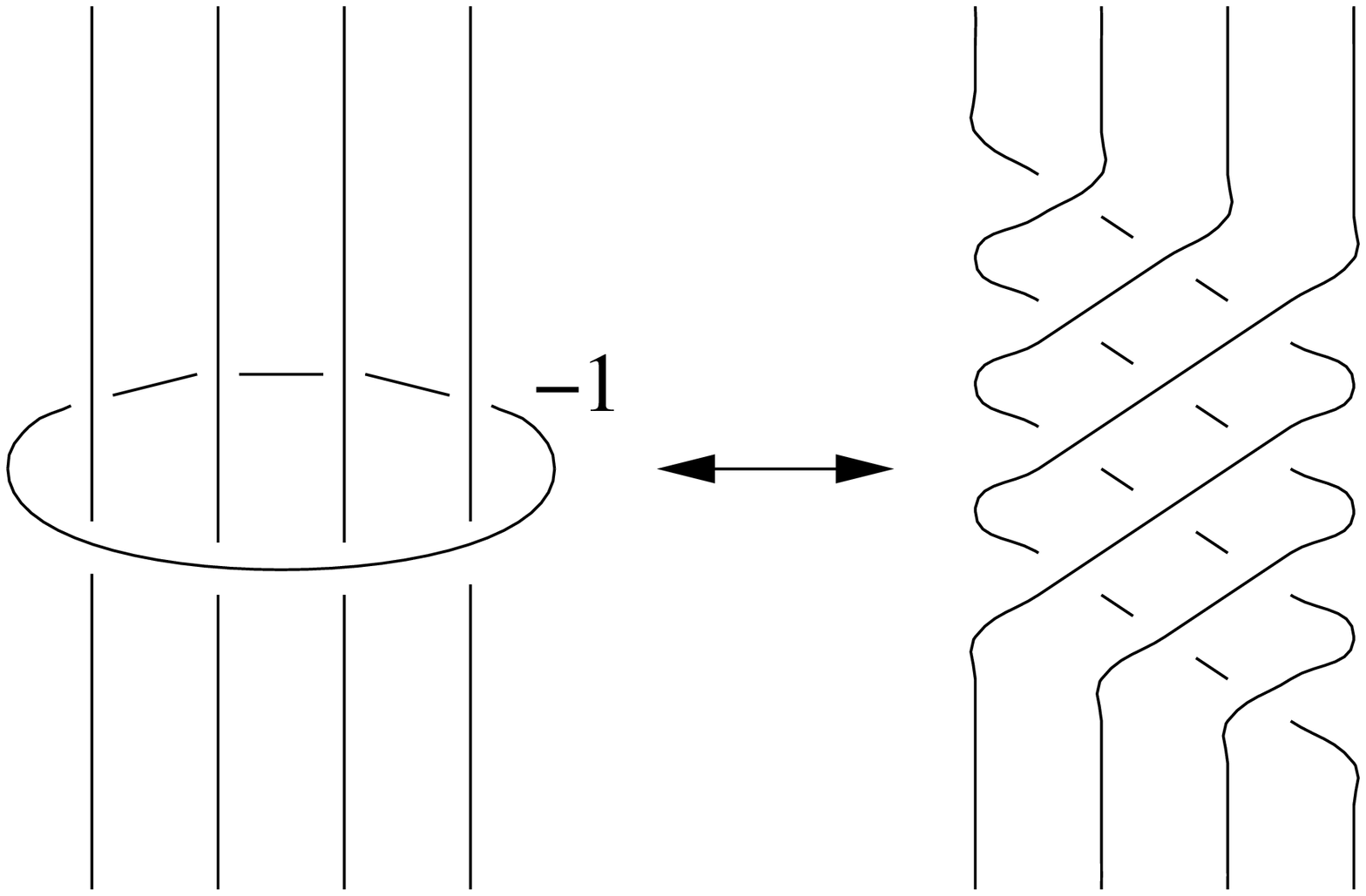} \]
\caption{\label8}
\end{figure}
Therefore, to give an invariant $\tau(M)\in k$ for 3-manifolds $M=M_L$
amounts to give an invariant $\tau(L)$ for framed links $L$ in $S^3$,
which would not change under Kirby or Kirby--Fenn--Rourke moves. We may
and we shall consider framed links in $\R^3$ instead of $S^3$.

\subsection{An invariant determined by a 3-modular category}
Let $\CC$ be a 3-modular category. The reader might want to assume that
$\CC=H$-mod for simplicity. First of all we construct an invariant of
framed links in $\R^3$.

Let $L$ be a framed tame link in $\R^3$. It can be deformed to a smooth
link so that its projection $p(L)$ to the standard plane $\R^2$ were a
smooth generic link diagram $D_L$ and the normal vector field on $L_i$
determining the framing were parallel to $\R^2$. Choose a straight line
$\R^1$ in $\R^2$ so that $D_L$ lies in one half-plane with repect to
this line (we shall draw it as a lower half-plane). Choose a point
$e_i\in L_i$ for each $i$ and connect it in $\R^3$ with a point $x_i$ of
$\R^1$ by a curve $\gamma_i$, which projects to the lower half-plane.
We assume $p(\gamma_i)$ smooth and transversal to $p(L_i)$ in the point
$p(e_i)$. Make it generic, so that $x_i\ne x_j$ and the curves $\gamma_j$
don't intersect each other and $L$ except at the ends. Draw the
corresponding plane diagram $\bar D_L$ assigning necessary signs of
overcrossing or undercrossing to each double point of the projection to
$\R^2$. Duplicate the projections of the curves $\gamma_i$, replacing them
by two parallel curves $\gamma_i^-$ and $\gamma_i^+$ (e.g. parts of the
boundary of a small neighbourhood $V_i$ of $p(\gamma_i)$). Remove the
connected component of $L_i\cup V_i$ containing $e_i$. The result will
be a diagram of a tangle looking like \figref9.
\def\epsfsize#1#2{0.4#1}
\begin{figure}[htbp]
\[ \epsfbox{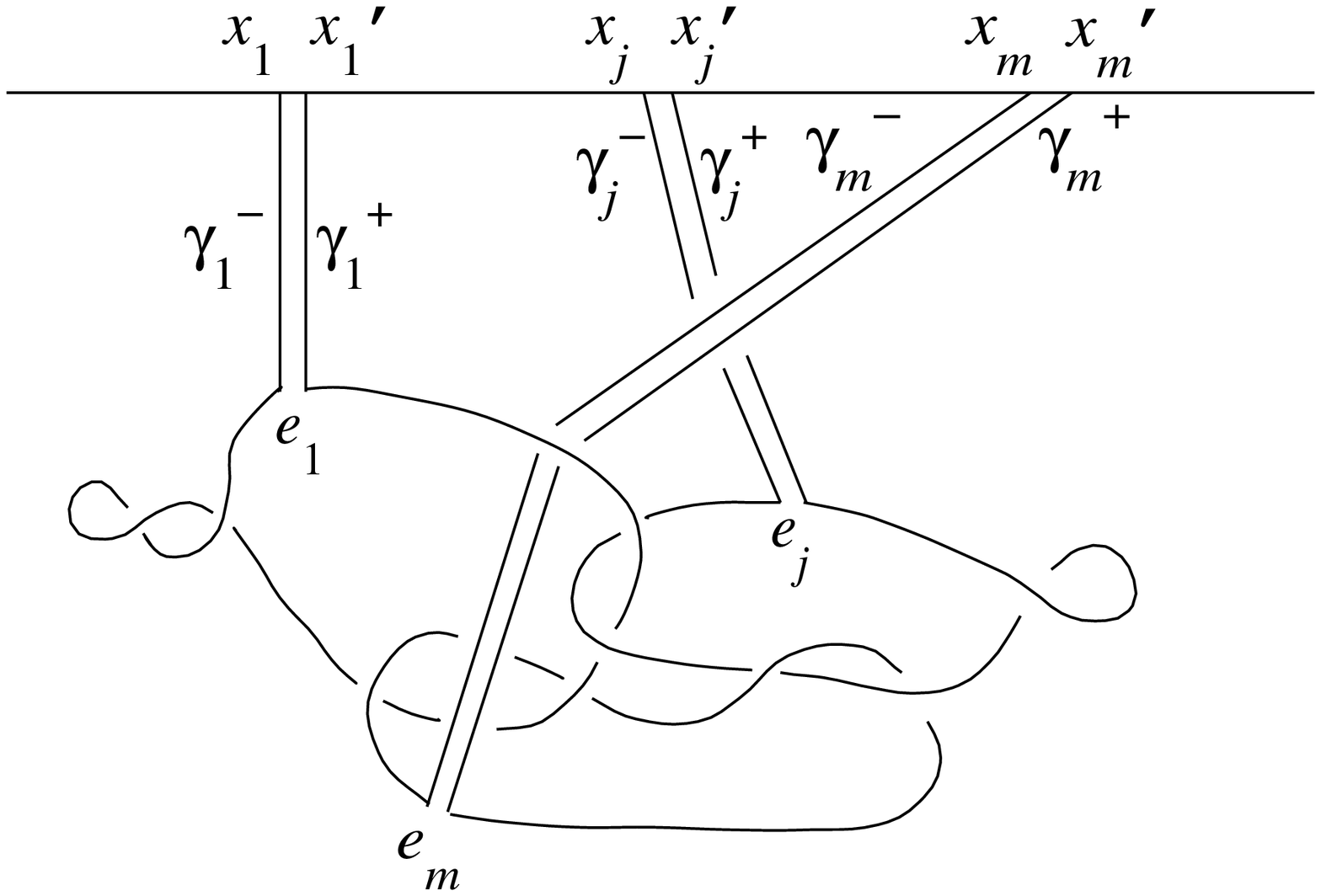} \]
\caption{\label9}
\end{figure}
Make this diagram into a $\CC$-tangle $T_L$ \cite{Lyu:mod} assigning a
color $X_i\in\Ob\CC$ to the point $x_i$ and $X_i\pti$ to the point $x_i'$
and inserting to $\gamma_i^-$ such a morphism
$u_0^{2a}: X_i\to X_i^{(2a}{}\pti{}^)$, $a\in\Z$, as consistency
requires. The tangle $T_L$ represents a morphism in $\CC$
\[ \Phi(T_L;X_1,\dots,X_m) : X_1\tens X_1\pti\tens X_2\tens X_2\pti\tens
\dots\tens X_m\tens X_m\pti\to I .\]

Fix all objects $X_i,X_i\pti$ except for $i=j$ and vary $X_j,X_j\pti$.
We see that the following diagram commute for any morphism
$f:Y_j\to Z_j\in\CC$
\[
\begin{CD}
{\scriptstyle
X_1\tens X_1\pti\tens\dots\tens Y_j\tens Z_j\pti\tens\dots\tens
X_m\tens X_m\pti} @>\dots\tens Y_j\tens f^t\tens\dots>>
\nquad {\scriptstyle
X_1\tens X_1\pti\tens\dots\tens Y_j\tens Y_j\pti\tens\dots\tens
X_m\tens X_m\pti} \nqqquad \\
@V\dots\tens f\tens Z_j\pti\tens\dots VV
@VV\Phi(T_L;X_1,\dots,Y_j,\dots,X_m)V \\
{\scriptstyle
X_1\tens X_1\pti\tens\dots\tens Z_j\tens Z_j\pti\tens\dots\tens
X_m\tens X_m\pti} @>\Phi(T_L;X_1,\dots,Z_j,\dots,X_m)>> {\scriptstyle I}
\end{CD}
\]
Indeed, insert a morphism $f$ to a point of $\gamma_j^-$ in the
\figref9 and push it along the curve through all crossings. It will appear
on $\gamma_j^+$ as $f^t$.

Commutativity of the above diagram means that $\Phi(T_L;\dots)$ factorizes
through the canonical mappings $i_{X_j}:X_j\tens X_j\pti \to F$ on
$j^{\text{th}}$ place (compare with \eqref{coend}).
Therefore, it defines a morphism
\[ \Phi(T_L): F^{\tens m} \to I .\]

Take an invariant element of $F$, that is, a morphism $\al:I\to F\in\CC$.
Define a number in $k$
\[ \phi(T_L,\al) : I=I^{\tens m} @>\al^{\tens m}>>
F^{\tens m} @>\Phi(T_L)>> I .\]

\begin{proposition}
Assume that $\gamma_F(\al)=\al$. Then the number $\phi(T_L,\al)\in k$
depends only on $L$ and $\al$ and not on choices made in constructing $T_L$.
\end{proposition}

\begin{notation}
$\tau(L,\al) = \phi(T_L,\al)$.
\end{notation}

\begin{pf}
Since the source of $\al$ is a unity object, we can change all overcrossings
with $\gamma_j$ to undercrossings and vice versa without changing the
value of $\phi(T_L,\al)$. Since all tensor factors in $\al^{\tens m}$
coincide, the result does not depend on the order of $x_j=\gamma_j(1)$ on
the line. Indeed, an interchange of the two neighbour ends $x_i$ and $x_j$
is equivalent to adding of a crossing which is interpreted as
$c:I\tens I \to I\tens I$, and this is an identity morphism.

The tangent vector $p(\gamma_i)'(0)$ can be chosen normal to $p(L_i)$ at
the point $p(e_i)$. The following diagram explains that the change from
one normal vector to another is equivalent to composing $\al$ with
$\gamma_F$ or $\gamma_F^{-1}$ on $j^{\text{th}}$ place (\figref{10}).
\def\epsfsize#1#2{0.4#1}
\begin{figure}[htbp]
\[ \epsfbox{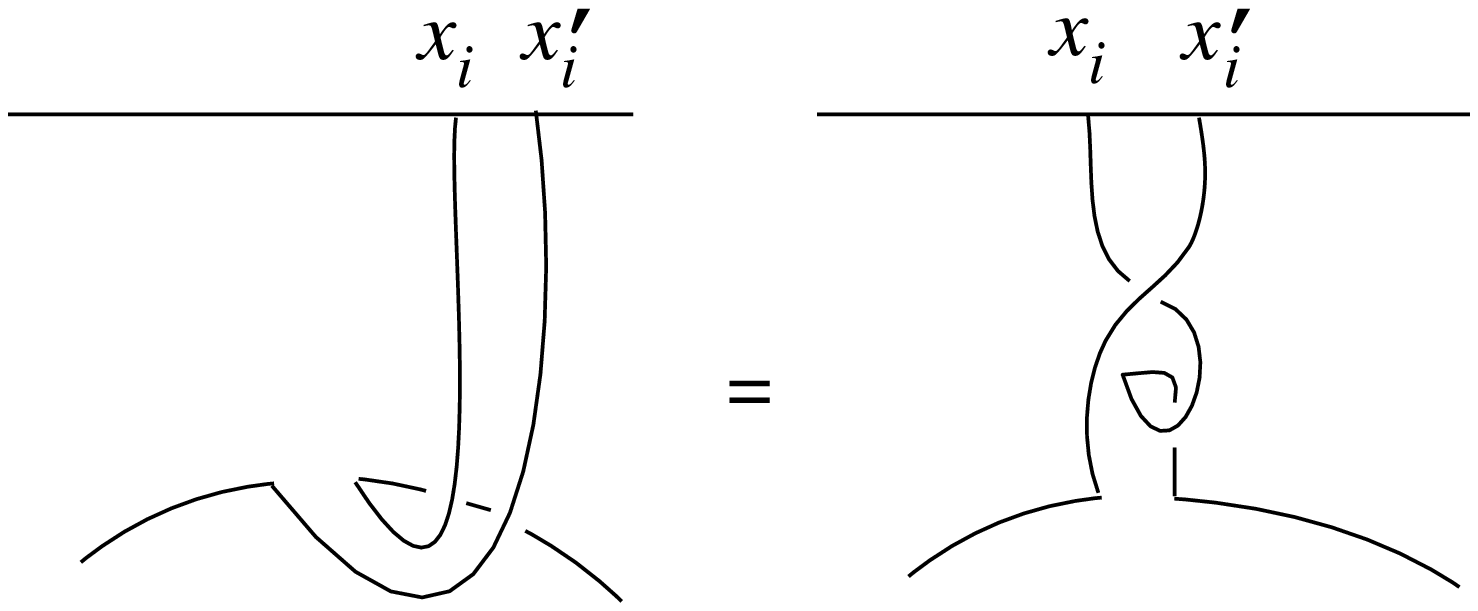} \]
\caption{\label{10}}
\end{figure}
Since $\gamma_F\al = \al = \gamma_F^{-1} \al$ the resulting $\phi(T_L,\al)$
will not change.

Also the value of $\phi(T_L,\al)$ does not depend on the choice of $e_i$.
Indeed, make $e_i$ slide along $L_i$ and deform $\gamma_i$ simultaneously.
This will not change $\Phi(T_L;X_1,\dots,X_m)$ and a forteriori
$\phi(T_L,\al)$.

Therefore, $\phi(T_L,\al)$ depends only on the plane diagram $\bar D_L$
and $\al$. It is invariant under the three Reidemeister moves performed
on $\bar D_L$, namely, $\Omega2$, $\Omega3$ \cite{Rei} and

$\Omega1$F \hspace*{3cm}
\unitlength=0.8mm
\raisebox{-9mm}{
\begin{picture}(51.53,25)
\put(12.33,12){\oval(10,10)[r]}
\put(12.33,14){\oval(6,6)[lt]}
\put(37.33,14){\oval(6,6)[rt]}
\put(9.33,14){\line(-2,-3){9.33}}
\put(0.33,24){\line(2,-3){6.67}}
\put(51.33,0){\line(-4,5){8}}
\put(12.33,10.50){\oval(6,7)[lb]}
\put(37.33,12){\oval(12,10)[l]}
\put(36.83,10){\oval(7,6)[rb]}
\put(40.33,10){\line(4,5){11.20}}
\put(24.33,12){\makebox(0,0)[cc]{=}}
\end{picture}
}

\noindent Indeed, these moves are local, and we can always choose the points
$e_i\in L_i$ outside the changed pieces, hence, the corresponding moves
for $\CC$-tangles apply without changing the morphism
$\Phi(T_L;X_1,\dots,X_m)$. For $\Omega1$F move notice that
$u_1^2:X\to X\pti\pti$ (resp. $u_1^{-2}:X\to \pti\pti X$) in the left
(resp. right) hand side will be accompanied by a power of $u_0^2$. Thus,
invariance under $\Omega1$F follows from the equation
\[ u_0^{-2} u_1^2 = \nu = u_0^2 u_1^{-2} : X\to X .\]
Since the set of equivalence classes of plane diagrams under $\Omega1$F,
$\Omega2$, $\Omega3$ is the same as the set of equivalence classes of framed
links in $\R^3$ (\cite{Lyu:tan} after \cite{Rei}), we deduce that
$\phi(T_L,\al)$ depends only on $L$ and $\al$.
\end{pf}

\subsubsection{A 3-manifold invariant}
By the very definition of a 3-modular category $\CC$ the Hopf algebra
$F\in\CC$ has a two-sided integral $\si:I\to F$. In the following we
shall consider the link invariant $\tau(L,\si) = \phi(T_L,\si)$.

Denote by $s(L)$ the signature of the intersection form in $H^2(W_L;\R)$,
where $W_L$ is the 4-manifold obtained by surgery along $L$ and
$M_L=\pa W_L$. It is the same as the signature of the linking matrix of
$L$ \cite{Kir:4M}.

\begin{theorem}\label{3-inv}
The number
\[ \tau(M_L) = \lambda^{-s(L)} \tau(L,\si) \]
is an invariant of 3-manifolds, that is, it depends only on the
homeomorphism class of $M_L$.
\end{theorem}

\begin{pf}
The equations \eqref{nulambda}, \eqref{nu-lambda-} show that application
of one of the Kirby--Fenn--Rourke moves multiplies $\tau(L,\si)$ by
$\lambda^{\pm1}$. Simultaneously $s(L)$ increases by $\pm1$.
\end{pf}

It is worth understanding why $\tau(M_L)$ is invariant under Kirby moves.
The answer is because $\sigma$ is an integral in $F$. Graphically this
property is expressed by the following equation
\be\label{intexi}
\unitlength=0.6mm
\raisebox{-18mm}{
\begin{picture}(101,71)
\put(1,13){\framebox(30,8)[cc]{$\xi$}}
\put(8,21){\line(5,4){25}}
\put(24,21){\line(3,4){15}}
\put(31,41){\framebox(10,8)[cc]{$\si$}}
\put(29,21){\line(0,1){5}}
\put(29,29){\line(0,1){7}}
\put(29,40){\line(0,1){20}}
\put(3,21){\line(0,1){39}}
\put(36,49){\line(0,1){11}}
\put(8,62){\makebox(0,0)[cb]{$X$}}
\put(24,62){\makebox(0,0)[cb]{$X\pti$}}
\put(40,62){\makebox(0,0)[cb]{$I$}}
\put(8,13){\line(0,-1){11}}
\put(24,2){\line(0,1){11}}
\put(16,2){\makebox(0,0)[ct]{$F$}}
\put(51,32){\makebox(0,0)[cc]{=}}
\put(67,21){\line(5,4){25}}
\put(83,21){\line(3,4){15}}
\put(90,41){\framebox(10,8)[cc]{$\si$}}
\put(95,49){\line(0,1){11}}
\put(67,62){\makebox(0,0)[cb]{$X$}}
\put(83,62){\makebox(0,0)[cb]{$X\pti$}}
\put(99,62){\makebox(0,0)[cb]{$I$}}
\put(75,60){\oval(20,22)[b]}
\put(67,21){\line(0,-1){19}}
\put(83,2){\line(0,1){19}}
\put(75,2){\makebox(0,0)[ct]{$F$}}
\end{picture}
} : X\tens X\pti\tens I \to F .
\end{equation}
Here the multiplication morphism $\xi$ is determined from the commutative
diagram
\[ \begin{CD}
X\tens Y\tens Y\pti\tens X\pti @>X\tens i_Y\tens X\pti>>
X\tens F\tens X\pti \\
@V\wr VV   @VV\xi V \\
(X\tens Y) \tens (X\tens Y)\pti @>i_{X\tens Y}>>  F
\end{CD} \]

\begin{figure}[htbp]
\[ \epsfbox{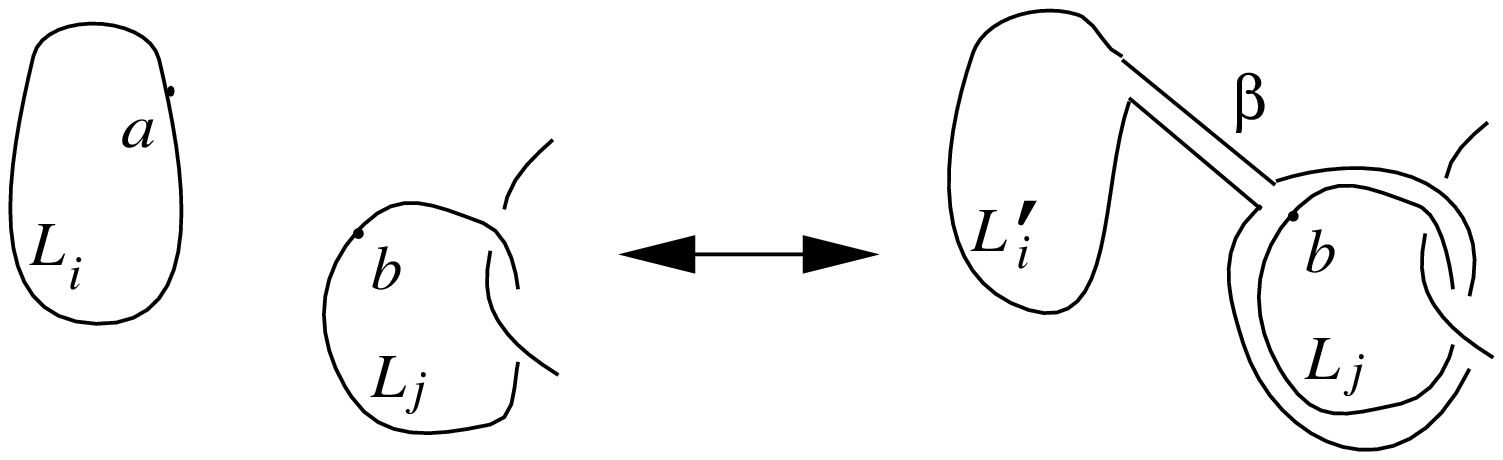} \]
\caption{\label{11}}
\end{figure}

\begin{figure}[htbp]
\[ \epsfbox{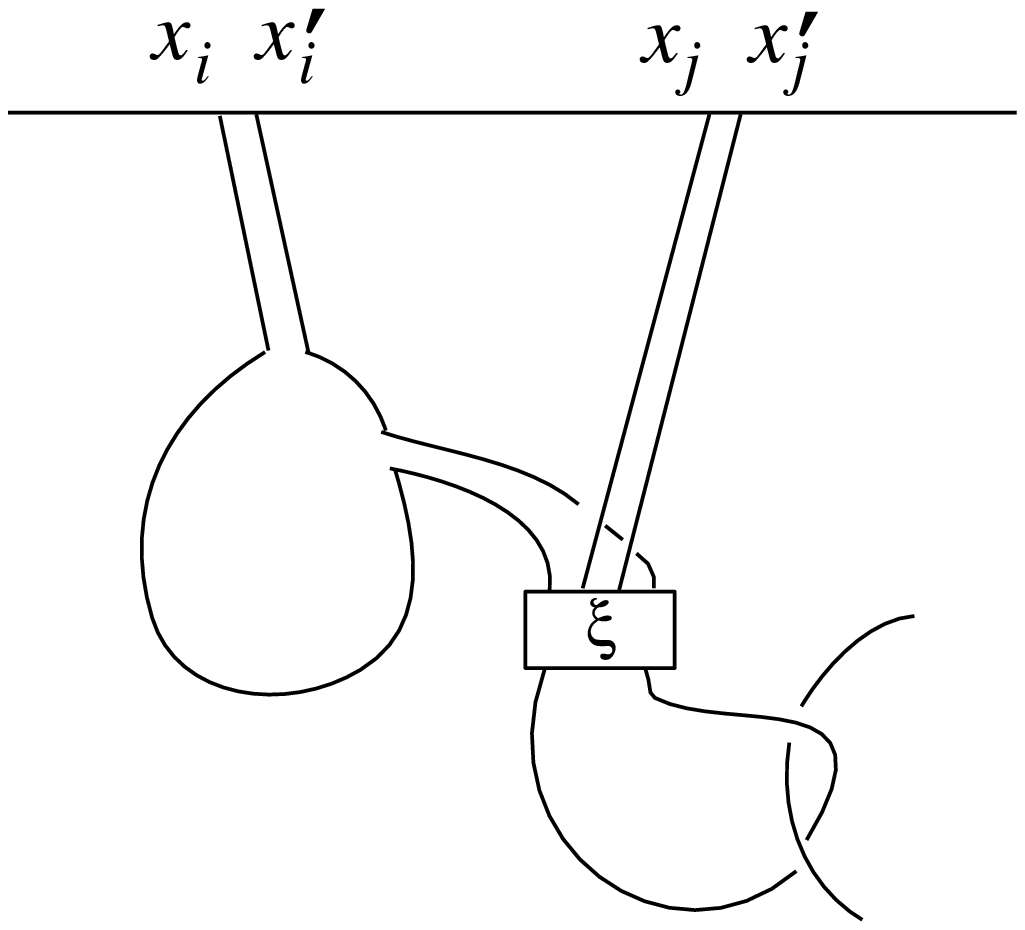} \]
\caption{\label{12}}
\end{figure}

Let us give another proof of \thmref{3-inv}, using only the Kirby moves.
For particular categories related with $SL(2)$ this was done by
Lickorish \cite{Lic:ske} through a different approach.
Choose a pair of components
$L_i,L_j$ in $L$ and perform a Kirby band move as shown at \figref{11}.
\def\epsfsize#1#2{0.579#1}
The points $a\in L_i$ and $b\in L_j$ become ``connected'' by a double line
$\beta$. Take the point $b$ as $e_j$ and choose $\gamma_j$ so that
$\gamma_j'(0)$ pointed out in the direction of $\beta$. Duplicate all
$\gamma_k$ as required by the recipee. Since the curves
$L_j$ and a new-made part of $L_i'$ go parallelly, we can insert the
morphism $\xi$ to this $\CC$-$F$-tangle without changing the morphism
$\dots\tens X\tens X\pti\tens\dots\tens F\tens\dots \to k$ it represents
(see \figref{12}).
\def\epsfsize#1#2{0.4#1}
In the vicinity of $e_j$ the picture looks like the left hand side of
\eqref{intexi}. Therefore we can change it to the right hand side without
changing $\tau(L',\si)$. The resulting tangle at \figref0
\begin{figure}[htbp]
\[ \epsfbox{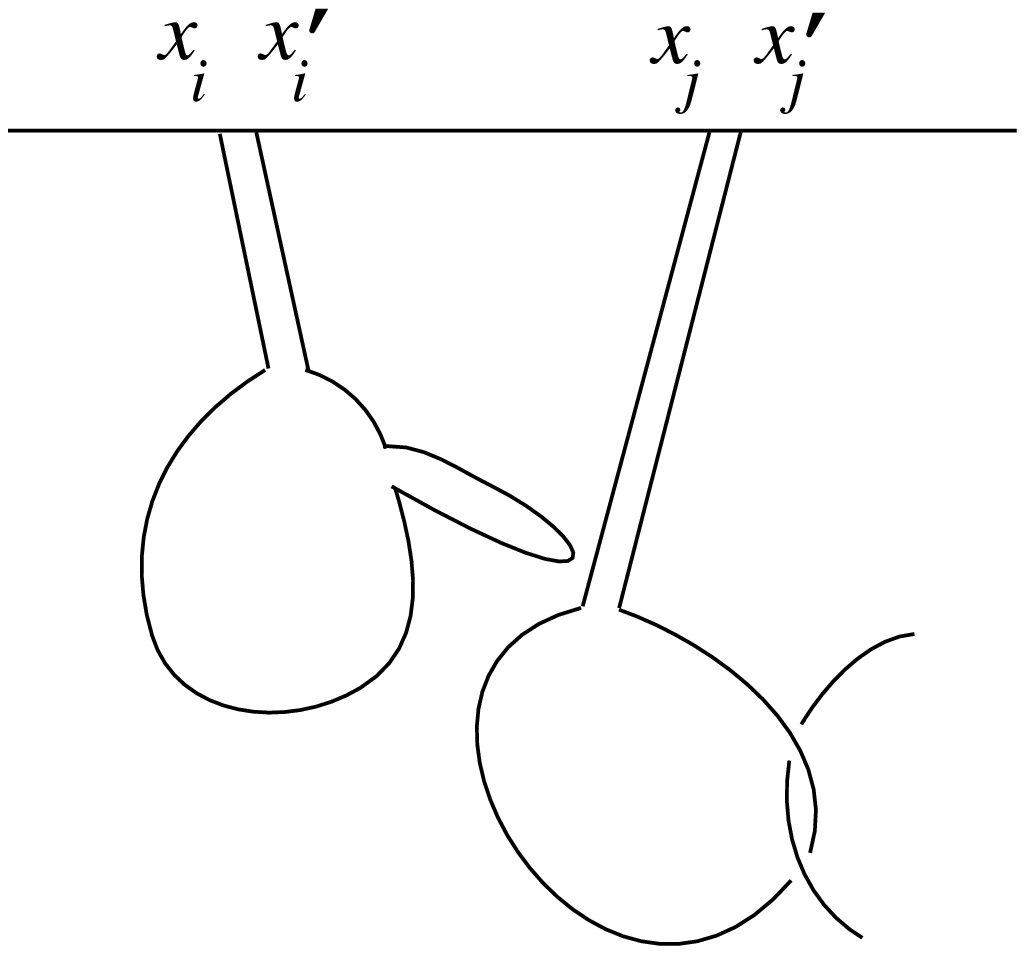} \]
\caption{\label0}
\end{figure}
is isotopic to the initial one, so its value $\tau(L',\si)$ equals to
$\tau(L,\si)$. Finally, the signature $s(L)$ will not change under Kirby
band move \cite{Kir:4M}, whence \thmref{3-inv} follows.

\subsection{Lens spaces}
We calculate as an example the invariant $\tau$ for lens spaces
$L(p,q)$. Consider the $n$-component chain link $L\subset S^3$ and the
value of its invariant
\def\epsfsize#1#2{0.579#1}
\be\label{chain}
\tau\left(\ \raisebox{-25mm}[25mm][25mm]{\epsfbox{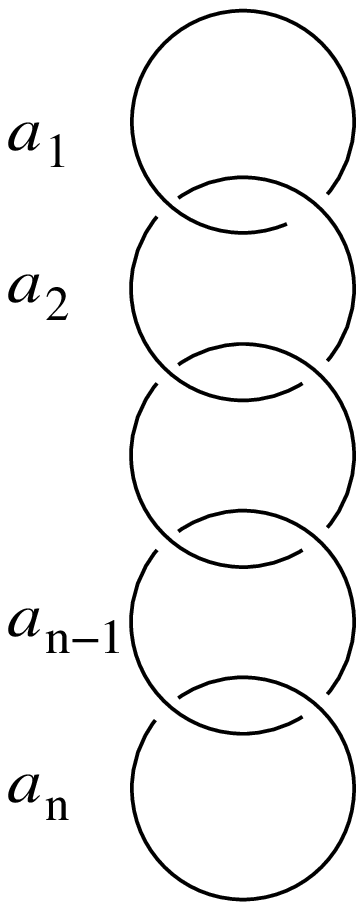}}\ ,
\sigma\right) = \ \raisebox{-25mm}{\epsfbox{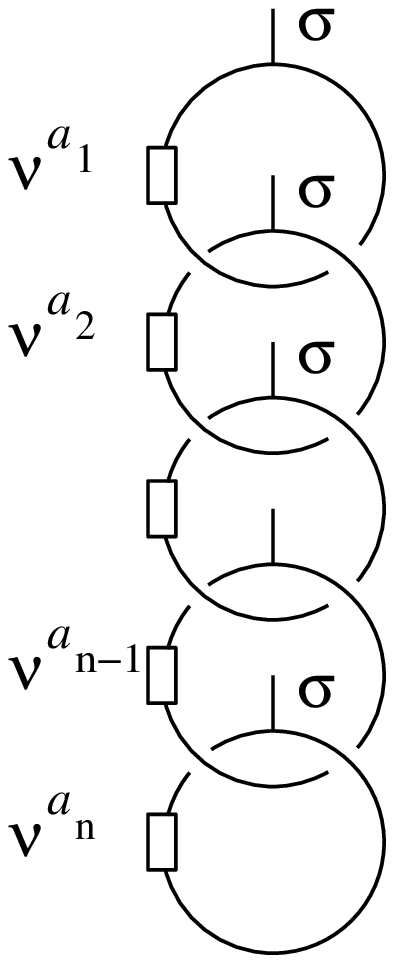}} \ .
\end{equation}
The surgery at $L$ gives the lens space $L(p,q)$, where prime to each other
$p,q\in\Z$ satisfy
\[ \frac pq = a_n -
\cfrac1{a_{n-1}-
 \cfrac1{\ddots-
  \cfrac1{a_2-
   \cfrac1{a_1
}}}}
\]
(see Rolfsen \cite{Rol:lin}). The right hand side of \eqref{chain}
can be presented through the maps $S,T:F\to F$ as
\begin{align*}
\tau(L,\si) &= \e(T^{a_n} S T^{a_{n-1}} S \dots T^{a_2} S T^{a_1} S(1)) \\
&= \<1, T^{a_n} S T^{a_{n-1}} S \dots T^{a_2} S T^{a_1} S(1)\> \\
&= \<\,^tS \,^tT^{a_1} \,^tS \,^tT^{a_2} \dots \,^tS \,^tT^{a_{n-1}}
\,^tS (\nu^{a_n}) ,1\> \\
&= \int'( \,^tT^{a_1} \,^tS \,^tT^{a_2} \dots \,^tS \,^tT^{a_{n-1}}
\,^tS (\nu^{a_n}) )
\end{align*}
The last formula uses only the maps $^tT$, $^tS={}^tS_-(\si)$ in $U$,
determined by \eqref{tT(u)} and \eqref{tS-sigma}. We can also use
$^tS_+(\si)$ (see \eqref{tS+}) instead of $^tS$.

In particular case $n=1$, $a_1=0$ we get $\tau(L,\si) = \e(\mu) = \int'1$,
which vanishes for all $u_q(\g)$, hence, $\tau(S^2\times S^1)=0$.
This contrasts sharply with the case of a perfect modular semisimple
$\CC$, where $\e(\mu) =\sqrt{\sum_i (\dim_\CC X_i)^2}$ is invertible
(see \cite{Lyu:pre,Tur:q3}). Here $X_i$ runs over the set of isomorphism
classes of simple objects of $\CC$. Their categorical dimensions
\[ \dim_\CC X_i = \bigl( I @>\coev>> X_i\tens\pti X_i @>1\tens u_0^2>>
X_i\tens X_i\pti @>\ev>> I \bigr) \]
are real numbers if the ground field is $\End I=\C$ \cite{Lyu:pre},
so $\e(\mu)$ is positive.

Finally, $s(L)$ is the signature of the linking matrix
\[
\begin{pmatrix}
a_1& 1 \\
 1 &a_2&  1   &      &   0 \\
   & 1 &\ddots&\ddots \\
   &   &\ddots&\ddots&   1 \\
   & 0 &      &    1 &a_{n-1}& 1 \\
   &   &      &      &   1   &a_n
\end{pmatrix}
\]
which gives $\tau(L(p,q))$.

\subsection{The Hennings invariant}
Let $H$ be a finite dimensional 3-modular Hopf algebra (see
\thmref{3-modthm}). Construct a 3-manifold invariant $\tau(M)$ taking
the 3-modular category $\CC=H\modul$ as data. Calculating $\tau(L,\si)$
for some link $L$ we get an expression involving $R$-matrices
(as many as there are crossings in $\bar D_L$), elements $\si$ and
the powers of $\kappa$ (as many as there are components in $L$) under
the sign of counity. It turns out that the result of calculation
coincides with the Hennings invariant \cite{Hen:3} defined in
unoriented setting by Kauffman and Radford~\cite{KauRad:3} up to
change of conventions. Indeed, the calculation of $\tau(L,\si)$
can be performed using the graphical rules of Hennings--Kauffman--Radford
as follows:

\noindent-- change all crossings in $\bar D_L$ to the composite
of $R$-matrix and the permutation
\[
\unitlength=0.4mm
\linethickness{0.4pt}
\begin{picture}(295,30)
\put(0,0){\line(1,1){30}}
\put(0,30){\line(1,-1){12}}
\put(18,12){\line(1,-1){12}}
\put(40,15){\vector(1,0){15}}
\put(85,30){\line(1,-1){30}}
\put(115,30){\line(-1,-1){30}}
\put(93,22){\circle*{2}}
\put(107,22){\circle*{2}}
\put(83,18){\makebox(0,0)[cc]{$R'$}}
\put(117,18){\makebox(0,0)[cc]{$R''$}}
\put(136,13){\makebox(0,0)[cc]{,}}
\put(160,30){\line(1,-1){30}}
\put(190,30){\line(-1,-1){12}}
\put(160,0){\line(1,1){12}}
\put(200,15){\vector(1,0){15}}
\put(240,30){\line(1,-1){30}}
\put(270,30){\line(-1,-1){30}}
\put(248,8){\circle*{2}}
\put(262,8){\circle*{2}}
\put(231,10){\makebox(0,0)[cc]{$\gamma(R')$}}
\put(275,10){\makebox(0,0)[cc]{$R''$}}
\put(295,13){\makebox(0,0)[cc]{;}}
\end{picture}
\]
\noindent-- add to each component such power of the morphism
$ u^2_0 = v^2_0 \cdot \kappa = \kappa \cdot v^2_0$ or its inverse
$ u^{-2}_0 = v^{-2}_0 \cdot \kappa^{-1}$
\[
\unitlength=0.40mm
\linethickness{0.4pt}
\begin{picture}(120,31)
\put(28,9){\oval(16,20)[lt]}
\put(28,21){\oval(16,20)[lb]}
\put(28,15){\oval(8,8)[r]}
\put(2,15){\makebox(0,0)[cc]{$v_0^2\ =$}}
\put(39,13){\makebox(0,0)[cc]{,}}
\put(87,15){\makebox(0,0)[cc]{$v_0^{-2}\ =$}}
\put(120,29){\line(0,-1){8}}
\put(120,9){\line(0,-1){8}}
\put(112,21){\oval(16,20)[rb]}
\put(112,9){\oval(16,20)[rt]}
\put(112,15){\oval(8,8)[l]}
\put(20,9){\line(0,-1){8}}
\put(20,21){\line(0,1){8}}
\end{picture}
\]
that after cancellations of $v^2_0$'s with $v^{-2}_0$'s the diagram
becomes a disjoint union of unknots;

\noindent-- slide all elements put on the diagram to the left of
the chosen points $e_i$ via the rule
\[
\unitlength=0.40mm
\linethickness{0.4pt}
\begin{picture}(256,20)
\put(20,20){\oval(20,40)[b]}
\put(10,12){\circle*{2}}
\put(4,12){\makebox(0,0)[cc]{$x$}}
\put(45,10){\vector(-1,0){7}}
\put(45,10){\vector(1,0){7}}
\put(70,20){\oval(20,40)[b]}
\put(80,12){\circle*{2}}
\put(94,12){\makebox(0,0)[cc]{$\gamma(x)$}}
\put(106,8){\makebox(0,0)[cc]{,}}
\put(146,0){\oval(20,40)[t]}
\put(156,8){\circle*{2}}
\put(163,8){\makebox(0,0)[cc]{$x$}}
\put(177,10){\vector(-1,0){7}}
\put(177,10){\vector(1,0){7}}
\put(196,8){\makebox(0,0)[cc]{$\gamma(x)$}}
\put(218,0){\oval(20,40)[t]}
\put(208,8){\circle*{2}}
\put(256,8){\makebox(0,0)[cc]{,\ \ \ $x\in H$ .}}
\end{picture}
\]

Now each component looks alike
\[
\unitlength=0.40mm
\linethickness{0.4pt}
\begin{picture}(24,31)
\put(17,12){\oval(14,24)[]}
\put(10,16){\circle*{2}}
\put(10,8){\circle*{2}}
\put(17,24){\line(0,1){7}}
\put(22,30){\makebox(0,0)[cc]{$\sigma$}}
\put(2,16){\makebox(0,0)[cc]{$x_1$}}
\put(2,8){\makebox(0,0)[cc]{$x_n$}}
\end{picture}
\]
with $x_1,\dots,x_n\in H$ attached to it, and its contribution is
\[ \e(\und{x_n\dots x_1\tens1} (\si)) =
\e(\<x_n\dots x_1, \si\one\> \si\two) =  \<x_n\dots x_1, \si\> \]
(see \secref{3-modcase}). We can view $\si$ as the left integral
$\si:H\to k$ on the algebra $H$. So
\[ \tau(L,\si) = \sum \si( x_n\dots x_1) \dots \si(z_m\dots z_1) \]
where the arguments are either powers of $\kappa$ or come from
$R$-matrices acted upon by some powers of the antipode. This is the
knot invariant used in the definition  of Hennings invariant
\cite{Hen:3} up to change of the sign of the braiding.


\bibliographystyle{amsplain}


\end{document}